\documentclass[11pt]{cernrep}
\usepackage{graphicx}
\usepackage{floatflt}
\usepackage{color}
\usepackage{colortbl}

\newcommand{\Pt}{{P_t}}
\newcommand{\dphi}{\Delta\phi}
\newcommand{\phigj}{\phi_{(\gamma,~jet)}}

\newcommand{\Ptgj}{$\Pt^{\gamma}$ and $\Pt^{Jet}$~~}
\newcommand{\ptgj}{$~\Pt^{\gamma}-\Pt^{Jet}~$}

\newcommand{\la}{\langle}
\newcommand{\ra}{\rangle}
\newcommand{\gpj}{~``$\gamma+Jet$''~}
\newcommand{\gpp}{~``$\gamma+parton$''~}
\newcommand{\rrr}{\to} 

\newcommand{\pth}{\hat{p}_{\perp}^{\;min}}
\newcommand{\Db}{\Pt(O\!+\!\eta>4.2)}
\newcommand{\Ptg}{\Pt^{\gamma}}
\newcommand{\ptg}{$\Pt^{\gamma}$}
\newcommand{\Fptgj}{(\Pt^{\gamma}\!-\!\Pt^{Jet})/\Pt^{\gamma}}

\newcommand{\coltab}{0.69}

\newcommand{\aaa}{\hspace*{.39cm}}

\newcommand{\hmm}{\hspace*{-1.3mm}}

\newcommand{\Gvc}{\footnotesize{$(GeV/c)$} }

\newcommand{\gJ}{(\Pt^{\gamma}\!-\!\Pt^{J})/\Pt^{\gamma}}
\newcommand{\gpart}{(\Pt^{\gamma}\! - \! \Pt^{part})/\Pt^{\gamma}}
\newcommand{\Jpart}{(\Pt^{J}\! - \! \Pt^{part})/\Pt^{J}}

\newcommand{\sgmgj}{\sigma(Db[\gamma,J])}
\newcommand{\sgmgp}{\sigma(Db[\gamma,part])}

\newcommand{\lt}{\!<\!}
\newcommand{\gt}{\!>\!}

\suppressfloats[!]

\def\baselinestretch{1.0}

\begin{document}


\vskip2cm

%
\noindent                                                                      
{\bf \Large \gpj process application for setting the absolute  scale of \\[4pt]
\hspace*{15mm} jet energy and determining the gluon distribution \\[4pt]
\hspace*{45mm} at the Tevatron Run II.}

\thispagestyle{empty}

\vskip3cm

\centerline{\sf\large D.V.~Bandurin, N.B.~Skachkov}
~\\[-3mm]

\centerline{\it Laboratory of Nuclear Problems}
~\\[-5mm]
\centerline{\it Joint Institute for Nuclear Research, Dubna, Russia}
~\\[15mm]
\centerline{\sf\large D0 Note 3948}

\vskip3cm
\hspace*{6cm}{\bf Abstract}\\[15pt]
\noindent
We study the effect of application of new set of cuts, proposed in our 
previous works, on the improvement of accuracy of the jet energy calibration with 
``$p\bar{p}\to\gamma+Jet+X$'' process at Tevatron.
Monte Carlo events produced by the PYTHIA 5.7 generator are used for this aim.
The selection criteria for \gpj event samples that would provide a good balance of
$\Pt^{\gamma}$ with $\Pt^{Jet}$  and would allow to reduce the background are described.
The distributions of these events over \ptg ~and $\eta^{Jet}$
are presented. The features of \gpj events in the central calorimeter region of
the D0 detector ($|\eta|\lt0.7$) are exposed.
The efficiency of the cuts used for background suppression is demonstrated.

It is shown that the samples of \gpj events, gained with the cuts for the jet energy calibration,
may have enough statistics for determining the gluon distribution inside a proton in the region 
of $x\geq10^{-3}$ and of $Q^2$ by one order higher than that studied at HERA.

\newpage

\tableofcontents
\thispagestyle{empty}

\newpage

\setcounter{page}{1}

\section{INTRODUCTION.} 

Setting an absolute energy scale for a jet, detected mostly by hadronic and
electromagnetic calorimeters (HCAL and ECAL), is an important task for any of
$p\bar{p}$ and $pp$  collider experiments (see e.g. [1--8]). 

The main goal of this work is to find out the selection criteria for ``$p\bar{p}\to\gamma+Jet+X$''
events (we shall use in what follows the abreviation \gpj for them) that would lead to 
the most precise determination of the transverse momentum of a jet (i.e. $\Pt^{Jet}$) via assigning a
photon $\Ptg$ to a signal produced by a jet. Our study is based on the \gpj events generated by 
using PYTHIA 5.7 \cite{PYT}. Their analysis was done on the ``particle level'' (in the terminology of [1]), 
i.e. without inclusion of the detector effects.
The information provided by this generator is analyzed to track
starting from the parton level (where parton-photon balance is supposed to take place in a case of 
initial state radiation absence) all possible sources that may lead to the \ptgj disbalance 
in a final state. We use here the methods applied in \cite{9}--\cite{BKS_GLU}  
(see also \cite{DMS}) and \cite{GMS}, \cite{GMS_NN} for analogous task at LHC energy.
The corresponding cuts on physical variables,  introduced in \cite{9}--\cite{BKS_P5},
are applied here. Their efficiency is estimated at the particle level of simulation
at Tevatron energy with account of D0 geometry. The results of further analysis of \gpj events planned to be 
done at the level of the full event reconstruction after the detector responce simulation with
GEANT-3 \cite{GEA} based  package  D0GSTAR \cite{D0GSTAR}
will be presented in our following publications.

We consider here the case of the Tevatron Run~II luminosity 
$L=10^{32}~ cm^{-2}s^{-1}$. It will be shown below that its value is quite sufficient for selecting 
the event samples of large enough volume for application of much more restrictive cuts as well as of new 
physical variables introduced in \cite{9}--\cite{BKS_P5}.
Our aim is to select the samples of topologically clean
\gpj events with a good balance of $\Ptg$ and $\Pt^{Jet}$ and to use them for further 
modeling of the jet energy calibration procedure within D0GSTAR. In this way one can estimate
a jet energy calibration accuracy that can be achieved with the proposed cuts in the experiment.

Section 2~ is a short introduction into the physics connected with the discussed problem.
General features of \gpj processes at Tevatron energy are presented here.
We review the possible sources of the $\Pt^{\gamma}$ and $\Pt^{Jet}$ disbalance and the ways of
selecting those events where this disbalance has a minimal value
on the particle level.

In Section 3.1 the definitions  are given for the transverse momenta of
different physical objects that we have introduced as a part of \gpj production event
and that we suppose to be important for studying the physics connected with a jet calibration
procedure. These values of transverse momenta 
enter into the $\Pt$-balance equation that reflects the
total $\Pt$ conservation law for the $p\bar{p}$-collision event as a whole.

Section 3.2 describes the criteria we have chosen to select \gpj events
for the jet energy  calibration procedure. The ``cluster'' (or mini--jet) suppression criterion 
($\Pt^{clust}_{CUT}$) which was formulated in an evident form in our previous publications 
\cite{9}--\cite{BKS_GLU} is used here
\footnote{We use here, as in \cite{BKS_P1}--\cite{BKS_GLU}, the LUCELL subroutine from PYTHIA as well as
two jetfinders UA1 and UA2 from the CMS program of fast simulation CMSJET \cite{CMJ} for defining jets 
in an event.}. 
(Its important role for selection of events with a good balance of \Ptgj
will be illustrated in Sections 5--8.)
\footnote{The analogous  third jet cut thresholds 
$E_T^3$ (varying from 20 to 8 $GeV$) for improving a single jet energy resolution
in di-jet events were used in \cite{Bert}.}
These clusters have a physical meaning of a part 
of another new experimentally measurable quantity,
introduced in \cite{9}--\cite{BKS_GLU} for the first time, namely,
the sum of $\vec{\Pt}$ of those particles that are {\it out} of the \gpj system
(denoted as $\Pt^{out}$) and are detectable
in the whole pseudorapidity $\eta$ region covered by the detector ($|\eta|\lt4.2$ for D0).
The vector and scalar forms of the total $\Pt$ balance equation, used for the $p\bar{p}-$event
as a whole, are given in Sections 3.1 and 3.2 respectively.

Another new thing is a use of a new physical object, proposed also in \cite{9}--\cite{BKS_GLU}
and named an ``isolated jet''. This jet is contained in the cone of radius $R=0.7$ in the $\eta-\phi$ 
space and it does not have any noticeable $\Pt$ activity in some ring around.
The width of this ring is taken to be of $\Delta R=0.3$ (or approximately of the width of 3 calorimeter towers).
In other words, we will select a class of events having  a total $\Pt$ activity inside 
the ring around this ``isolated jet'' within $3-5\%$ of jet $\Pt$.
(It will be shown in Sections 6, 7 and Appendices 2--5  that the number of events
with such a clean topological structure would not be small at Tevatron energy.)

Section 4 is devoted to the estimation of the size of
the non-detectable neutrino contribution to $\Pt^{Jet}$.
The correlation of the upper cut value, imposed onto $\Pt^{miss}$, with the mean value 
of $\Pt$ of neutrinos belonging to the jet $\Pt$, i.e. $\la \Pt_{(\nu)}^{Jet}\ra$,
is considered. The detailed results of this section are presented
in the tables of Appendix 1. They also include 
the ratios of the gluonic events $qg\to q+\gamma$ containing the information about
the gluon distribution inside a proton. In the same tables
the expected number of events
(at $L_{int}= 300 ~pb^{-1}$) having charm ($c$) and beauty ($b$) quarks 
in the initial state of the gluonic subprocess are also given.

Since the jet energy calibration is rather a practical than an academic task,
in all the following sections we present the rates obtained with the cuts 
varying from strict to weak because their choice would be
a matter of step-by-step statistics collection during the data taking.

Section 5 includes the results of studying the dependence of the initial state radiation (ISR) $\Pt$-spectrum 
 on the cut imposed on the clusters $\Pt$ ($\Pt^{clust}_{CUT}$) and on the angle 
between the transverse momenta vectors of a jet and a photon.
We also present the rates for four different types of \gpj events,
in which jet fits completely in one definite region 
of the calorimeter: in Central Calorimeter (CC) with $|\eta|\lt0.7$ or in
Intercryostat Calorimeter (IC) with $0.7<|\eta|\lt1.8$ or in End Calorimeter (EC) with 
$1.8\lt|\eta|\lt2.5$ or, finally, in Forward Calorimeter (FC) with $2.5\lt|\eta|\lt4.2$.

Starting with Section 6 our analysis is concentrated on  the ``$\gamma+1~jet$''
events having a jet entirely contained within the central calorimeter region.
The dependence of spectra of different physical variables
\footnote{mostly those that have a strong influence on the \ptgj balance in an event.}
(and among them those appearing in the $\Pt$ balance equation of event as a whole) 
on $\Pt^{clust}_{CUT}$, as well as the dependence  on it of the spatial 
distribution of $\Pt$ activity inside a jet as well as outside it is shown in Figs.~8--11.

The dependence of the number of events (for $L_{int}=300~pb^{-1}$) on $\Pt^{clust}_{CUT}$ 
as well as the dependence on it of the fractional $\Fptgj$ disbalance
is studied in Section 7. The details of this study are presented in the tables of
Appendices 2--5 that together with the corresponding Figs.~12--18  can serve to justify
the variables and cuts introduced in Section 3. 
Figs.~15--18 as well as Tables 13--18 of Appendices 2 -- 5 demonstrate the influence of the jet isolation
criterion. The impact of $\Pt^{out}_{CUT}$ on the fractional $\Fptgj$ disbalance is shown in 
Figs.~19 and 20.


In Section 8 we present the estimation of the efficiency of background suppression
(that was one of the main guidelines to establish the selection rules proposed in Section 3) for
different numerical values of cuts. 

The importance of the simultaneous use of the above-mentioned new parameters 
$\Pt^{clust}_{CUT}$ and $\Pt^{out}_{CUT}$ and also of the ``isolated jet'' criterion for background 
suppression (as well as for improving the value of the \Ptgj balance)
is demonstrated in Tables 14--17 of Section 8 as well as in the tables of Appendix 6
that show  the dependence of selected events on $\Pt^{clust}_{CUT}$ and $\Pt^{out}_{CUT}$
for various $\Ptg$ intervals. 
The tables of Appendix 6 include  
the fractional disbalance values $\Fptgj$ that are found with an additional (as compared with tables of Appendix 2--5) 
account of the $\Pt^{out}$ cut. In this sense the tables of Appendix 6 contain the final (and {\it first main}) 
result (as they include the background contribution) 
of our study of setting the absolute scale of  the jet energy at the particle level defined by generation with 
PYTHIA.

In Section 9 we show the tables and some plots that demonstrate a possible influence of the
intrinsic transverse parton momentum $k_t$ parameter variation
(including, as an illustration, some extreme $k_t$ values) on the \ptgj disbalance.

Section 10 contains the {\it second main} result of our  study of \gpj events 
at the Tevatron energy. Here we investigate the possibility of using the same sample
of the topologically clean \gpj events, obtained with the described cuts,
for determining the gluon distribution in a proton (as it was done earlier for LHC energy in
\cite{BKS_GLU}, \cite{DMS}). The kinematic plot presented
here shows what a region of $x$ and $Q^2$ variables 
(namely: $10^{-3}\leq x\leq 1.0$ and $1.6\cdot 10^{3}\leq Q^2\leq2\cdot10^{4} ~(GeV/c)^2$)
can be covered at Tevatron energies, with a sufficient number of events for this aim. The comparison
with the kinematic regions covered by other experiments
where parton distributions were studied is also shown in the same plot (see Fig.~29).

About the Summary. We tried to write it in a way allowing a dedicated reader,
who is interested in result rather than in method, to pass directly to it after this sentence.

Since the results presented here were obtained with the PYTHIA simulation,
we are planning to carry out analogous estimations with another event generator like HERWIG, for example,
in subsequent papers.

\section{GENERALITIES OF THE \gpj PROCESS.}
\it\small
\hspace*{9mm}
The useful variables are introduced for studying the effects of its on initial and final state
radiation basing on the simulation in the framework of PYTHIA. Other effects of non-perturbative nature
like primordial parton $k_{~t}$ effect, parton-to-jet hadronization that may
lead to \ptgj disbalance within the physical models used in PYTHIA are also discussed.
\rm\normalsize
\vskip4mm

\subsection{Leading order picture.}        

\setcounter{equation}{1}

The idea of absolute jet energy scale setting (and hadronic calorimeter (HCAL)
calibration) by means of the physical process ``$p\bar{p}\rrr \gamma+Jet+X$''
was realized many times in different experiments
(see [1--8] and references therein).
It is based on the parton picture where two partons ($q\bar{q}$ or
$qg$), supposed to be moving in different colliding nucleons with
zero transverse momenta
(with respect to the beam line), produce a photon called the ``direct photon''.
This process is described by the leading order (LO) Feynman
diagrams shown in Fig.~1 (for the explanation of the numeration of lines see
Section 2.2) for the ``Compton-like'' subprocess (ISUB=29 in PYTHIA) \\[-20pt]
\begin{eqnarray}
\hspace*{6.04cm} qg\to q+\gamma \hspace*{7.5cm} (1a)
\nonumber
\end{eqnarray}
\vspace{-4mm}
and for the ``annihilation'' subprocess (ISUB=14)
~\\[-8pt]
\begin{eqnarray}
\hspace*{6.02cm} q\overline{q}\to g+\gamma.  \hspace*{7.4cm} (1b)
\nonumber
\end{eqnarray}

As the initial partons were supposed to have zero transverse momenta,
$\Pt$ of the ``$\gamma$+parton'' system produced in the final state
should be also equal to zero, i.e. one can write the following $\Pt$ balance equation for 
photon and final parton \\[-15pt]
\begin{eqnarray}
\vec{\Pt}^{\gamma+part}=\vec{\Pt}^{\gamma}+\vec{\Pt}^{part} = 0.
\end{eqnarray}
Thus, one could expect that the transverse momentum
of the jet produced by the final state parton ($q$ or $g$), having 
$\vec{\Pt}^{part}=-\vec{\Pt}^{\gamma}$, will
be close in magnitude, with a reasonable precision, 
to the transverse momentum of the final state photon,
 i.e. $\vec{\Pt}^{Jet}\approx-\vec{\Pt}^{\gamma}$.

It allows the absolute jet energy scale to be determined 
(and the HCAL to be calibrated) in the experiments
with a well-calibrated electromagnetic calorimeter (ECAL). To put it simpler,
one can assign to the part of the jet transverse energy $E_t^{Jet}$ 
deposited in the HCAL the value of the difference between the values of
the transverse energy deposited in the ECAL in the photon direction
(i.e. $E_t^{\gamma}$) and  the transverse energy deposited in the ECAL 
in the jet direction. 
~\\[-6.1cm]
\begin{center}
\begin{figure}[h]
  \hspace{10mm} \includegraphics[width=13cm,height=8.3cm]{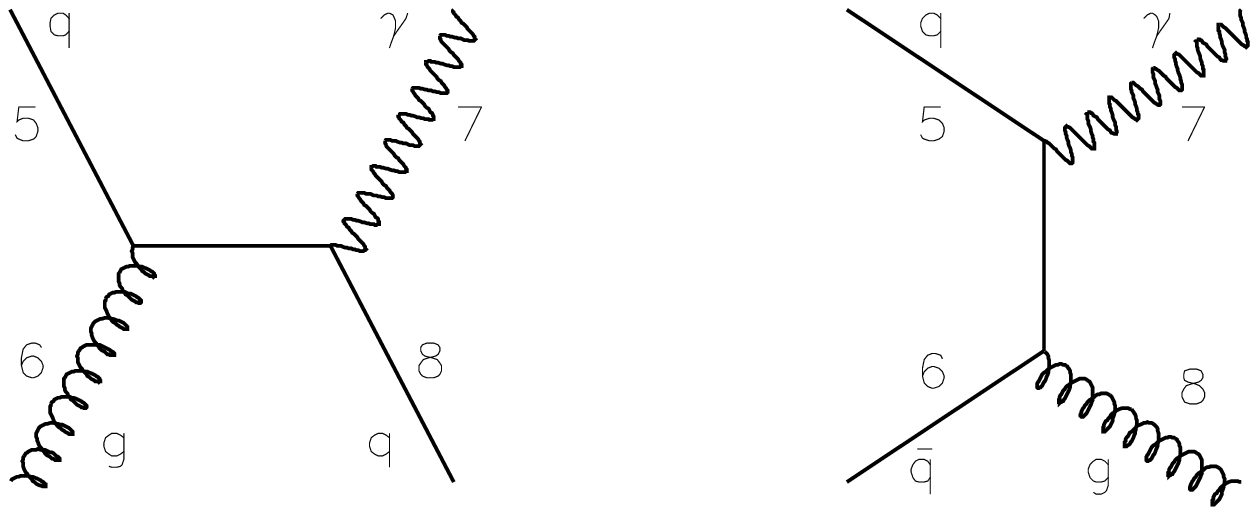} 
\vskip-11mm
\hspace*{42mm} {\small(a)} \hspace*{47mm} {\small(b)}
 \vskip-2mm
\caption{\hspace*{0.0cm} Some of the leading order Feynman diagrams for direct
photon production.} \label{fig1:LO}
\end{figure}
\vskip-10.5mm
\end{center}

\subsection{Initial state radiation.}                           

Since we believe in the perturbation theory, the leading
order (LO) picture described above is expected to be dominant and
to determine the main contribution to the cross section. The Next-to-Leading Order
(NLO) approximation (see some of the NLO diagrams in Figs.~2 and 4) introduces some
deviations from a rather straightforward LO-motivated idea of jet energy calibration.
A gluon ~radiated in the initial state (ISR), as it is seen from Fig.~2, 
can have its own non-zero transverse momentum
$\Pt^{gluon}\equiv \Pt^{ISR}\neq 0$. Apart of a problem of appearance of extra jets
(or mini-jets and clusters), that will be discussed in what follows, it leads
to the non-zero transverse momenta of partons that appear in the initial state
of fundamental $2\rrr2$ QCD subprocesses (1a) and (1b). As a result of
the transverse momentum conservation there arises a disbalance between 
the transverse momenta of a photon $\Pt^{\gamma}$ and of a parton $\Pt^{part}$ produced 
in the fundamental $2\to 2$ process $5+6\to 7+8$ shown in Fig.~2 (and in Fig.~3)
and thus, finally, the disbalance between $\Ptg$ and $\Pt$ of a jet produced by this parton.
~\\[-6cm]
\begin{center} \begin{figure}[h]
  \hspace{15mm} \includegraphics[width=13cm,height=8.3cm,angle=0]{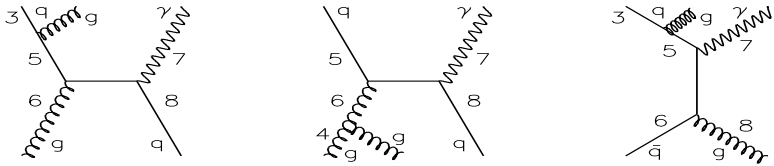}
  \vspace{-12mm}
  \caption{\hspace*{0.0cm} Some of Feynman diagrams of direct photon
production including gluon radiation in the initial state.}
    \label{fig2:NLO}
  \end{figure}
\end{center}
\vspace{-1.0cm}

Following \cite{BKS_P1}--\cite{BKS_P5} and \cite{GPJ} we choose  the modulus of the vector sum of
the transverse momentum vectors $\vec{\Pt}^{5}$ and $\vec{\Pt}^{6}$
of the incoming  into $2\rrr 2$ fundamental QCD subprocesses $5+6\to 7+8$
partons (lines 5 and 6 in Fig.~2) and the sum of their modulus
as two quantitative measures \\[-5pt]
\vspace{-7mm}
\begin{eqnarray}
\Pt^{5+6}=|\vec{\Pt}^5+\vec{\Pt}^6|, \qquad \Pt{56}=|\Pt^5|+|\Pt^6|
\end{eqnarray}
to estimate the $\Pt$ disbalance caused by ISR
\footnote{The variable $\Pt^{5+6}$ was used in analysis in \cite{9}--\cite{BKS_P1}.}. 
The modulus of the vector sum \\[-5mm]
\begin{eqnarray}
\Pt^{\gamma+Jet}=|\vec{\Pt}^{\gamma}+\vec{\Pt}^{Jet}|
\end{eqnarray}
was also used as an estimator of the final state  $\Pt$
disbalance in the \gpj system in \cite{BKS_P1}--\cite{BKS_P5}.

The numerical notations in the Feynman diagrams  (shown in Figs. 1 and 2)
and in formula (3)  are chosen to be in correspondence with those
used in the PYTHIA event listing for description of the parton--parton subprocess
displayed schematically in Fig.~3. The ``ISR'' block describes the initial
state radiation process that can take place before the fundamental
hard $2\to 2$ process.
\begin{center}
  \begin{figure}[h]
  \vspace{-0.8cm}
   \hspace{3cm} \includegraphics[width=10cm,height=5cm,angle=0]{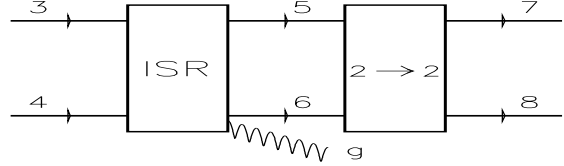}
  \vspace{-2.5cm}
 \caption{\hspace*{0.0cm} PYTHIA ``diagram'' of~ $2\to2$ process (5+6$\to$7+8)
following the block (3+4$\to$5+6) of initial state radiation (ISR), drawn here to illustrate the PYTHIA 
event listing information.}
    \label{fig4:PYT}
  \end{figure}
\end{center}

~\\[-2.3cm]

\subsection{Final state radiation.}                            
\begin{center}
\begin{figure}[htbp]
  \hspace{15mm} \includegraphics[width=13cm,height=8.4cm,angle=0]{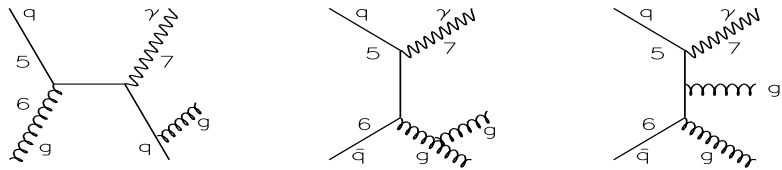}
  \vspace{-12mm}
  \caption{\hspace*{0.0cm} Some of Feynman diagrams of direct photon
production including gluon radiation in the final state.}
    \label{fig4:NLO}
  \end{figure}
\end{center}

Let us consider fundamental subprocesses in which there is
no initial state radiation but instead  final state radiation
(FSR) takes place. These subprocesses are described in the quantum
field theory by the NLO diagrams like those shown in Fig.~\ref{fig4:NLO}. It
is clear that appearance of an extra gluon leg in the final
state may lead to appearance of two (or more) jets or an
intense jet and a weaker jet (mini-jet or cluster) in an event as it happens 
in the case of ISR described above.
So, to suppress FSR (manifesting itself as some extra
jets or clusters) the same tools as for reducing ISR should be
used. But due to the string model of fragmentation used in
PYTHIA it is much more difficult to deduce basing on the PYTHIA event listing information
the variables (analogous to (3) and (4)) to describe the disbalance between 
$\Pt$ of a jet parent parton and $\Ptg$. That is why, keeping in mind a close
analogy of the physical pictures of ISR and FSR (see
Figs.~\ref{fig2:NLO} and \ref{fig4:NLO}), we shall concentrate 
in the following sections on the initial state radiation
supposing it to serve in some sense as a quantum field theory perturbative
model of the final state radiation mechanism.
\\[-6cm]
\vspace{-1.0cm}

\subsection{Primordial parton $k_t$ effect.}                   

Now after considering the disbalance sources connected with the
perturbative corrections to the leading order diagrams let us
mention the physical effects of the non-perturbative nature. Thus,
a possible non-zero value of the intrinsic transverse parton
velocity inside a colliding proton may be another source of the
$\Pt^{\gamma}$ and $\Pt^{part}$ disbalance in the final state.
Nowadays this effect can be described mainly in the
phenomenological way. Its reasonable value is supposed to lead to
the value $k_t \leq \,1.0 ~GeV/c$. Sometimes in the literature the total
effect of ISR and of the intrinsic parton transverse momentum is denoted by a common
symbol ``$k_t$''. Here we follow the approach and the
phenomenological  model used in PYTHIA where these two sources of the \Ptgj
disbalance, having different nature, perturbative and non-perturbative, 
can be switched on separately by different keys (MSTP(61) for ISR and PARP(91), PARP(93), MSTP(91) for
intrinsic parton transverse momentum $k_t$). In what follows we shall keep the value 
of $k_t$ mainly to be fixed  by the PYTHIA default value $\langle k_t \rangle=0.44~ GeV/c$. 
The dependence of the disbalance between $\Pt^{\gamma}$ and $\Pt^{Jet}$
on possible variation of $k_t$ will be discussed in detail in Section 9.
 The general conclusion from here is that any variation of $k_t$ within
 reasonable boundaries (as well as slightly beyond them) does not produce a
 large effect in the case when the initial state radiation is
 switched on. The latter makes a dominant contribution.

 \subsection{Parton-to-jet hadronization.}                      

Another non-perturbative effect that leads to the \ptgj disbalance
is connected with hadronization (or fragmentation into hadrons) of the parton
produced in the fundamental $2\to 2$ subprocess into a jet. The  hadronization of the parton into a jet 
is described in PYTHIA within the Lund string fragmentation model. 
The mean values of the fractional $\Pt^{Jet}-\Pt^{parton}$ disbalance
will be presented in the tables of Appendices 2 -- 5 for three different
jetfinders. As it will be shown in Section 7 (see also tables of Appendices 2--5)
the hadronization effect has approximately the contribution into \ptgj disbalance of the same size 
as that of ISR.

\section{CHOICE OF MEASURABLE PHYSICAL VARIABLES FOR THE \gpj
PROCESS AND THE CUTS FOR BACKGROUND REDUCTION.}               

\it\small
\hspace*{9mm}
The classification of different physical objects that participate in \gpj events and that may give
a noticeable contribution into the total $\Pt$-balance in the event as a whole is done.

Two new physical observables, namely, $\Pt$ of a cluster and $\Pt$ of all detectable particles beyond 
\gpj system, as well as the definion of isolated jet, proposed for studying \ptgj disbalance in 
\cite{9}--\cite{BKS_P5}, are discussed.

The selection cuts for physical observables of \gpj events are given.

The  $\Pt$-balance equation for the event as a whole is written in scalar form that allow to express
the \ptgj disbalance in terms of the considered physical variables.
\rm\normalsize
\vskip4mm

Apart from (1a) and (1b), other QCD subprocesses with large cross sections, 
by orders of magnitude larger than the
cross sections of (1a) and (1b), can also lead to high $\Pt$ photons
and jets in final state. So, we face the problem of selecting
signal \gpj events from a large QCD background.
Here we shall discuss the choice of physical variables that would be
useful, under some cuts on their values, for separation of the
desirable processes with direct photon (``$\gamma^{dir}$'') 
from the background events. The possible ``$\gamma^{dir}-$candidate''
may originate from the $\pi^0,~\eta,~\omega$ and $K^0_s$ meson decays 
\cite{GMS}, \cite{GMS_NN} or may be caused by a bremsstrahlung photon 
or by an electron (see Section 8).

We take the D0 ECAL size to be limited by$|\eta|\!\!\leq \!\!2.5$ and 
the calorimeter to be limited by $|\eta| \leq 4.2$ and
to consist of CC, IC, EC, FC parts, where $\eta = -ln~(tan~(\theta/2))$ is a pseudorapidity
defined in terms of a polar angle $\theta$ counted from the beam line. In the plane
transverse to the beam line the azimuthal angle $\phi$ defines the
directions of $\vec{\Pt}^{Jet}$ and $\vec{\Pt}^{\gamma}$. \\

\subsection{Measurable physical observables and the $\Pt$ vector balance equation.}

In $p\bar{p}\to \gamma + Jet + X$ events we are going to study
the main physical object will be a high $\Pt$ jet to be detected
in the $|\eta|\lt4.2$ region and a direct photon registered by the
ECAL up to $|\eta|\lt2.5$. In these events there will be a set of
particles mainly caused by beam remnants, i.e. by spectator parton
fragments, that are flying mostly in the direction of a non-instrumented
volume ($|\eta|>4.2$) in the detector. Let us denote the total transverse
momentum of these non-observable particles ($i$) as \\[-4mm]
\begin{equation}
\sum\limits_{i
\in |\eta| \gt4.2} \vec{\Pt}^i \equiv \vec{\Pt}^{|\eta| \gt4.2}.  \label{eq:sel1}
\end{equation} 

Among the particles  with $|\eta|\lt4.2$ there may also be neutrinos.
We shall denote their total momentum as
\begin{equation}
\sum\limits_{i \in |\eta|\lt4.2} \vec{\Pt}_{(\nu)}^i
 \equiv \vec{\Pt}_{(\nu)}.
\label{eq:sel2}
\end{equation}
\vspace{-2mm}

\noindent
The sum of the transverse momenta of these two kinds of non-detectable particles
will be denoted as $\Pt^{miss}$
\footnote{This value is a part of true missing $\Pt$ in an experiment that includes the detector
effects (see [1, 2]).}:
\vspace{-3mm}
\begin{eqnarray}
 \vec{\Pt}^{miss} = \vec{\Pt}_{(\nu)} + \vec{\Pt}^{|\eta|>4.2}.
\end{eqnarray}

\vspace{-2mm}

A high-energy jet may also contain neutrinos that may carry
part of the total jet energy and of $\Pt^{Jet}$. The average
values of these neutrino parts can be estimated from simulation.

From the total jet transverse momentum $\vec{\Pt}^{Jet}$  we shall
separate the part that can be measured in the detector, i.e. in
the ECAL+HCAL calorimeter system and in the muon system. Let us denote
this detectable part as $\vec{\Pt}^{jet}$ (small ``j''!).
 So, we shall
present the total jet transverse momentum $\vec{\Pt}^{Jet}$ as a sum of three
parts:  

1. $\vec{\Pt}^{Jet}_{(\nu)}$, containing the
contribution of neutrinos that belong to the jet, i.e.
a non-detectable part of jet $\Pt$ ($i$ - neutrino):
\vspace{-4mm}
\begin{eqnarray}
 \vec{\Pt}^{Jet}_{(\nu)} = \sum\limits_{i \in Jet} \vec{\Pt}_{(\nu)}^i.
\end{eqnarray}
~\\[-20pt]

2. $\vec{\Pt}^{Jet}_{(\mu)}$, containing the
contribution of jet muons to $\vec{\Pt}^{Jet}$ ($i$ - muon):
\vspace{-2mm}
\begin{eqnarray}
 \vec{\Pt}^{Jet}_{(\mu)} = \sum\limits_{i \in Jet} \vec{\Pt}_{(\mu)}^i.
\end{eqnarray}
~\\[-20pt]

These muons make a weak signal in the calorimeter
but their energy can be measured, in principle, in the muon system (in the
region of $|\eta|\lt2.5$ in the case of D0 geometry).
Due to the absence of the muon system and the tracker
beyond the $|\eta|\lt2.5$ region,
there exists a part of $\Pt^{Jet}$ caused by muons with $|\eta|>2.5$.
We denote this part as $\Pt^{Jet}_{(\mu,|\eta|>2.5)}$. It
can be considered, in some sense, as the analogue of $\Pt^{Jet}_{(\nu)}$ 
since the only trace of its presence would be weak MIP signals in calorimeter towers.

As for both points 1 and 2, let us say in advance that
the estimation of the average values of neutrino and muon
contributions to $\Pt^{Jet}$ (see Section 4 and Tables 1--12 of Appendix 1) have shown
that they are quite small: about $0.30\%$ of $\la\Pt^{Jet}\ra_{all}$ is due to neutrinos 
and about $0.33\%$ of $\la\Pt^{Jet}\ra_{all}$ is due to muons,
where ``$all$'' means averaging over all events including those without neutrinos 
and/or muons in jets. So, they together may cause  approximately about $0.63\%$ of 
the \Ptgj disbalance if muon signal is lost.

3. And finally, as we have mentioned before,
we use $\vec{\Pt}^{jet}$ to denote the part of $\vec{\Pt}^{Jet}$ which
includes all detectable particles of the jet
\footnote{We shall consider the issue of charged particles
contribution with small $\Pt$ 
into the total jet $\Pt$ while discussing the results of the full GEANT
 simulation (with account of the magnetic field effect) in our forthcoming
papers.} , i.e. the sum of $\Pt$ of jet particles that may produce a signal
in the calorimeter and muon system (calo=ECAL+HCAL signal)
\vspace{-2mm}
\begin{eqnarray}
 \vec{\Pt}^{jet} =  \vec{\Pt}^{Jet}_{(calo)} +\vec{\Pt}^{Jet}_{(\mu)} ,
\quad |\eta^{\mu}|\lt2.5.
\end{eqnarray}

Thus, in the general case we can write for any $\eta$ values:
\vspace{-1.5mm}
\begin{eqnarray}
 \vec{\Pt}^{Jet}
=\vec{\Pt}^{jet}
+\vec{\Pt}^{Jet}_{(\nu)}
+\vec{\Pt}^{Jet}_{(\mu,|\eta^{\mu}|>2.5)}.
\end{eqnarray}

In the case of $p\bar{p}\to \gamma + Jet + X$ events
the particles detected in the $|\eta|\lt4.2$ region may originate
from the fundamental subprocesses (1a) and (1b) corresponding to LO diagrams shown in Fig.~1, 
as well as from the processes corresponding to NLO diagrams (like those in Figs.~2, 4 that include ISR and FSR),
and also from the ``underlying'' event [1], of course.

As was already mentioned in Section 2,
the final states of fundamental subprocesses (1a) and (1b) may contain
additional jets due to the ISR and final state radiation (FSR) caused by the higher
 order QCD corrections to the LO Feynman diagrams given in Fig.~1. To understand and then 
to realize the jet energy calibration procedure, we need to use the event generator
to find the criteria for selection of events with a good
balance of $\vec{\Pt}^{\gamma}$ with the $\vec{\Pt}^{jet}$ part measurable in
the detector. It means that to make a reasonable simulation of the calibration procedure,
we need to have a selected sample of generated events having a small $\Pt^{miss}$ (see Section 4) 
contribution and use as a model.
We also have to find a way to select 
events without additional jets or with jets suppressed down to the level of mini-jets or clusters
having very small $\Pt$.

So, for any event we separate the particles in the $|\eta|\lt4.2$ region into two subsystems. 
The first one consists of the particles belonging to the 
``$\gamma +Jet$'' system (here ``$Jet$'' denotes the jet with the highest
$\Pt \geq 30 ~GeV/c$) having the total transverse momentum $\vec{\Pt}^{\gamma +Jet}$ 
(large ``Jet'', see (4)). The second subsystem involves all other ($O$) particles
beyond the ``$\gamma +Jet$'' system in the  region, covered by the detector, i.e. $|\eta|\lt4.2$.
Let us mention that the value of $\vec{\Pt}^{\gamma +Jet}$ 
may be different from the value of observable:\\[-18pt]
\begin{eqnarray}
\vec{\Pt}^{\gamma +jet} =
\vec{\Pt}^{\gamma} + \vec{\Pt}^{jet} \quad {\rm (small ``jet'')},
\end{eqnarray}
~\\[-10pt]
\noindent
in the case of non-detectable particles presence in a jet.
The total transverse momentum of this
$O$-system are denoted as $\Pt^{O}$ and it is a sum of
$\Pt$  of additional mini-jets (or clusters) and $\Pt$ of
single hadrons, photons and leptons in the $|\eta| \lt 4.2$ region. Since a part of
neutrinos are also present among these leptons, 
the difference of $\vec{\Pt}_{(\nu)}$ and $\vec{\Pt}^{Jet}_{(\nu)}$
gives us the transverse momentum \\[-20pt]
\begin{eqnarray}
 \vec{\Pt}^{O}_{(\nu)} = \vec{\Pt}_{(\nu)} - \vec{\Pt}^{Jet}_{(\nu)} \quad
|\eta^{\nu}|\lt4.2,
\end{eqnarray}

\noindent
carried out by the neutrinos that do not belong to the jet but are
contained in the $|\eta|\lt4.2$ region.

We denote by $\vec{\Pt}^{out}$ a part of $\vec{\Pt}^O$ that can be measured, in principle, in the detector.
Thus, $\vec{\Pt}^{out}$ is a sum of $\Pt$ of other mini-jets or, generally,
clusters (with $\Pt^{clust}$ smaller than $\Pt^{Jet}$) and $\Pt$ of single
hadrons ($h$), photons ($\gamma$) and electrons ($e$) with $|\eta| \lt 4.2$
and muons ($\mu$) with $|\eta^\mu| \lt 2.5$ that are out of the \gpj system.
For simplicity these mini-jets and clusters will be called ``clusters''
\footnote{As was already mentioned in  Introduction, these clusters are found
by the LUCELL jetfinder with the same value of the cone radius as for jets: $R^{clust}=R^{jet}=0.7$.}.
So, for our \gpj events $\vec{\Pt}^{out}$ is the following sum 
(all $\{h,~\gamma,~e,~\mu \} \not\in$ Jet):\\[-17pt]
\begin{eqnarray}
 \vec{\Pt}^{out} =
\vec{\Pt}^{clust}
+\vec{\Pt}^{sing}_{(h)}
+\vec{\Pt}^{nondir}_{(\gamma)}
+\vec{\Pt}^{}_{(e)}+\vec{\Pt}^{O}_{(\mu, |\eta^\mu|\lt2.5)}; \quad  |\eta|\lt4.2.
\end{eqnarray}

\noindent
And thus, finally, we have:\\[-18pt]
\begin{eqnarray}
 \vec{\Pt}^{O} =
\vec{\Pt}^{out}+\vec{\Pt}^{O}_{(\nu)}+\vec{\Pt}^{O}_{(\mu, |\eta^\mu|>2.5)}.
\end{eqnarray}

\noindent
With these notations we come to the following vector form \cite{BKS_P1} of
the $\Pt$- conservation law for the ``$\gamma + Jet$'' event
(where $\gamma$ is a direct photon) as a whole (supposing that the jet and the photon are contained
in the corresponding detectable regions):\\[-14pt]
\begin{eqnarray}
\vec{\Pt}^{\gamma} +
\vec{\Pt}^{Jet} +
\vec{\Pt}^{O}+
\vec{\Pt}^{|\eta|>4.2} = 0
\end{eqnarray}

\noindent
with last three terms defined correspondingly by (11), (15) and (5) respectively.

\subsection{Definition of selection cuts for physical variables and
the scalar form of the $\Pt$ balance equation.}

\noindent
1. We shall select the events with one jet and one ``$\gamma^{dir}$-candidate''
(in what follows we shall designate it as $\gamma$ and call the
``photon'' for brevity and only in Section 8, devoted to the backgrounds, we shall denote
$\gamma^{dir}$-candidate by $\tilde{\gamma}$) with\\[-11pt]
\begin{equation}
\Pt^{\gamma} \geq 40~ GeV/c~\quad {\rm and} \quad \Pt^{Jet}\geq 30 \;GeV/c.
\label{eq:sc1}
\end{equation}
The ECAL signal can be considered as a candidate for a direct photon
if it fits inside one D0 calorimeter tower having size $0.1\times0.1$ in 
the $\eta-\phi$ space.

For most of our applications in Sections 4, 5 and 6 mainly the PYTHIA
jetfinding algorithm LUCELL will be used.
The jet cone radius R in the $\eta-\phi$ space counted from the jet initiator cell (ic) is
taken to be $R_{ic}=((\Delta\eta)^2+(\Delta\phi)^2)^{1/2}=0.7$.
Below in Section 6 we shall also consider the jet radius counted
from the center of gravity (gc) of the jet, i.e. $R_{gc}$.
Comparison with the UA1 and UA2 jetfinding algorithms (taken from the CMSJET
program of fast simulation \cite{CMJ}) is presented in Sections 6 and 7.

\noindent
2. To suppress the contribution of background processes, i.e. to select mostly the events with ``isolated''
direct photons and to discard the events with fake ``photons'' (that
may originate as $\gamma^{dir}$-candidates from meson decays, for instance), we restrict

a) the value of the scalar sum of $\Pt$ of hadrons and other particles surrounding
a ``photon'' within a cone of $R^{\gamma}_{isol}=( (\Delta\eta)^2+(\Delta\phi)^2)^{1/2}=0.7$
(``absolute isolation cut'')
\footnote{We have found that $S/B$ ratio with $R^{\gamma}_{isol}\!=\!0.7$ is in about 1.5 times better
than with $R^{\gamma}_{isol}\!=\!0.4$ what is accompanied by only $10\%$ of additional loss of the number of
signal events.}
\\[-7pt]
\begin{equation}
\sum\limits_{i \in R} \Pt^i \equiv \Pt^{isol} \leq \Pt_{CUT}^{isol};
\label{eq:sc2}
\end{equation}
\vspace{-2.6mm}

b) the value of a fraction (``fractional isolation cut'')\\[-7pt]
\begin{equation}
\sum\limits_{i \in R} \Pt^i/\Pt^{\gamma} \equiv \epsilon^{\gamma} \leq
\epsilon^{\gamma}_{CUT}.
\label{eq:sc3}
\end{equation}

\noindent
3. To be consistent with the application condition of the NLO
formulae, one should avoid an infrared dangerous region and take care of
$\Pt$ population in the region close to a $\gamma^{dir}$-candidate
(see \cite{Fri}, \cite{Cat}). In accordance with \cite{Fri} and \cite{Cat}, 
we also restrict the scalar sum of $\Pt$ of particles
 around a ``photon'' within a cone of a smaller radius $R^{\gamma}_{singl}=0.2$.

Due to this cut,\\[-4mm]
\begin{equation}
\sum\limits_{i \in R^{\gamma}_{singl}} \Pt^i \equiv \Pt^{singl} \leq 2~ GeV/c
~~~~~(i\neq ~\gamma^{dir}),
\label{eq:sc4}
\end{equation}
an ``isolated'' photon with high $\Pt$ also becomes a ``single'' one within
an area of 8 calorimeter towers (of size 0.1$\times$0.1  according to D0 geometry)
which surround the tower fired by it, 
i.e. a tower with the highest $\Pt$.

\noindent
4. We accept only the events having no charged tracks (particles)
with $\Pt>5~GeV/c$ within the $R=0.4$ cone around the $\gamma^{dir}$-candidate.

\noindent
5. 
To suppress the background events with photons resulting from
$\pi^0$, $\eta$, $\omega$
and $K_S^0$ meson decays, we require the absence of a high $\Pt$ hadron
in the tower containing the $\gamma^{dir}$-candidate:\\[-10pt]
\begin{equation}
\Pt^{hadr} \leq 7~ GeV/c.
\label{eq:sc5}
\end{equation}

\noindent
At the PYTHIA level of simulation this cut may effectively take into account 
the imposing of an upper cut on the HCAL signal in the towers behind
the ECAL tower fired by the direct photon. 

\noindent   
6. We select the events with the vector $\vec{\Pt}^{Jet}$ being ``back-to-back'' to
the vector $\vec{\Pt}^{\gamma}$ (in the plane transverse to the beam line)
within $\dphi$ defined by the equation:\\[-12pt]
\begin{equation}
\phigj=180^\circ \pm \Delta\phi,
\label{eq:sc7}
\end{equation}
where $\phigj$ is the angle
between the \Ptgj vectors: 
$\vec{\Pt}^{\gamma}\vec{\Pt}^{Jet}=\Pt^{\gamma}\Pt^{Jet}\cdot cos(\phigj)$,
$\Pt^{\gamma}=|\vec{\Pt}^{\gamma}|,~~\Pt^{Jet}=|\vec{\Pt}^{Jet}|$.
The cases $\Delta\phi \leq 17^\circ, 11^\circ, 6^\circ$ are considered in this paper
($6^\circ$ is approximately one D0 calorimeter tower  size in $\phi$).

\noindent
7. The initial and final state radiations (ISR and FSR) manifest themselves most clearly
as some final state mini-jets or clusters activity.
To suppress it, we impose a new cut condition that was not formulated in
an evident form in previous experiments: we choose the \gpj events
that do not have any other
jet-like or cluster high $\Pt$ activity  by selecting the events with the values of
$\Pt^{clust}$ (the cluster cone $R_{clust}(\eta,\phi)=0.7$), being lower than some threshold
$\Pt^{clust}_{CUT}$ value, i.e. we select the events with\\[-10pt]
\begin{equation}
\Pt^{clust} \leq \Pt^{clust}_{CUT}
\label{eq:sc8}
\end{equation}
($\Pt^{clust}_{CUT}=15, 10, 5 ~GeV/c$ are most effective as will be shown in Sections 6--8).
Here, in contrast to \cite{BKS_P1}--\cite{BKS_P5}, the clusters are found
by one and the same jetfinder LUCELL while three different jetfinders UA1, UA2 and LUCELL
are used to find the jet ($\Pt^{Jet}\geq 30~ GeV/c$) in the event.

\noindent
8. Now we pass to another new quantity (proposed also for the first time in \cite{BKS_P1}--\cite{BKS_P5}) 
that can be measured at the experiment.
We limit the value of the modulus of the vector sum of $\vec{\Pt}$ of all
particles, except those of the \gpj system, that fit into the region $|\eta|\lt4.2$ covered by
the ECAL and HCAL, i.e., we limit the signal in the cells ``beyond the jet and photon'' region
by the following cut:
\begin{equation}
\left|\sum_{i\not\in Jet,\gamma-dir}\!\!\!\vec{\Pt}^i\right| \equiv \Pt^{out} \leq \Pt^{out}_{CUT},
~~|\eta^i|\lt4.2.
\label{eq:sc9}
\end{equation}

\noindent
The importance of $\Pt^{out}_{CUT}$ and $\Pt^{clust}_{CUT}$
for selection of events with a good balance of \Ptgj and for
the background reduction will be demonstrated in Sections 7 and 8.

Below the set of selection cuts 1 -- 8 will be referred to as
``Selection 1''. The last two of them, 7 and 8, are new criteria \cite{BKS_P1}
not used in previous experiments. 

9. In addition to them one more new object, introduced in 
\cite{BKS_P1} -- \cite{BKS_P5} and named an ``isolated jet'',  will be used in our analysis.
i.e. we shall require the presence of a ``clean enough'' (in the sense of limited $\Pt$
activity) region inside the ring of $\Delta R=0.3$ width (or approximately of a size of 
three calorimeter towers) around the jet.  Following this picture, we restrict the ratio of the scalar sum
of transverse momenta of particles belonging to this ring, i.e.\\[-5pt]
\begin{equation}
\Pt^{ring}/\Pt^{jet} \equiv \epsilon^{jet}\leq\epsilon^{jet}_0, \quad {\rm where ~~~~ }
\Pt^{ring}=\!\!\!\!\sum\limits_{\footnotesize i \in 0.7\lt R \lt1.0} \!\!\!\!|\vec{\Pt}^i|.
\label{eq:sc10}
\end{equation}
~\\[-4pt]
($\epsilon^{jet}_0$ is chosen to be $3-5\%$, see Sections 7 and 8).

The set of cuts 1 -- 9 will be called in what follows ``Selection 2''.

\noindent
10. In  the following we shall consider also ``Selection 3'' where we shall keep only those events in which
one and the same jet  is found simultaneously by every of three jetfinders used here: UA1, UA2 and LUCELL 
(i.e. up to good accuracy having the same values of $\Pt^{Jet}, ~R^{jet}$ and $\Delta\phi$).
For these jets (and also clusters in the same event) we require the following conditions:\\[-5mm]
\begin{eqnarray}
\Pt^{Jet}>30~GeV/c, \qquad \Pt^{clust}\lt \Pt^{clust}_{CUT},\qquad
\dphi\lt17^\circ (11^\circ,~6^\circ), \qquad \epsilon^{jet} \leq 3-5\%
\end{eqnarray}

The exact values of the cut parameters $\Pt^{isol}_{CUT}$,
$\epsilon^{\gamma}_{CUT}$, $\epsilon^{jet}$, $\Pt^{clust}_{CUT}$, $\Pt^{out}_{CUT}$
 will be specified below, since they may be
different, for instance, for various $\Pt^{\gamma}$ intervals
(being looser for higher  $\Pt^{\gamma}$).

\noindent 
11. As we have already mentioned in Section 3.1, one can
expect reasonable results of the jet energy calibration procedure
modeling and subsequent practical realization
 only if one uses a set of selected events with small $\Pt^{miss}$. So, we also use
the following cut:\\[-17pt]
\begin{eqnarray}
\Pt^{miss}~\leq \Pt^{miss}_{CUT}.
\label{eq:sc11}
\end{eqnarray}
For this reason we shall study in the next Section 4 the influence of
$\Pt^{miss}$ parameter on the selection of events with a reduced value
of the total sum of neutrino contribution into $\Pt^{Jet}$, i.e. 
$\Pt^{Jet}_{(\nu)}$.
The aim of the event selection with small $\Pt^{Jet}_{(\nu)}$
is quite obvious: we need a set of events with a reduced
$\Pt^{Jet}$ uncertainty due to a possible presence of a non-detectable
particle contribution to a jet
\footnote{In Section 8 we also underline the importance of this cut for reduction of $e^\pm$
events contribution to the background to the signal $\gamma^{dir}+jet$ events.}.

To conclude this section, let us write the basic $\Pt$-balance equation (16) of the previous section
with the notations introduced here in the form
more suitable to present the final results. For this
purpose we shall write equation (16) in the following scalar form (see also \cite{BKS_P1}, \cite{GPJ} and
\cite{QCD_talk}): \\[-15pt]
\begin{eqnarray}
\frac{\Pt^{\gamma}-\Pt^{Jet}}{\Pt^{\gamma}}=(1-cos\dphi) 
+ \Db/\Pt^{\gamma}, \label{eq:sc12}
\label{eq:sc_bal}
\end{eqnarray}
where
$\Db\equiv (\vec{\Pt}^{O}+\vec{\Pt}^{|\eta|>4.2)})\cdot \vec{n}^{Jet}$ ~~~~ with ~~
$\vec{n}^{Jet}=\vec{\Pt}^{Jet}/\Pt^{Jet}$.

As will be shown in Section 7, the first term on the
right-hand side of equation (\ref{eq:sc_bal}), i.e. $(1-cos\dphi)$ is negligibly
small as compared with the second term (in a case of Selection 1) 
and tends to decrease fast with  growing $\Pt^{Jet}$. 
So, in this case the main contribution to the $\Pt$ disbalance in the
\gpj system is caused by the term $\Db/\Pt^{\gamma}$.

\section{ESTIMATION OF A NON-DETECTABLE PART OF $\Pt^{Jet}$.}       

\it\small
\hspace*{9mm}
It is shown that by imposing an upper cut on  the missing transverse momentum
$\Pt^{miss}\lt10~GeV/c$ one can reduce  the correction to the measurable part of $\Pt^{jet}$ due 
to neutrino contribution down to the value of
$\Delta_\nu=\la\Pt^{Jet}_{(\nu)}\ra_{all\; events}=0.1~GeV/c$
in all intervals of $\Ptg$. 
\rm\normalsize
\vskip4mm

In Section 3.1 we have divided the transverse momentum of a jet,
i.e. $\Pt^{Jet}$, into two parts, a detectable $\Pt^{jet}$ and non-detectable
($\Pt^{Jet}-\Pt^{jet}$), consisting of $\Pt^{Jet}_{(\nu)}$
and $\Pt^{Jet}_{(\mu,|\eta|>2.5)}$ (see (11)).
In the same way, according to equation (15), we divided
the transverse momentum $\Pt^O$ of ``other particles'' that are out of $\gamma^{dir}+jet$ system
into a detectable part $\Pt^{out}$
and a non-detectable part consisting of the sum of $\Pt^O_{(\nu)}$
and $\Pt^O_{(\mu,|\eta|>2.5)}$ (see (15))
\footnote{But in a real experiment non-detectable part of $\Pt^{Jet}$ may be also conditioned by energy 
leakage due to constructive/material features of a detector.}.

We shall estimate here what part of $\Pt^{Jet}$ may be carried out by non-detectable particles 
(mainly neutrinos originating from weak decays)
\footnote{In \cite{BKS_P5} and \cite{QCD_talk} it was  shown that main source of high $\Pt$ 
neutrinos in background processes 
are $W^\pm$ decays, which also contain $e^\pm$ that in its turn may fake direct photons.}.

We shall consider the case of switched-on decays of $\pi^{\pm}$ and $\;K^{\pm}$ mesons 
\footnote{According to the PYTHIA default agreement, $\pi^{\pm}$ and $\;K^{\pm}$ mesons are stable.}.
Here $\pi^{\pm}$ and $K^{\pm}$ meson decays are allowed inside the solenoid volume with the barrel 
radius $R_B=80~ cm$ and the distance from the interaction vertex to Endcap along the $z$-axis 
$L=130~cm$ (D0 geometry).

For this aim we shall use the bank of the signal \gpj events, i.e. caused by subprocesses
(1a) and (1b), generated for three $\Ptg$ intervals:
$40\lt\Ptg\lt50$, $70\lt\Ptg\lt90$ and $90\lt\Ptg\lt140~GeV/c$
and selected with conditions (17) -- (24) (Selection 1) and the following cut values:
\begin{equation}
 \Pt_{CUT}^{isol}=4 ~GeV/c,~~\epsilon^{\gamma}_{CUT}=7\%,
~~\dphi\lt17^\circ,~~\Pt^{clust}_{CUT}=30~ GeV/c.
\end{equation}
Here the cut $\Pt^{clust}_{CUT}=30~ GeV/c$ has the meaning of a very weak restriction
on mini-jets or clusters activity.
No restriction was imposed on the $\Pt^{out}$ value.
The results of analysis of these events, based on the application of LUCELL jetfinder,
are presented in Fig.~\ref{fig20-22}, while more detailed tables of Appendix 1
contain the results found with UA1 and UA2 jetfinding algorithms as well.

The first row of Fig.~\ref{fig20-22} contains  $\Pt^{miss}$ spectra in the \gpj events for different
$\Ptg$ intervals and demonstrates (to a good accuracy) their practical independence on $\Ptg$.

In the second row of Fig.~\ref{fig20-22} we present the spectra of
$\Pt^{miss}$ for those events (denoted as $\Pt^{Jet}_{(\nu)}>0$) which
contain jets having neutrinos, i.e. having  a non-zero $\Pt^{Jet}_{(\nu)}$
component of  $\Pt^{Jet}$. 
These figures also show the practical independence of the $\Pt^{miss}$ spectrum
on the direct photon $\Pt^{\gamma}$ (approximately equal to $\Pt^{Jet}$):
the peak position remains in the region of $\Pt^{miss}\lt2.5 ~GeV/c$. Comparison of the
number of entries in the second row plots of Fig.~\ref{fig20-22} with
those in the first row allows to conclude that the part of events with the jet
having the non-zero neutrinos contribution ($\Pt^{Jet}_{(\nu)}>0$) has practically the same
size of about $15\%$ in all $\Ptg$ intervals.

The same spectra of $\Pt^{miss}$ for events with $\Pt^{Jet}_{(\nu)}\gt0$
show how many of these events would remain after imposing a cut on
$\Pt^{miss}$ in every $\Ptg$ interval.
The important thing is that reduction of the number of events
with $\Pt^{Jet}_{(\nu)}\gt0$~ in every $\Ptg$ interval leads to reduction of 
the mean value of the $\Pt^{Jet}_{(\nu)}$, i.e. the value averaged over all
collected events $\la\Pt^{Jet}_{(\nu)}\ra_{all\; events}$.
This value, found from PYTHIA generation, serves as a model correction
$\Delta_\nu$ and it has to be estimated for proper determination
of the total $\Pt^{Jet}$ from the measurable part $\Pt^{jet}$. Thus, the complete jet $\Pt$
can be defined as: 
$\Pt^{Jet}= \Pt^{jet}+\Delta_\nu +\Delta_\mu(|\eta^\mu|\gt2.5)$
\footnote{With account of real processes in the detector, as we mentioned above,
$\Pt^{jet}$ should be also corrected by energy leakage from the detectable volume of
the detector.},
where $\Delta_\nu=\la\Pt^{Jet}_{(\nu)}\ra_{all\; events}$ and 
$\Delta_\mu=\la\Pt^{Jet}_{(\mu,|\eta>2.5|)}\ra_{all\; events}$.
(As we plan to use in this paper only events with jets belonging to CC part of calorimeter,
$\Delta_\mu$ is not important for our analysis.)

The effect of imposing general $\Pt^{miss}_{CUT}$  in each event of our
sample is shown in the third row of Fig.~\ref{fig20-22}. The upper cut
$\Pt^{miss}_{CUT}=1000 ~GeV/c$, as is seen from the comparison with the second row
pictures, means the absence of any upper limit for $\Pt^{Jet}_{(\nu)}$.
The most important illustrative fact that in the absence of any restriction
on $\Pt^{miss}$ the total neutrino $\Pt$ inside the jet
averaged over all events can be as large as $\Pt^{Jet}_{(\nu)}\approx 0.32~GeV/c$
at $90\lt\Ptg\lt140 ~GeV/c~  GeV/c$ comes from the right-hand plot of the third row in 
Fig.~\ref{fig20-22}.
In the $40\lt\Pt^{Jet}\lt50~GeV/c$ interval, which is less dangerous from the point of view of the
neutrino $\Pt$ content in a jet, we have already
a very small mean value of $\Pt^{Jet}_{(\nu)}$ equal to $0.12 ~GeV/c$
even without imposing any $\Pt^{miss}_{CUT}$. From the same plots of the third row of
Fig.~\ref{fig20-22} we see that with general $\Pt^{miss}_{CUT}=10~GeV/c$
the average correction due to neutrino contribution is $\Delta_\nu=0.1~GeV/c$
in all three intervals of $\Ptg$. 

At the same time, as it was demonstrated in \cite{BKS_P5} and \cite{QCD_talk},
this cut essentially reduces  the admixture of the $e^\pm$-events, in which $e^\pm$, mainly 
originating from 
the $W^\pm\to e^\pm\nu$ weak decays, may fake the direct photon signal. 
These events are characterized by big values of $\Pt^{miss}$ (it is higher, on the average, 
by about one order of magnitude than in the signal ``$\gamma^{dir}+jet$'' events) that may worsen 
the jet calibration accuracy.

The analogous (to neutrino) situation holds for the $\Pt^{Jet}_{(\mu)}$ contribution.

The detailed information about the values of non-detectable
$\Pt^{Jet}_{(\nu)}$ averaged over all events
(no cut on $\Pt^{miss}$ was used) as well as about mean $\Pt$ values
of muons belonging to jets $\la \Pt^{Jet}_{(\mu)}\ra$ is presented
in Tables 1--12 of Appendix 1 for the sample of events with jets which are
entirely contained in the central region of the calorimeter
($|\eta^{jet}|\lt0.7$) and found  by UA1, UA2 and LUCELL jetfinders.           
In these tables the ratio of the number of events with non-zero $\Pt^{Jet}_{(\nu)}$
to the total number of events is denoted by $R^{\nu \in Jet}_{event}$ and
the ratio of the number of events with non-zero $\Pt^{Jet}_{(\mu)}$
to the total number of events is denoted by $R^{\mu \in Jet}_{event}$.

The quantity $\Pt^{miss}$ in events with $\Pt^{Jet}_{(\nu)}\gt0$ is denoted in these tables as
$\Pt^{miss}_{\nu\in Jet}$ and is given there for four $\Ptg$ intervals
($40\lt\Ptg\lt50$, $50\lt\Ptg\lt70$, $70\lt\Ptg\lt90$ and $90\lt\Ptg\lt140$)
and other $\Pt^{clust}_{CUT}$ values ($\Pt^{clust}_{CUT}=20,15,10,5 ~GeV/c$)
complementary to $\Pt^{clust}_{CUT}=30 ~GeV/c$ used for the second row plots
\footnote{Please, note that the values of $\Pt^{miss}$ and $\Pt^{miss}_{\nu\in Jet}$ in the plots of 
Fig.~\ref{fig20-22}
are slightly different from those of Appendix 1 as the numbers in from Fig.~\ref{fig20-22}
were found for events in the whole $|\eta|\lt4.2$ region.}
of Fig.~\ref{fig20-22}. From Tables 1--3 we see that the averaged value of
$\Pt^{miss}$ calculated by using only the events with $\Pt^{Jet}_{(\nu)}\gt0$, i.e.
$\la\Pt^{miss}_{\nu \in Jet}\ra$, is about 2.3--2.4 $GeV/c$ for the $40\lt\Ptg\lt50 ~GeV/c$
interval. It increases to about 3.4--3.5 $GeV/c$ for the $90\lt\Ptg\lt140 ~GeV/c$ interval 
(see Tables 10--12). It should be noted that the averaged values of the modulus of
$\Pt^{Jet}_{(\nu)}$ (see formula (8)) presented in the
third lines of Tables 1--12 from Appendix 1 coincide with the averaged values of the difference 
$\la\Pt^{Jet}\!-\!\Pt^{jet}\ra \equiv \Delta_\nu$ (see Section 3.2 and second lines of Tables 1--12 ) 
to three digits, i.e. $<\!\!\Pt^{Jet}_{(\nu)}\!\!>=\Delta_\nu$. This is because the 
$\vec{\Pt}^{Jet}$ and $\vec{\Pt}^{jet}$ vectors are practically collinear 
and because we consider here the ``CC-events'' in which all jet muons may also be
detected by the central muon system.

Let us mention that the 11-th lines of Tables 1--12 
show the ratio (``$29sub/all$'') of the number of events due to the gluonic subprocess (1a) 
only (see also Section 10) to the number of events due to the sum of subprocesses (1a) and (1b).
It is seen that this ratio drops with $\Ptg$ growth (see also Table 21).
Two upper lines 9, 10 contain an additional information on the 
numbers of \gpj events, i.e. $Nevent_{(c)}$ and $Nevent_{(b)}$, 
produced in a case of the gluonic subprocess (1a) and
having in the final state jets that originate from $c$ and $b$ quarks.
These numbers correspond to the integrated luminosity
$L_{int}=300~ pb^{-1}$ and vary for different $\Pt^{Jet}$($\approx\Ptg$) intervals.
The averaged jet radii
$<\!\!R_{jet}\!\!>$ and the number of entries are shown in last two lines.

\begin{figure}[htbp]
\vspace{-1.9cm}
\hspace{-10mm} \includegraphics[width=19cm,height=19.7cm]{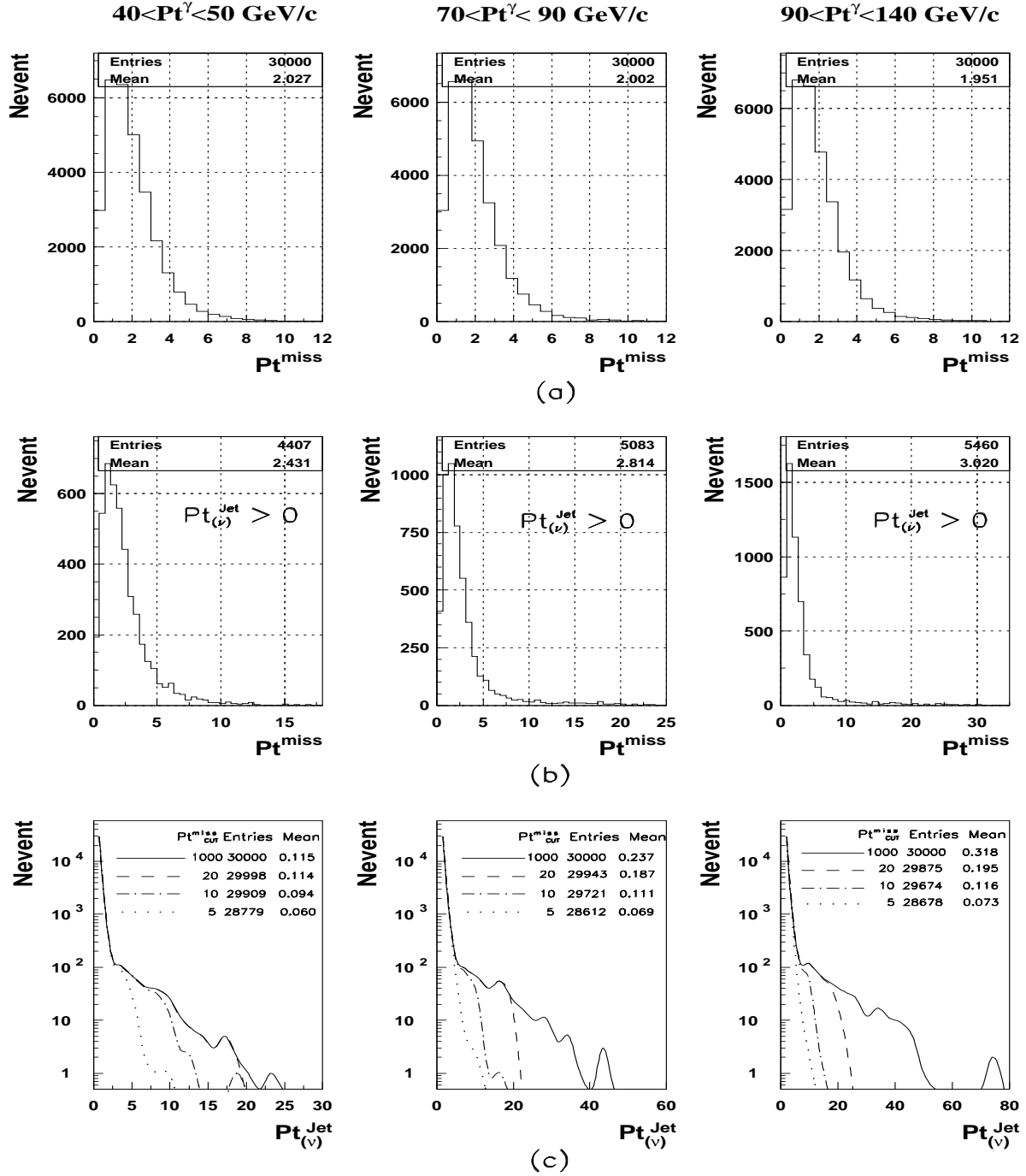}
\vspace{-0.5cm}
\caption{\hspace*{0.0cm}~ a) $\Pt^{miss}$ spectra in all events; ~~
b) $\Pt^{miss}$ spectra in events having jets with non-zero $\Pt$
neutrinos,~ i.e. 
$\Pt^{Jet}_{(\nu)}\gt0$;~ c) $\Pt^{Jet}_{(\nu)}$ spectra
and their mean values dependence on the values 
of $\Pt^{miss}_{CUT}$ in various $\Ptg(\approx\Pt^{Jet})$ intervals.
$\pi^{\pm}$ and $K^{\pm}$ meson decays are allowed inside the solenoid of $R=80~cm$ and $L=130~cm$
($\Pt^{clust}_{CUT}=30 ~GeV/c$).}
\label{fig20-22}
\end{figure} 

\newpage
~\\[-9mm]


\section{EVENT RATES FOR DIFFERENT $\Pt^{\gamma}$ AND $\eta^{Jet}$ INTERVALS.}

\it\small
\hspace*{9mm}
The number of \gpj events distribution over $\Pt^{\gamma}$ and $\eta^{\gamma}$ is studied here.
It is found that in each interval of the $\Delta\Pt^{\gamma}=10~ GeV/c$ width 
the rates decrease by a factor more than 2.
The number of events with jets which transverse momentum are completely (or with $5\%$ accuracy)
contained in CC, IC, EC and FC regions are presented in Tables \ref{tab:sh1}--\ref{tab:sh4}
for integrated luminosity $L_{int}=300~pb^{-1}$.
\rm\normalsize
\vskip4mm

\subsection{Dependence of distribution of the number of events
on the ``back-to-back'' angle $\phigj$ and on $\Pt^{ISR}$. }            

The definitions of the physical variables introduced in
Sections 2 and 3  allow to study
a possible way to select the events with a good $\Pt^{\gamma}$ and $\Pt^{Jet}$ balance.
Here we shall be interested to get (by help of PYTHIA generator and the theoretical models 
therein) an idea about the form of the spectrum of the variable $\Pt{56}$ 
(which is approximately proportional to $\Pt^{ISR}$ up to the value of intrinsic parton
transverse momentum $k_t$ inside a proton) at different values of $\Ptg$.
For this aim four samples of \gpj events were generated by using PYTHIA
with 2 QCD subprocesses (1a) and (1b) being included simultaneously. In what follows we shall call these events as 
``signal events''. The generations were done with the values of the PYTHIA parameter CKIN(3)($\equiv\pth$)
 equal to  20, 25, 35 and 45 $GeV/c$ in order to cover
four $\Pt^{\gamma}$ intervals: 40--50, 50--70, 70--90 and 90--140 $GeV/c$, respectively.
Each sample in these $\Ptg$ intervals had a size of $5\cdot 10^6$ events.
The cross sections for the two subprocesses were found to be as given in Table~\ref{tab:cross4}.\\[-9mm]
\begin{table}[h]
\begin{center}
\caption{The cross sections (in $microbarn$) of the $qg\to q+\gamma$ and $q\overline{q}\to g+\gamma$ subprocesses
for four $\Pt^{\gamma}$ intervals.}
\normalsize
\vskip.1cm
\begin{tabular}{||c||c|c|c|c|}                  \hline \hline
\label{tab:cross4}
Subprocess& \multicolumn{4}{c|}{$\Pt^{\gamma}$ interval ($GeV/c$)} \\\cline{2-5}
   type   & 40 -- 50 & 50 -- 70 & 70 -- 90  & 90 -- 140 \\\hline \hline
$qg\to q+\gamma$           & 0.97$\cdot10^{-2}$& 4.78$\cdot10^{-3}$& 1.36$\cdot10^{-3}$& 4.95$\cdot10^{-4}$ \\\hline
$q\overline{q}\to g+\gamma$& 0.20$\cdot10^{-2}$& 0.96$\cdot10^{-3}$& 0.35$\cdot10^{-3}$& 1.56$\cdot10^{-4}$ \\\hline
Total                      & 1.17$\cdot10^{-2}$& 5.75$\cdot10^{-3}$& 1.71$\cdot10^{-3}$& 6.51$\cdot10^{-4}$ \\\hline
\end{tabular}
\end{center}
\vskip-7mm
\end{table}

 For our analysis  we used ``Selection 1''  (formulae (17)--(24))
defined in Sections 3.2 and the values of cut parameters (29).

In Tables \ref{tab:pt56-1}, \ref{tab:pt56-2} and \ref{tab:pt56-4},
\ref{tab:pt56-5}  we present $\Pt56$ spectra for
two most illustrative cases of
$\Pt^{\gamma}$ intervals $40\lt\Pt^{\gamma}\lt50 ~GeV/c$ (Tables 2 and 5) and
$70\lt\Pt^{\gamma}\lt90 ~GeV/c$ (Tables 3 and 6). The distributions of the
 number of events for the integrated luminosity $L_{int}=300\,pb^{-1}$
in different $\Pt56$ intervals ($\la k_t\ra$ was taken to be
fixed at the PYTHIA default value, i.e. $\la k_t\ra=0.44\,GeV/c$) and for different ``back-to-back'' angle intervals
$\phigj=180^\circ \pm \dphi~$ ($\dphi\leq17^\circ,\,11^\circ$ and $6^\circ$ as well as without 
any restriction on $\dphi$, i.e. for the whole $\phi$ interval $\dphi\leq180^\circ$)
\footnote{The value $\Delta\phi=6^\circ$ approximately
coincides with one D0  HCAL tower size in the $\phi$-plane.
}
are given there. The LUCELL jetfinder was used for determination of jets and clusters
\footnote{More details connected with UA1 and UA2 jetfinders application can be found in Section 7
and Appendices 2--5 for a jet contained in CC region.}.
Tables \ref{tab:pt56-1} and \ref{tab:pt56-2} correspond to $\Pt^{clust}\lt30\,GeV/c$ 
and serve as an illustration since it is rather a weak cut condition, while
Tables~\ref{tab:pt56-4} and \ref{tab:pt56-5}  correspond to a more
restrictive selection cut $\Pt^{clust}_{CUT}=5\,GeV/c$
(which leads to about twofold reduction of the number of events for $\dphi\leq17^\circ$; 
see summarizing Tables \ref{tab:pt56-3} and \ref{tab:pt56-6}).

First, from the last summary lines of Tables
\ref{tab:pt56-1}, \ref{tab:pt56-2} and \ref{tab:pt56-4}, \ref{tab:pt56-5}
we can make a general conclusion about the $\dphi$-dependence
of the event spectrum.
Thus, in the case of weak restriction  $\Pt^{clust}\lt30 ~GeV/c$
we can see from Table \ref{tab:pt56-1} that for the $40\leq \Pt^{\gamma}\leq 50 ~GeV/c$
interval about 76$\%$ of events are concentrated
in the $\Delta\phi\lt17^\circ$ range, while 41$\%$ of events are
in the $\Delta\phi\lt6^\circ$ range.
At the same time the analogous summary line of Table \ref{tab:pt56-2}
shows us that for higher $\Ptg$ interval $70\leq \Pt^{\gamma}\leq90\, GeV/c$ the $\Pt56$ spectrum
for the same restriction $\Pt^{clust}\!\lt30 ~GeV/c$
moves (as compared with low $\Ptg$ intervals) to the small $\dphi$ region: 
more than 80$\%$ of events have
$\Delta\phi\lt17^\circ$ and 50$\%$ of them have $\Delta\phi\lt6^\circ$.
\def\baselinestretch{0.98}
\begin{table}[htbp]
\begin{center}
\vskip-1.2cm
\small
\caption{Number of events dependence on $\Pt56$ and
$\dphi_{max}$ for}
\vskip-3pt
{\footnotesize $40\leq Pt^{\gamma}\leq 50 \, GeV/c$ and
$\Pt^{clust}_{CUT}= 30 \, GeV/c$ for $L_{int}=300~ pb^{-1}$.}
\vskip.2cm
\begin{tabular}{||c||r|r|r|r||} \hline \hline
\label{tab:pt56-1}
 $\Pt{56}$  &\multicolumn{4}{c||}{ $\dphi_{max}$} \\\cline{2-5}
 $(GeV/c)$  &\aaa $180^\circ$\aaa&\aaa$17^\circ$\aaa&\aaa$11^\circ$\aaa&\aaa$6^\circ$\aaa \\\hline\hline
    0 --   5 &     25859 &     24732 &     23583 &     20035 \\\hline
    5 --  10 &     28712 &     27371 &     24597 &     14409 \\\hline
   10 --  15 &     18899 &     15989 &     10903 &      5422 \\\hline
   15 --  20 &     11830 &      6729 &      4399 &      2157 \\\hline
   20 --  25 &      7542 &      2784 &      1825 &       900 \\\hline
   25 --  30 &      5496 &      1642 &      1159 &       629 \\\hline
   30 --  40 &      8506 &      2636 &      1800 &       952 \\\hline
   40 --  50 &      3297 &       856 &       523 &       254 \\\hline
   50 -- 100 &       550 &       175 &       133 &        72 \\\hline
  100 -- 300 &         0 &         0 &         0 &         0 \\\hline
  300 -- 500 &         0 &         0 &         0 &         0 \\\hline
    0 -- 500 &    110691 &     82913 &     68921 &     44830 \\\hline   
\end{tabular}
\vskip0.2cm
\small
\caption{Number of events dependence on $\Pt56$ and $\dphi_{max}$ for}
\vskip-3pt
{\footnotesize $70\leq Pt^{\gamma}\leq 90 \, GeV/c$ and
$\Pt^{clust}_{CUT}= 30 \, GeV/c$ for $L_{int}=300~ pb^{-1}$.}
\vskip0.2cm
\begin{tabular}{||c||r|r|r|r||} \hline \hline
\label{tab:pt56-2}
 $\Pt{56}$  &\multicolumn{4}{c||}{ $\dphi_{max}$} \\\cline{2-5}
 $(GeV/c)$  &\aaa $180^\circ$\aaa&\aaa$17^\circ$\aaa&\aaa$11^\circ$\aaa&\aaa$6^\circ$\aaa \\\hline\hline
    0 --   5 &      2734 &      2627 &      2506 &      2224 \\\hline
    5 --  10 &      3318 &      3203 &      3059 &      2444 \\\hline
   10 --  15 &      2356 &      2258 &      2081 &      1138 \\\hline
   15 --  20 &      1680 &      1570 &      1187 &       579 \\\hline
   20 --  25 &      1288 &      1072 &       735 &       354 \\\hline
   25 --  30 &      1013 &       678 &       437 &       210 \\\hline
   30 --  40 &      1150 &       626 &       421 &       216 \\\hline
   40 --  50 &       545 &       265 &       192 &       109 \\\hline
   50 -- 100 &       768 &       427 &       300 &       144 \\\hline
  100 -- 300 &         1 &         0 &         0 &         0 \\\hline
  300 -- 500 &         0 &         0 &         0 &         0 \\\hline
    0 -- 500 &     14853 &     12727 &     10919 &      7418 \\\hline    
\end{tabular}
\vskip0.2cm
\small
\caption{Number of events dependence on $\dphi_{max}$ and on
$\Pt^{\gamma}$ for $L_{int}=300~ pb^{-1}$,}
\vskip-3pt
{\footnotesize $\Pt^{clust}_{CUT}=30 ~GeV/c$ (summary).}
\vskip0.2cm
\begin{tabular}{||c||r|r|r|r||} \hline \hline
\label{tab:pt56-3}
 $\Pt{56}$  &\multicolumn{4}{c||}{ $\dphi_{max}$} \\\cline{2-5}
 $(GeV/c)$  &\aaa $180^\circ$\aaa&\aaa$17^\circ$\aaa&\aaa$11^\circ$\aaa&\aaa$6^\circ$\aaa \\\hline\hline
 40 -- 50  &    110691 &     82913 &     68921 &     44830 \\\hline   
 50 -- 70  &     71075 &     55132 &     45716 &     29692 \\\hline 
 70 -- 90  &     14853 &     12727 &     10919 &      7418 \\\hline   
 90 -- 140 &      5887 &      5534 &      4974 &      3655 \\\hline 
\end{tabular}
\end{center}
\end{table}

\def\baselinestretch{0.98}
\begin{table}[htbp]
\begin{center}
\vskip-1.2cm
\small
\caption{Number of events dependence on $\Pt56$ and $\dphi_{max}$ for}
\vskip-3pt
{\footnotesize $40\leq Pt^{\gamma}\leq 50 \, GeV/c$ and
$\Pt^{clust}_{CUT}= 5 \, GeV/c$ for $L_{int}=300~ pb^{-1}$.}
\vskip.2cm
\begin{tabular}{||c||r|r|r|r||} \hline \hline
\label{tab:pt56-4}
 $\Pt{56}$  &\multicolumn{4}{c||}{ $\dphi_{max}$} \\\cline{2-5}
 $(GeV/c)$  &\aaa $180^\circ$\aaa&\aaa$17^\circ$\aaa&\aaa$11^\circ$\aaa&\aaa$6^\circ$\aaa \\\hline\hline
    0 --   5 &     18462 &     18457 &     18361 &     16529 \\\hline
    5 --  10 &     14774 &     14722 &     13881 &      8622 \\\hline
   10 --  15 &      3195 &      3008 &      2298 &      1230 \\\hline
   15 --  20 &       562 &       481 &       409 &       266 \\\hline
   20 --  25 &       217 &       217 &       207 &       150 \\\hline
   25 --  30 &       121 &       113 &       106 &        81 \\\hline
   30 --  40 &       165 &       160 &       145 &       108 \\\hline
   40 --  50 &        69 &        67 &        54 &        30 \\\hline
   50 -- 100 &        10 &        10 &        10 &         7 \\\hline
  100 -- 300 &         0 &         0 &         0 &         0 \\\hline
  300 -- 500 &         0 &         0 &         0 &         0 \\\hline
    0 -- 500 &     37576 &     37235 &     35473 &     27025 \\\hline       
\end{tabular}
\vskip0.2cm
\small
\caption{Number of events dependence on $\Pt56$ and $\dphi_{max}$ for}
\vskip-3pt
{\footnotesize $70\leq Pt^{\gamma}\leq 90 \, GeV/c$ and
$\Pt^{clust}_{CUT}= 5 \, GeV/c$ for $L_{int}=300~ pb^{-1}$.}
\vskip0.2cm
\begin{tabular}{||c||r|r|r|r||} \hline \hline
\label{tab:pt56-5}
 $\Pt{56}$  &\multicolumn{4}{c||}{ $\dphi_{max}$} \\\cline{2-5}
 $(GeV/c)$  &\aaa $180^\circ$\aaa&\aaa$17^\circ$\aaa&\aaa$11^\circ$\aaa&\aaa$6^\circ$\aaa \\\hline\hline
    0 --   5 &      1671 &      1671 &      1670 &      1640 \\\hline
    5 --  10 &      1553 &      1553 &      1552 &      1379 \\\hline
   10 --  15 &       407 &       407 &       399 &       264 \\\hline
   15 --  20 &        70 &        70 &        63 &        40 \\\hline
   20 --  25 &        24 &        23 &        21 &        19 \\\hline
   25 --  30 &        12 &        12 &        12 &        10 \\\hline
   30 --  40 &        18 &        18 &        18 &        18 \\\hline
   40 --  50 &         9 &         9 &         8 &         8 \\\hline
   50 -- 100 &        11 &        11 &        11 &         8 \\\hline
  100 -- 300 &         0 &         0 &         0 &         0 \\\hline
  300 -- 500 &         0 &         0 &         0 &         0 \\\hline
    0 -- 500 &      3773 &      3773 &      3755 &      3387 \\\hline      
\end{tabular}
\vskip0.2cm
\small
\caption{Number of events dependence on $\dphi_{max}$ and on
$\Pt^{\gamma}$ for $L_{int}=300~ pb^{-1}$,}
\vskip-3pt
{\footnotesize $\Pt^{clust}_{CUT}=5 ~GeV/c$ (summary).}
\vskip0.2cm
\begin{tabular}{||c||r|r|r|r||} \hline \hline
\label{tab:pt56-6}
 $\Pt{56}$  &\multicolumn{4}{c||}{ $\dphi_{max}$} \\\cline{2-5}
 $(GeV/c)$  &\aaa $180^\circ$\aaa&\aaa$17^\circ$\aaa&\aaa$11^\circ$\aaa&\aaa$6^\circ$\aaa \\\hline\hline
 40 -- 50  &  37576 &     37235 &     35473 &     27025 \\\hline  
 50 -- 70  &  19056 &     19017 &     18651 &     15149 \\\hline      
 70 -- 90  &   3773 &      3773 &      3755 &      3387 \\\hline 
 90 -- 140 &   1525 &      1525 &      1524 &      1468 \\\hline     
\end{tabular}
\end{center}
\end{table}      

\def\baselinestretch{1.0}

A tendency of distributions of the number of signal \gpj events to be
very rapidly concentrated in a rather narrow
back-to-back angle interval $\Delta\phi\lt17^\circ$ as $\Pt^{\gamma}$ grows
becomes more distinct with a more restrictive cut
(see Tables \ref{tab:pt56-4}, \ref{tab:pt56-5} and
\ref{tab:pt56-6}). From the last summary line of Table \ref{tab:pt56-4} we see that in the first interval
 $40\leq \Pt^{\gamma}\leq 50\, GeV/c~$ more than $99\%$ of the events, selected 
with  $\Pt^{clust}_{CUT}= 5\,GeV/c$, have $\Delta\phi\lt17^\circ$, 
while $~72\%$ of them are in the $\Delta\phi\lt6^\circ$ range. 
It should be mentioned that after application of this cut only about $40\%$ of events remain.
For $70\leq \Pt^{\gamma}\leq 90\, GeV/c$ (see Table \ref{tab:pt56-5}) more than  $90\%$ of the events, 
subject to the cut $\Pt^{clust}_{CUT}= 5 ~GeV/c$, have $\Delta\phi\lt6^\circ$. 
It means that while suppressing cluster or mini-jet activity by imposing $\Pt^{clust}_{CUT}=5 ~GeV/c$
we can select the sample of events with a clean ``back-to-back'' (within 17$^\circ$ in $\phi$) 
topology of $\gamma$ and jet orientation.
(Unfortunately, as it will be discussed below basing on the information from Tables 5 and 6,
it does not mean that $\Pt^{clust}_{CUT}$ allows to suppress completely the ISR
as is seen from Tables 5 and 6.)
\footnote{See also the event spectra over $\Pt^{clust}$ in Fig.~7 of the following Section 6.}.

So, one can conclude that PYTHIA simulation predicts that
at Tevatron  energies most of the \gpj events (more than $75\%$)
may have the vectors $\vec{\Pt}^{\gamma}$ and $\vec{\Pt}^{jet}$ being back-to-back
within $\Delta\phi\lt17^\circ$ after imposing $\Pt^{clust}_{CUT}=30~GeV/c$.  
The cut $\Pt^{clust}_{CUT}=5~GeV/c$ significantly improves
\footnote{An increase in \ptg produces the same effect, as is seen
from comparison of Tables \ref{tab:pt56-1} and \ref{tab:pt56-2} and
will be demonstrated in more detail in Section 6 and Appendices 2--5.}
this tendency.

It is worth mentioning that this picture
reflects the predictions of one of the generators
based on the approximate  LO values for
the cross section. It may change if the
next-to-leading order or soft physics
\footnote{We thank E.~Pilon and J.~Ph.~Jouliet for the information
about new Tevatron data on this subject and for clarifying the importance
of NLO corrections and soft physics effects.}
effects are included.

The other lines of Tables \ref{tab:pt56-1}, \ref{tab:pt56-2} and
\ref{tab:pt56-4}, \ref{tab:pt56-5} contain the information
about the $\Pt56$ spectrum
or, up to intrinsic transverse parton momentum $\langle k_t \rangle=0.44\;GeV/c$,
about $\Pt^{ISR}$ spectrum).

From Tables 2 and 3 one can see that in the case when there are no restrictions
on $\Pt^{clust}$ the $\Pt56$ spectrum becomes a bit wider for larger values of
$\Pt^{\gamma}$. 

At the same time, one can conclude from the comparison of Table \ref{tab:pt56-1}  with  Table \ref{tab:pt56-4}
that for lower $\Ptg$ intervals the width of the most populated part of the $\Pt56$ (or $\Pt^{ISR}$) 
spectrum reduces by about 40$\%$ with restricting $\Pt^{clust}_{CUT}$. So, for $\dphi_{max}=17^\circ$ we
see that it drops from $0\lt\Pt{56}\lt20\;GeV/c$ $\!$ for $\!$ $\Pt^{clust}_{CUT}=30\;GeV/c$ to a
$\!$ narrower interval $\,$ of $\!$
$0\lt\Pt{56}\lt10\,GeV/c$ $\!$ for $\!$ the $\Pt^{clust}_{CUT}=5\,GeV/c$.
At higher $\Ptg$ intervals (Tables \ref{tab:pt56-2} and \ref{tab:pt56-5})
for the same value $\dphi_{max}=17^\circ$ the reduction factor of the $\Pt56$ spectrum width 
(from the $0\lt\Pt56\lt30~ GeV/c$ interval for $\Pt^{clust}_{CUT}=30\,GeV/c$
to the $0\lt\Pt56\lt10-15 ~GeV/c$ interval for $\Pt^{clust}_{CUT}=5 ~GeV/c$)
is more than two. 

Thus, we can summarize that the PYTHIA  generator predicts
an increase in the $\Pt^{ISR}$ spectrum with growing $\Pt^{\gamma}$ (compare Tables 2 and 3), but this increase
can be reduced by imposing a restrictive cut on 
$\Pt^{clust}$ (for more details see Sections 6 and 7).

So, the $\Pt{56}$ spectra presented in Tables
\ref{tab:pt56-1}, \ref{tab:pt56-2} and \ref{tab:pt56-3},
\ref{tab:pt56-4} show PYTHIA prediction that the ISR effect is a large one at Tevatron
energies. Its $\Pt$ spectrum continues at least up to $\Pt{56}=
10 ~GeV/c$ in the case of $\Ptg$ (or $\Pt^{jet}$) $\approx 90 ~GeV/c$
(and up to higher values as $\Ptg$ grows) even for $\Pt^{clust}_{CUT}=5 ~GeV/c$. It cannot be
completely suppressed  by $\Delta\phi$ and $\Pt^{clust}$ cuts alone.
(In Section 8 the effect of the additional $\Pt^{out}_{CUT}$ will be discussed)
Therefore we prefer to use the $\Pt$ balance equation for
the event as a whole (see equations (16) and (28) of Sections 3.1 and 3.2), i.e. an equation
that takes into account the ISR and FSR effects,
rather than balance equation (2) for fundamental processes (1a) and (1b) as discussed in Section 2.1.
(In Section 6 we shall study a behavior of each term that enter equation (28) in order to find the criteria that 
would allow to select events with a good balance of \Ptgj).

Since the last lines in Tables \ref{tab:pt56-1}, \ref{tab:pt56-2} and
\ref{tab:pt56-4}, \ref{tab:pt56-5} contain an illustrative
information on $\Delta\phi$ dependence of the total number of events, we supply these tables with the
summarizing Tables \ref{tab:pt56-3} and \ref{tab:pt56-6}. They include more $\Pt^{\gamma}$ 
intervals and contain analogous numbers of events that can be collected
in different $\Delta\phi$ intervals for two different
 $\Pt^{clust}$ cuts at $L_{int}=300\,pb^{-1}$.

\subsection{$\Pt^{\gamma}$ and $\eta^{\gamma}$ dependence of event rates.}

~\\[-12mm]

\def\baselinestretch{1.0}
\begin{flushleft}
\parbox[r]{.5\linewidth}{
Here we shall present
the number of events for different $\Pt^{\gamma}$ and $\eta^{\gamma}$ intervals as
predicted by PYTHIA simulation with weak cuts defined mostly by (29)
with only change of $\Pt^{clust}_{CUT}$ value from 30 to 10$~GeV/c$.
The lines of Table \ref{tab:pt-eta} correspond to  $\Pt^{\gamma}$ intervals
and the columns to $\eta^{\gamma}$ intervals. The last column of this table contains the total number
of events  (at $L_{int}=300\,pb^{-1}$) in the whole ECAL $\eta^{\gamma}$-region
$|\eta^{\gamma}|\lt2.5$ for a given $\Pt^{\gamma}$ interval.
We see that the number of events  decreases fast
with growing $\Pt^{\gamma}$ (by more than 50$\%$ for each subsequent interval). For the fixed
$\Pt^{\gamma}$ interval the dependence on $\eta^{\gamma}$ 
is given in lines of Table 8 and illustrated by Fig.~6.
}
\end{flushleft}
\begin{flushright}
\begin{figure}[h]
\vskip-90mm
  \hspace{85mm} \includegraphics[width=105mm,height=78mm]{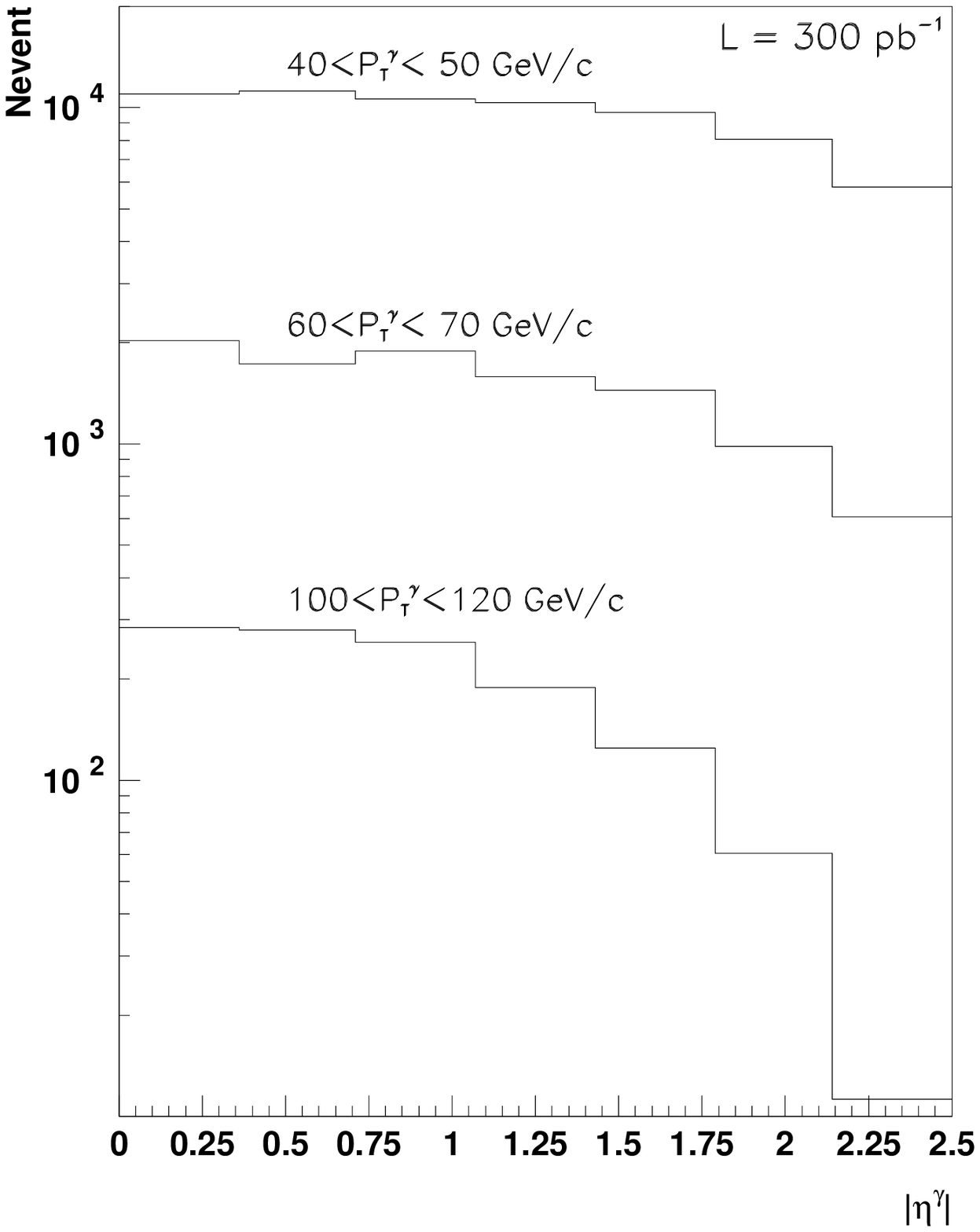}
  \label{fig:eta}
\end{figure}
\end{flushright}
\vspace{-1.7cm}
\hspace*{10cm} {\footnotesize {Fig.~6: $\eta$-dependence of rates for
different\\}}
\hspace*{11cm} {\footnotesize  {$\Pt^{\gamma}$ intervals.}}
\setcounter{figure}{5}
\def\baselinestretch{0.99}
\begin{table}[htbp]
~\\[-12mm]
\small
\begin{center}
\caption{Rates for $L_{int}=300\,pb^{-1}$ for different $\Pt^{\gamma}$ intervals and $\eta^{\gamma}$
($\Pt^{clust}_{CUT}= 10 \, GeV/c$ and $\Delta\phi \leq 17^\circ$).}
\vskip0.2cm
\begin{tabular}{||c||r|r|r|r|r|r|r||r||} \hline \hline
\label{tab:pt-eta}
$\Pt^{\gamma}$ &\multicolumn{7}{c||}{$\eta^{\gamma}$~~ intervals}
&all ~~$\eta^{\gamma}$ \\\cline{2-9}
$(GeV/c)$ & 0.0-0.4 & 0.4-0.7 & 0.7-1.1 & 1.1-1.4 & 1.4-1.8 & 1.8-2.1 & 2.1-2.5 &0.0-2.5
 \\\hline \hline
 40 -- 50  &   10978  &  11232  &  10604  &  10337  &   9662  &   8051 &    5806 &   66679 \\\hline  
 50 -- 60  &    4483  &   4210  &   4489  &   3938  &   3624  &   2814 &    1562 &   25121 \\\hline  
 60 -- 70  &    2028  &   1732  &   1890  &   1587  &   1442  &    984 &     607 &   10270 \\\hline  
 70 -- 80  &     949  &    931  &    937  &    753  &    637  &    392 &     170 &    4770 \\\hline  
 80 -- 90  &     508  &    513  &    469  &    363  &    309  &    180 &      62 &    2405 \\\hline  
 90 --100  &     302  &    287  &    252  &    201  &    149  &     80 &      25 &    1295 \\\hline  
100 --120  &     285  &    280  &    257  &    189  &    125  &     61 &      11 &    1207 \\\hline  
120 --140  &     134  &    121  &     98  &     63  &     38  &      9 &       1 &     465 \\\hline\hline    
 40 --140  &   19662  &  19302  &  18992  &  17427  &  15986  &  12571 &    8245 &  112216 \\\hline\hline      
\hline
\end{tabular}
\end{center}
\end{table}

\begin{table}[h]
~\\[-7mm]
\normalsize
\subsection{Estimation of \gpj event rates for different calorimeter regions.} 
Since a jet is a wide-spread object, the $\eta^{jet}$ dependence of rates
for different $\Pt^{\gamma}$ intervals will be presented in a different way than in Section 5.2.
Namely, Tables \ref{tab:sh1}--\ref{tab:sh4} include the rates of 
events ($L_{int}=300\,pb^{-1}$) 
for different $\eta^{jet}$ intervals, covered by the 
central, intercryostat, end and forward (CC, IC, EC and FC) parts of the calorimeter and 
for different $\Ptg(\approx\Pt^{Jet})$ intervals.
\begin{center}  
\small
\caption{Selection 1. $\Delta \Pt^{jet} / \Pt^{jet} = 0.00$~~~
($\Pt^{clust}_{CUT}= 10 \, GeV/c, ~~\Delta\phi \leq 17^\circ$ and $L_{int}=300\,pb^{-1}$ ).}
\vskip0.2cm
\begin{tabular}{||c||c|c||c|c||c|c||c|c||} \hline \hline
\label{tab:sh1}
$\Pt^{\gamma}$&$CC$&$CC\!\to \! IC$&$IC$&$IC\!\to \! CC,EC$&$EC$ &$EC\!\to\! IC,FC$&$FC$&$FC\!\to\! EC$  
\\\hline \hline
 40 -- 50  &    9965  &  13719  &   8152 &   22225  &    617 &    8854  &    554  &   1912\\\hline
 50 -- 60  &    4009  &   5597  &   3104 &    8791  &    207 &    2766  &    109  &    413\\\hline
 60 -- 70  &    1754  &   2515  &   1339 &    3615  &     71 &     979  &     14  &     93\\\hline
 70 -- 80  &     930  &   1195  &    651 &    1593  &     21 &     348  &      1  &     23\\\hline
 80 -- 90  &     503  &    596  &    328 &     811  &      9 &     136  &      0  &      6\\\hline
 90 -- 100 &     283  &    352  &    165 &     421  &      3 &      59  &      0  &      1\\\hline
100 -- 120 &     263  &    351  &    137 &     389  &      2 &      37  &      0  &      0\\\hline
120 -- 140 &     118  &    143  &     50 &     142  &      1 &       7  &      0  &      0\\\hline\hline
 40 -- 140 &   17822  &  24462  &  13927 &   37988  &    930 &   13184  &    678  &   2448 \\\hline\hline   
\end{tabular}
\vskip.4cm
\caption{Selection 1. $\Delta \Pt^{jet} / \Pt^{jet} \leq 0.05$~~~
($\Pt^{clust}_{CUT}= 10 \, GeV/c, ~~\Delta\phi \leq 17^\circ$ and $L_{int}=300\,pb^{-1}$ ).}
\vskip0.2cm
\begin{tabular}{||c||c|c||c|c||c|c||c|c||} \hline \hline
\label{tab:sh2}
$\Pt^{\gamma}$&$CC$&$CC\!\to \! IC$&$IC$&$IC\!\to \! CC,EC$&$EC$ &$EC\!\to\! IC,FC$&$FC$&$FC\!\to\! EC$  
\\\hline \hline
 40 -- 50  &   17951 &    5733 &   20631 &    9746 &    4174 &    5296 &    1280 &    1186\\\hline
 50 -- 60  &    7466 &    2141 &    8313 &    3583 &    1403 &    1570 &     253 &     269\\\hline
 60 -- 70  &    3405 &     863 &    3553 &    1401 &     492 &     558 &      39 &      68\\\hline
 70 -- 80  &    1699 &     426 &    1667 &     577 &     179 &     190 &       6 &      17\\\hline
 80 -- 90  &     902 &     197 &     838 &     301 &      75 &      71 &       3 &       4\\\hline
 90 --100  &     528 &     107 &     440 &     146 &      31 &      31 &       0 &       0\\\hline
100 --120  &     537 &      98 &     384 &     142 &      19 &      20 &       0 &       0\\\hline
120 --140  &     223 &      37 &     143 &      48 &       5 &       3 &       0 &       0\\\hline\hline
 40 --140  &   32701 &    9603 &   35971 &   15943 &    6377 &    7738 &    1582 &    1545\\\hline\hline 
\end{tabular}
\end{center}
\end{table}

\newpage
No restrictions on other parameters are used. The first columns of these tables $CC$ give the
number of events with the jets (found by the LUCELL jetfinding algorithm of PYTHIA),
all particles of which are comprised (at the particle level of simulation)
entirely (100$\%$) in the CC part and there is a
$0\%$ sharing of  $\Pt^{jet}$ ($\Delta \Pt^{jet}=0$) between the CC and
the neighboring IC part of the calorimeter. The second columns of the 
tables $CC\!\to\!IC$ contain the number of events in which $\Pt$ of the jet is shared
between the CC and IC regions. The same sequence of restriction
conditions takes place in the next columns. Thus, the 
$IC, EC$ and $FC$ columns include the number of events with jets
entirely contained in these regions, while the $EC\!\to\!IC,FC$ column gives the number of  events where
the jet covers both the EC and IC or EC and FC regions. From these tables we can see what number of events
can, in principle, most suitable for the precise jet energy absolute scale setting,
 carried out separately for the CC, EC and FC parts of the calorimeter in different $\Ptg$ 
intervals.
\newpage
\normalsize
\def\baselinestretch{1.0}
The selection cuts are as those of Section 3.2  specified by the following values of the cut
parameters: \\[-7mm]
\begin{eqnarray}
\Pt^{isol}_{CUT}=4\;GeV/c; \quad
{\epsilon}^{\gamma}_{CUT}=7\%; \quad
\Delta \phi\lt17^{\circ}; \quad
\Pt^{clust}_{CUT}=10\;GeV/c.
\end{eqnarray}
\begin{table}[h]
\vskip-10mm
\begin{center} 
 \small
\caption{Selection 2. $\Delta \Pt^{jet} / \Pt^{jet} = 0.00$~~~
($\epsilon^{jet}<3\%, \Pt^{clust}_{CUT}=10~GeV/c, \Delta\phi \leq17^\circ$ and $L_{int}=300\,pb^{-1}$ ).}
\vskip-0.2cm
\begin{tabular}{||c||c|c||c|c||c|c||c|c||} \hline \hline
\label{tab:sh3}
$\Pt^{\gamma}$&$CC$&$CC\!\to \! IC$&$IC$&$IC\!\to \! CC,EC$&$EC$ &$EC\!\to\! IC,FC$&$FC$&$FC\!\to\! EC$  
\\\hline \hline
 40- 50  &    4274  &   5119  &   3916  &   8287   &   321  &   3543 &     261  &    776\\\hline
 50- 60  &    2031  &   2472  &   1766  &   3879   &   121  &   1215 &      66  &    194\\\hline
 60- 70  &     989  &   1330  &    852  &   1834   &    52  &    503 &       9  &     41\\\hline
 70- 80  &     586  &    663  &    444  &    923   &    17  &    192 &       0  &     14\\\hline
 80- 90  &     338  &    367  &    241  &    505   &     8  &     81 &       0  &      3\\\hline
 90-100  &     207  &    233  &    126  &    287   &     3  &     43 &       0  &      0\\\hline
100-120  &     223  &    251  &    112  &    282   &     2  &     28 &       0  &      0\\\hline
120-140  &      97  &    110  &     42  &    115   &     0  &      7 &       0  &      0\\\hline\hline
 40-140  &    8743  &  10544  &   7499  &  16108   &   523  &   5611 &     337  &   1028\\\hline\hline      
\end{tabular}
\vskip0.4cm
\caption{Selection 2. $\Delta \Pt^{jet} / \Pt^{jet} \leq 0.05$~~~
($\epsilon^{jet}<3\%, \Pt^{clust}_{CUT}= 10~GeV/c, \Delta\phi\leq17^\circ$ and $L_{int}=300\,pb^{-1}$ ).}
\vskip0.2cm
\begin{tabular}{||c||c|c||c|c||c|c||c|c||} \hline \hline
\label{tab:sh4}
$\Pt^{\gamma}$&$CC$&$CC\!\to\!IC$&$IC$&$IC\!\to\! CC,EC$&$EC$ &$EC\!\to\! IC,FC$&$FC$&$FC\!\to\! EC$  
\\\hline \hline
 40- 50  &    7384  &   2009  &   8858  &   3344  &   1912  &   1952  &    557 &     480\\\hline
 50- 60  &    3689  &    813  &   4157  &   1488  &    729  &    606  &    150 &     110\\\hline
 60- 70  &    1907  &    412  &   1991  &    695  &    305  &    251  &     23 &      27\\\hline
 70- 80  &    1027  &    223  &   1040  &    326  &    116  &     93  &      3 &      11\\\hline
 80- 90  &     598  &    108  &    572  &    173  &     47  &     41  &      2 &       1\\\hline
 90-100  &     375  &     64  &    320  &     93  &     23  &     22  &      0 &       0\\\hline
100-120  &     408  &     66  &    295  &     89  &     14  &     15  &      0 &       0\\\hline
120-140  &     179  &     28  &    118  &     39  &      5  &      2  &      0 &       0\\\hline\hline
 40-140  &   15563  &   3724  &  17349  &   6247  &   3151  &   2983  &    736 &     630\\\hline\hline            \end{tabular}
\end{center}
\end{table}   
\normalsize
\def\baselinestretch{1.0}

\vskip-3mm
Less restrictive conditions, when up to $5\%$ of the jet $\Pt$
are allowed to be shared between the CC, EC and FC parts of the calorimeter, are given in 
Tables \ref{tab:sh2} and \ref{tab:sh4}.
Tables \ref{tab:sh1}  and \ref{tab:sh2} correspond to the case of Selection 1.
Tables \ref{tab:sh3} and \ref{tab:sh4} contain the  number of events collected
with Selection 2 criteria (defined in Section 3.2), i.e.
they include only the events with ``isolated jets'' satisfying the isolation criterion
$\epsilon^{jet}\lt3\%$. The reduction factor of 2 for the  number of  events can be 
found by comparing those tables with Tables 9, 10. This is a cost of passing to Selection 2.

Table \ref{tab:sh1} corresponds to the most restrictive selection
$\Delta\Pt^{jet}=0$ and gives the number of events most suitable for 
jet energy calibration. From its last summarizing line we see 
that for the entire interval $40\lt\Pt^{\gamma}\lt140\; GeV/c$~ PYTHIA predicts 
around 18000 events for CC and 1000 events for EC at integrated luminosity $L_{int}=300~pb^{-1}$.

An additional information on the number of ``CC-events'' 
(i.e. events, corresponding to $CC$ column of Table 11) 
with jets produced by $c$ and $b$ quarks in gluonic subprocess (1a), i.e. $Nevent_{(c)}$ and $Nevent_{(b)}$ 
(given for the integrated luminosity $L_{int}=300~ fb^{-1}$)
for different $\Pt^{Jet}$($\approx\Ptg$) intervals $40-50, 50-70, 70-90$ and $90-140~GeV/c$
are contained in Tables 1--12 of Appendix 1.
%
%
The ratio (``$29sub/all$'') of the number of events caused by gluonic subprocess (1a) $(=29sub)$, 
summed over quark flavours, to the number of events due to the sum of subprocesses (1a) and (1b) $(=all)$, 
also averaged over all quark flavours, is also shown there.

\section{FEATURES OF ~\gpj EVENTS IN THE CENTRAL CALORIMETER REGION.}

\it\small
\hspace*{9mm}
The influence of $\Pt^{clust}_{CUT}$ parameter (defining the upper limit on $\Pt$ of clusters or mini-jets 
in the event) on the variables characterizing the \ptgj balance  as well as on the 
$\Pt$ distribution in jets and out of them is studied.
\rm\normalsize
\vskip3mm

In this section we shall study the specific sample of events considered in 
the previous section that may be most suitable for the jet energy calibration in the CC region,
with jets entirely (100$\%$) contained in this region, i.e.
having 0$\%$ ~sharing of $\Pt^{jet}$ (at the PYTHIA particle level of simulation) with IC.
{\it Below we shall call them ''CC-events''}. The $\Ptg$ spectrum for this particular
set of events for $\Pt^{clust}=10 ~GeV/c$ was presented in the first column (CC) of Table \ref{tab:sh1}.
Here we shall use three different jetfinders, namely, LUCELL from PYTHIA
and UA1 and UA2 from CMSJET \cite{CMJ}. The  $\Pt^{clust}$ distributions for generated events found by 
the all three jetfinders in two $\Pt^{\gamma}$ intervals, $40\lt\Pt^{\gamma}\lt50~GeV/c$
and $\!$ $70\lt\Pt^{\gamma}\lt90~GeV/c$, are shown in Fig.~7 for $\Pt^{clust}_{CUT}=30\;GeV/c$.
It is interesting to note an evident similarity of the $\Pt^{clust}$ spectra with $\Pt56$ spectra
(for $\dphi\leq 17^\circ$) shown in Tables 2 and 3 (see also Figs.~8, 9), 
what support our intuitive picture of ISR and cluster connection described in  Section 2.2. \\[-0.6cm]

\begin{flushleft}
\begin{figure}[htbp]
 \vskip-19mm
 \hspace{-.5cm} \includegraphics[height=74mm,width=9.7cm]{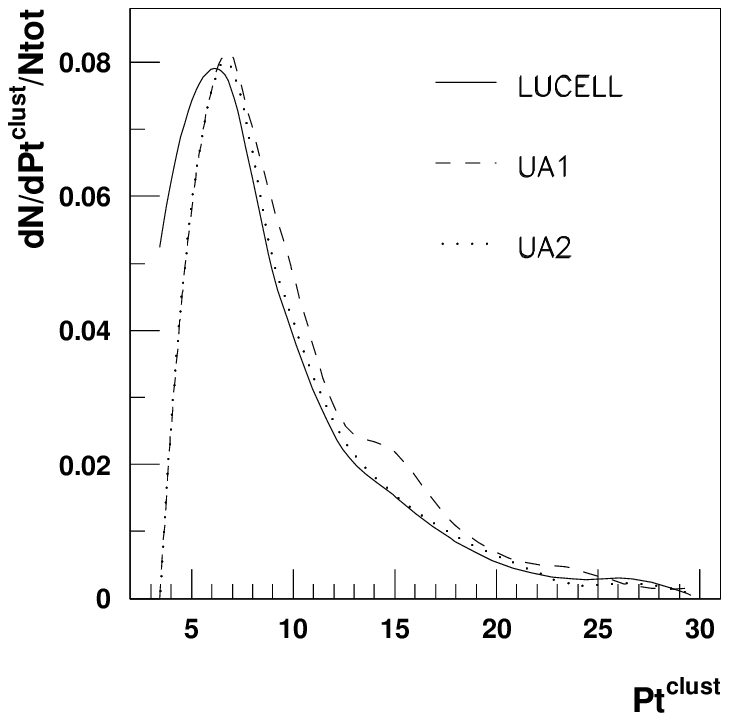}
    \label{fig9}
   \nonumber
  \end{figure}
\end{flushleft}
\begin{flushright}
\begin{figure}[htbp]
 \vskip-93.5mm
  \hspace{7.5cm} \includegraphics[height=74mm,width=9.7cm]{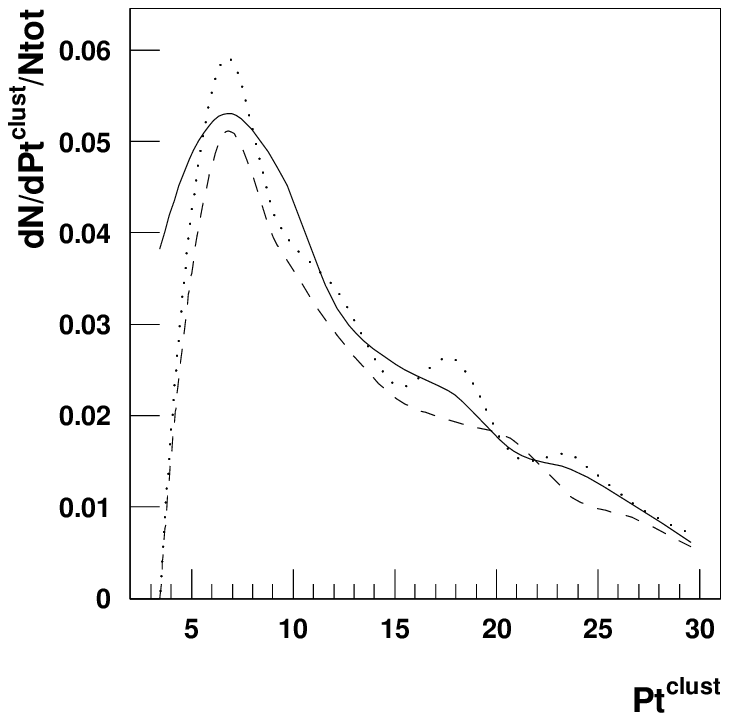}
   \nonumber
\label{fig7}
\vskip-12mm

\hspace*{4.2cm} (a) \hspace*{7.5cm} (b)\\[7pt]
\hspace*{0.3cm}{\footnotesize Fig.~7: $\Pt^{clust}$ distribution in \gpj
events from two $\Pt^{\gamma}$ intervals:
(a) $40<\Pt^{\gamma}<50\, GeV/c$ ~and \\
\hspace*{1.3cm} (b) $70<\Pt^{\gamma}<90\,GeV/c$~ with the same cut
$\Pt^{clust}_{CUT}=30\;GeV/c$ ($\dphi\leq 17^\circ$).}\\[-14mm]
\end{figure}
\end{flushright}

\subsection{Influence of the $\Pt^{clust}_{CUT}~$ parameter
on the photon and jet $\Pt$ balance and on the initial state radiation 
suppression.}

Here we shall study in more detail correlation of $\Pt^{clust}$ with $\Pt^{ISR}$ mentioned above.
The averaged value of intrinsic parton transverse momentum 
will be fixed at $\langle k_t \rangle= 0.44~ GeV/c$
\footnote{The influence of possible $\la k_t \ra$ variation on the
\ptgj~ balance is discussed in Section 9. See also \cite{BKS_P1}--\cite{BKS_P5}.}.

The banks of 1-jet \gpj events gained from the results of PYTHIA
generation of $5\cdot10^6$  signal \gpj events in each of four $\Pt^{\gamma}$
intervals (40 -- 50, 50 -- 70, 70 -- 90, 90 -- 140 $GeV/c$)
\footnote{they were discussed in Section 5}
will be used here. The observables defined in Sections 3.1 and 3.2  will be
restricted here by Selection 1 cuts (17) -- (24)
 of Section 3.2 and the cut parameters defined by (29).

We have chosen two of these intervals  to illustrate the influence of the
$\Pt^{clust}_{CUT}$ parameter on the distributions of physical variables,
 that enter the balance equation (28). These distributions are shown in Fig.~\ref{fig:40luc}
($40<\Pt^{\gamma}<50~ GeV/c$) and Fig.~\ref{fig:70luc} ($70<\Pt^{\gamma}<90~ GeV/c$). 
 In these figures, in addition to three variables $\Pt56$,
$\Pt^{|\eta|>4.2}$, $\Pt^{out}$, already explained in Sections 2.2, 3.1 and 3.2,
we present distributions of two other variables, $\Db$~ and $(1-cos\dphi)$, which
define the right-hand side of equation (28).
The distribution of the $\gamma$-jet back-to-back  angle $\dphi$ (see (22))
is also presented in Figs.~\ref{fig:40luc}, \ref{fig:70luc}.

The ISR describing variable $\Pt56$ (defined by formula (3))
and both components of the experimentally observable disbalance measure $\Fptgj$
(see (28)) as a sum of $(1-cos\dphi)$ and $\Db/\Ptg$, as well as two others, $\Pt^{out}$ and  $\dphi$,
show a tendency, to become smaller (the mean values and the widths)
with the restriction of the upper limit on the 
$\Pt^{clust}$ value (see Figs.~8, 9).
It means that the jet energy calibration precision may increase with decreasing
$\Pt^{clust}_{CUT}$, which justifies the intuitive choice of this new variable in Section 3.
The origin of this improvement becomes clear from the $\Pt{56}$ density plot, which demonstrates 
the decrease of $\Pt{56}$ (or $\Pt^{ISR}$) values with decrease of $\Pt^{clust}_{CUT}$.
In Section 2.3 we gave arguments why it may also influence FSR.

Comparison of Fig.~\ref{fig:40luc} (for $~40\!<\Pt^{\gamma}\!<50 ~GeV/c$) and Fig.~\ref{fig:70luc}
(for $~70\!<\Pt^{\gamma}\!<90 ~GeV/c$) also shows that the values of $\Delta\phi$ as a degree of
back-to-backness of the photon and jet $\Pt$ vectors in the $\phi$-plane
decreases with increasing $\Pt^{\gamma}$. At the same time $\Pt^{out}$ and $\Pt^{ISR}$ distributions 
become slightly wider. It is also seen that
the $\Pt^{|\eta|>4.2}$ distribution practically does not depend on
$\Pt^{\gamma}$ and $\Pt^{clust}$
\footnote{see also Appendices 2--5}.

It should be mentioned that the results presented in Figs.~\ref{fig:40luc} and \ref{fig:70luc} were
 obtained with the LUCELL jetfinder of PYTHIA
\footnote{The results obtained with all jetfinders and
\ptgj ~balance will be discussed in Section 7 in more detail.}.


\subsection{$\Pt$ distribution inside and outside of a jet.}

Now let us see what spatial distribution may have the $\Pt$ activity in the volume outside the jet
(i.e. in the calorimeter cells outside the jet cone) in these CC \gpj events collected with the Selection 1 cuts. 
For this purpose we calculate a vector sum $\vec{\Pt}^{sum}$ of individual 
transverse momenta of $\Delta \eta \times \Delta \phi$ cells included
by a jetfinder into a jet and of cells in a larger
volume that surrounds a jet. In the latter case this procedure can be viewed as 
straightforward enlarging of the jet radius in the $\eta -\phi$ space.

The figures that show the ratio  $\Pt^{sum}/{\Pt^{\gamma}}$
as a function of the distance $R( \eta,\phi)$ counted from the jet
gravity center towards its boundary and further into the space outside the jet
are shown in 
the left-hand columns of
Figs.~\ref{fig:40rjet} and \ref{fig:70rjet} for two different $\Ptg$ intervals ($40<\Ptg<50 ~GeV/c$ 
in Fig.~\ref{fig:40rjet} and the $70<\Ptg<90 ~GeV/c$ in Fig.~\ref{fig:70rjet}). 

From these figures we see that the space surrounding the jet 
in general, i.e. for Selection 1, is far from being an empty
in the case of \gpj events considered here. 
We also see that an average value of
the total $\Pt^{sum}$ increases with increasing volume around the jet
and it exceeds $\Pt^{\gamma}$ at $R=0.8-1.0$ (see Figs.~\ref{fig:40rjet} and \ref{fig:70rjet}).

From the right-hand columns of Figs.~\ref{fig:40rjet} and \ref{fig:70rjet} we also see that  
the disbalance measure (the analog of (4)) \\[-4mm]
\begin{equation}
\Pt^{\gamma+sum}=
\left|\vec{\Pt}^{\gamma}+\vec{\Pt}^{sum}\right|
~\\[-1mm]
\end{equation}

\noindent
achieves its minimum at $R\approx 0.9-1.1$ for all three jetfinding algorithms.

The value of $\Pt^{\gamma+sum}$ continues to grow  with increasing
$R$ after the point $R=1.0$ for $40<\Pt^{\gamma}<50~ GeV/c$
(see Figs.~\ref{fig:40rjet}), while for higher $\Pt^{\gamma}$ (see Figs.~\ref{fig:70rjet} for the
$70<\Pt^{\gamma}<90 ~GeV/c$ 
\setcounter{figure}{7}
\begin{figure}[htbp]
\vspace{-3.0cm}
  \hspace{-0mm} \includegraphics[width=16cm]{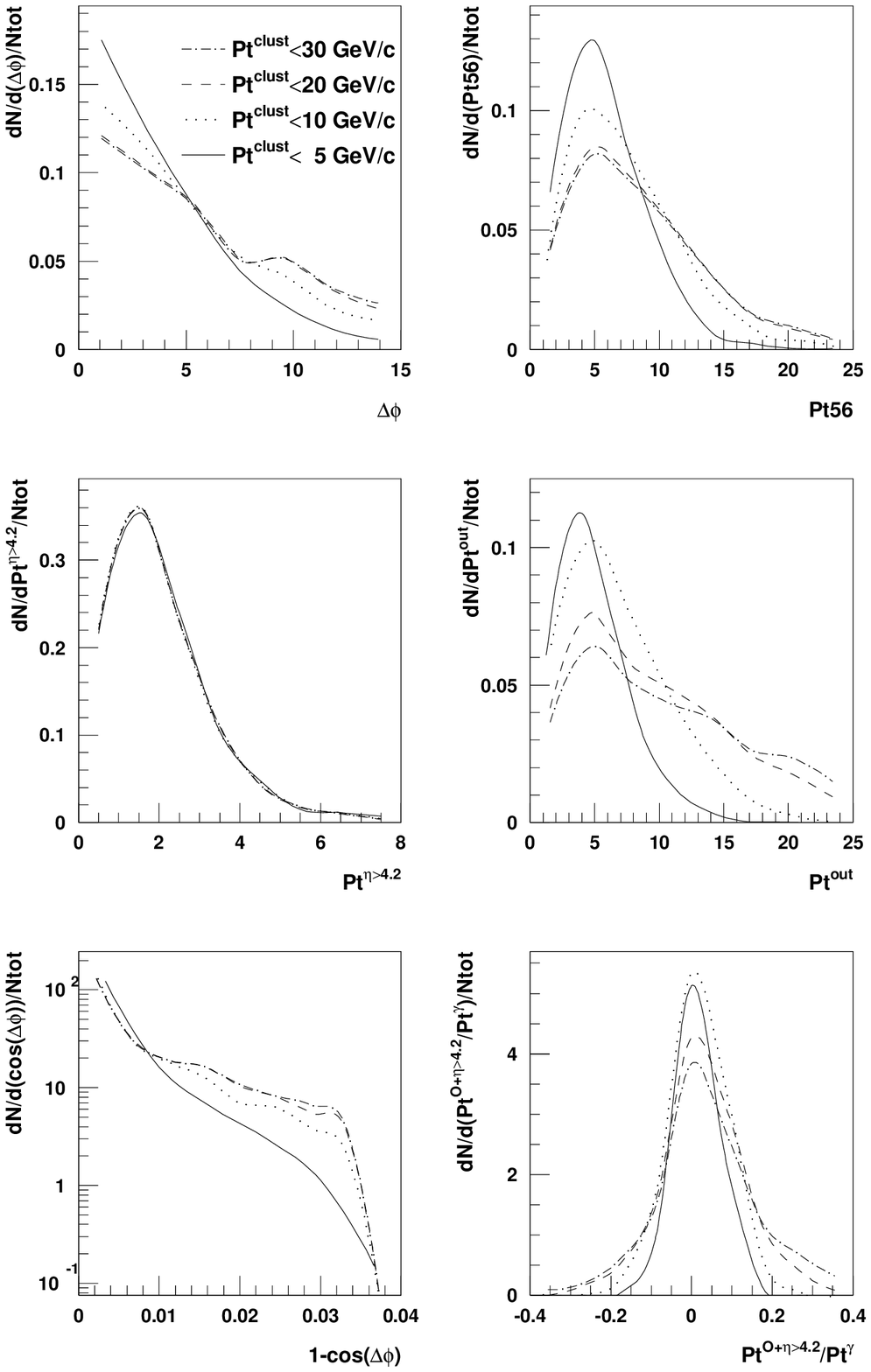} 
  \vspace{-0.5cm}
\caption{\hspace*{0.0cm} LUCELL algorithm, $\dphi<17^\circ$;~~
$40<\Pt^{\gamma}<50\, GeV/c$. Selection 1.}
    \label{fig:40luc}
\end{figure} 
\begin{figure}[htbp]
 \vspace{-3.0cm}
  \hspace{-0mm} \includegraphics[width=16cm]{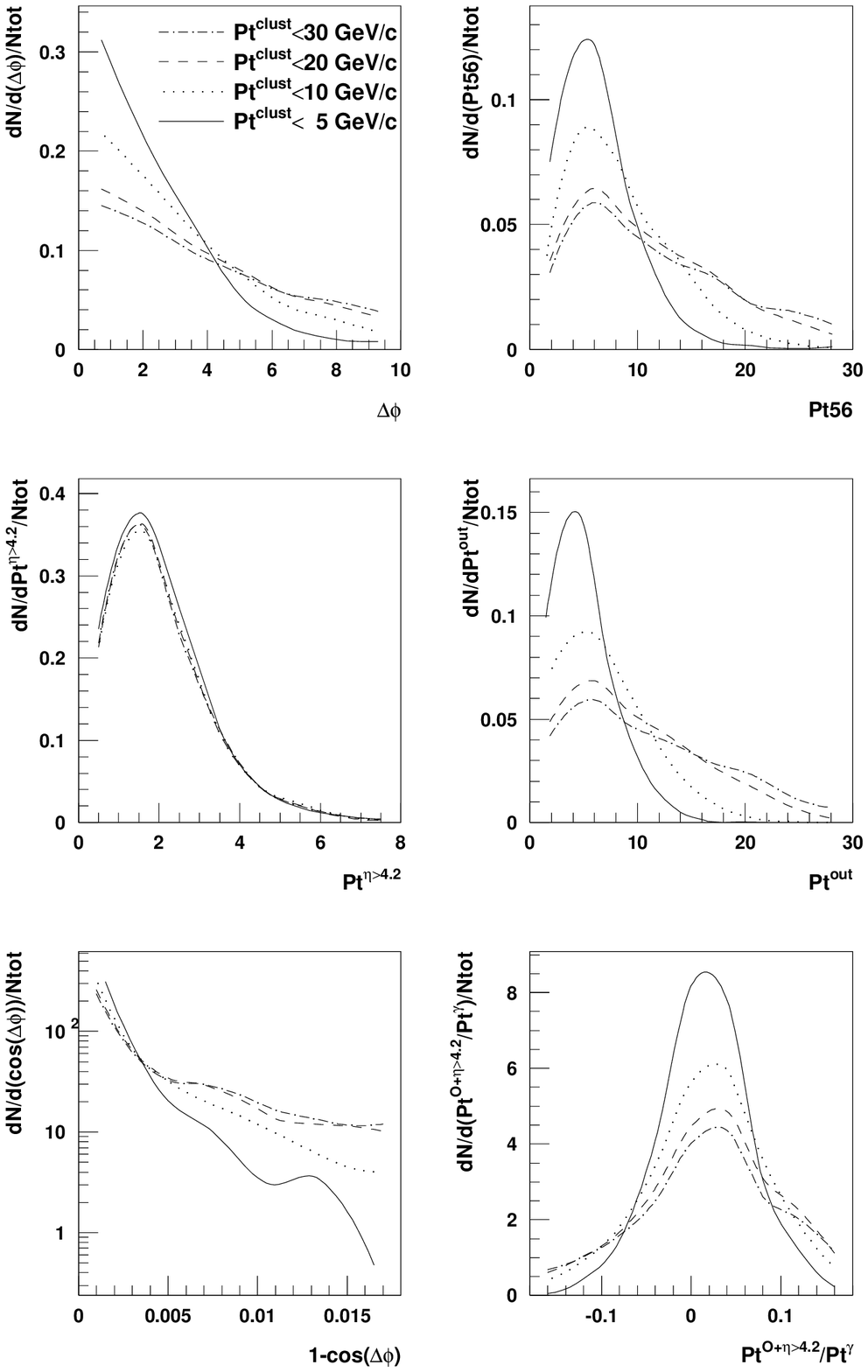}
  \vspace{-0.5cm}
\caption{\hspace*{0.0cm} LUCELL algorithm, $\dphi<17^\circ$;~~
$70<\Pt^{\gamma}<90\, GeV/c$. Selection 1.}
    \label{fig:70luc}
\end{figure} 
\begin{center}
\begin{figure}[htbp]
 \vspace{-3.0cm}
  \hspace{-0mm} \includegraphics[width=16cm]{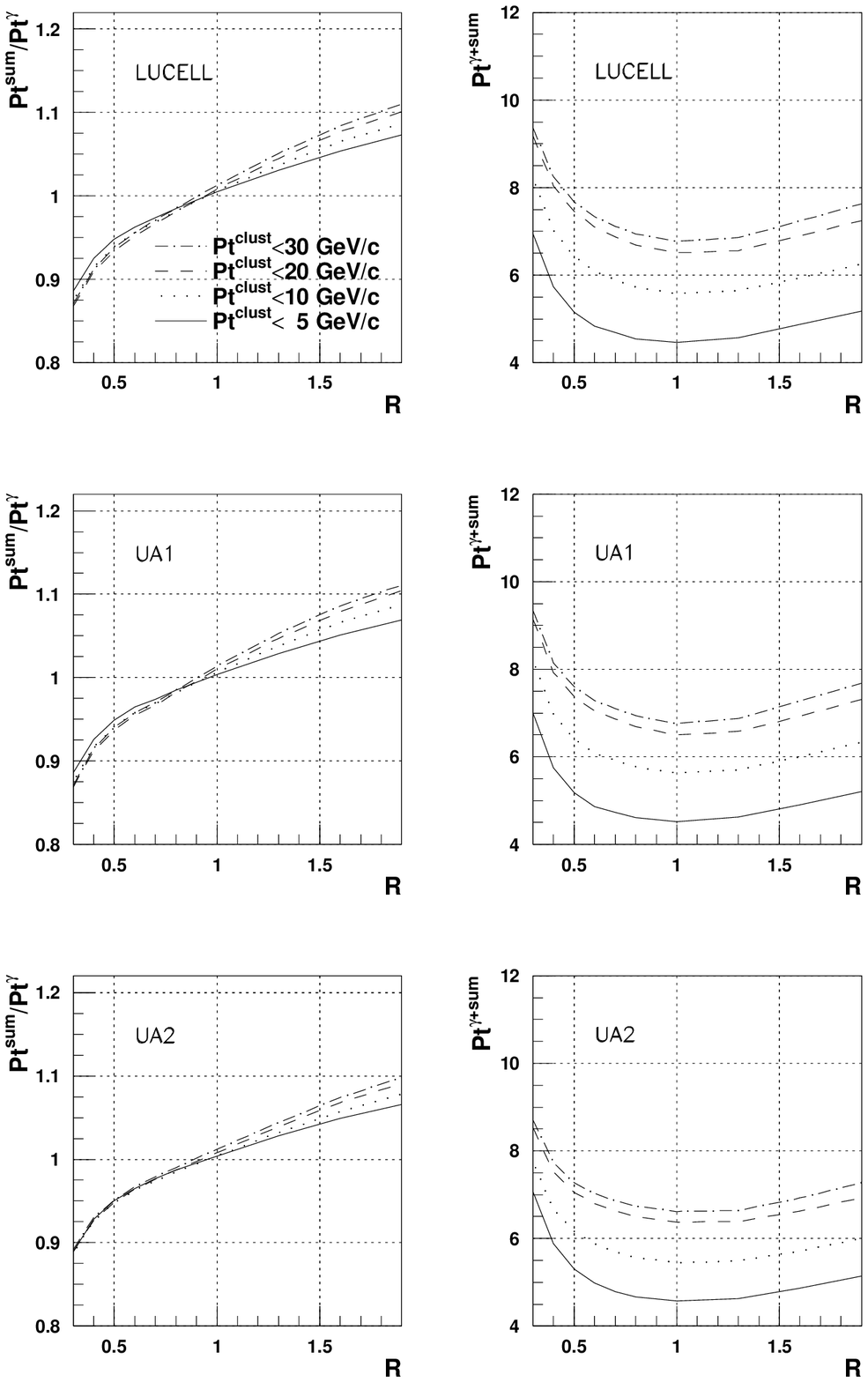} 
  \vspace{-0.5cm}
\caption{
\hspace*{0.0cm} LUCELL, UA1 and UA2 algorithms, $\dphi<17^\circ$;~~
$40<\Pt^{\gamma}<50\, GeV/c$. Selection 1.}
    \label{fig:40rjet} 
\end{figure}
\begin{figure}[htbp]
 \vspace{-3.0cm}
  \hspace{-0mm} \includegraphics[width=16cm]{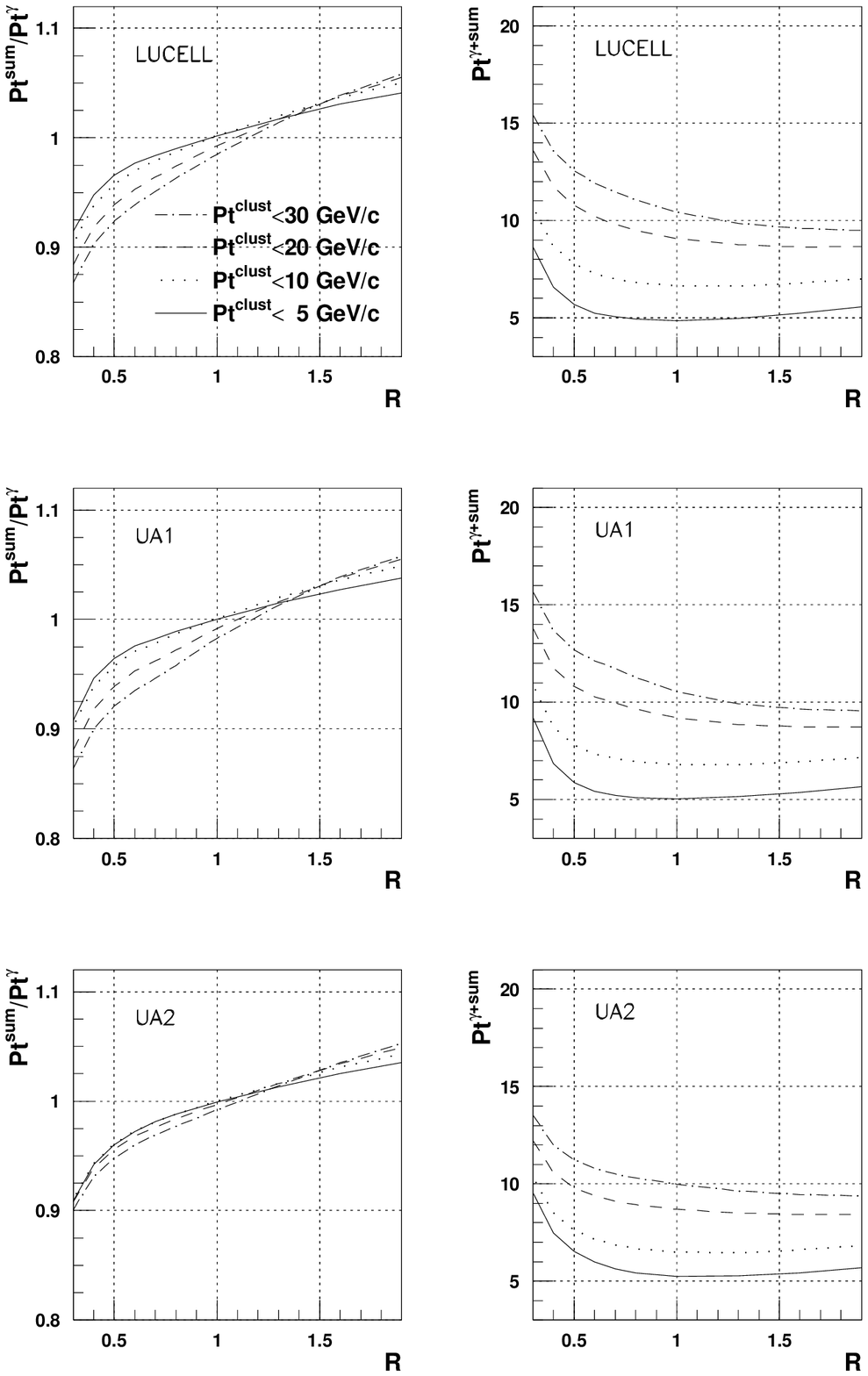}
  \vspace{-0.5cm}
    \caption{
\hspace*{0.0cm} LUCELL, UA1 and UA2 algorithms, $\dphi<17^\circ$;~~
$70<\Pt^{\gamma}<90\, GeV/c$. Selection 1.}
    \label{fig:70rjet}
  \end{figure}
\end{center}

\vskip-7mm
\noindent
interval) the ratio $\Pt^{sum}/\Pt^{\gamma}$ and the disbalance
measure $\Pt^{\gamma+sum}$ increase more slowly with 
increasing $R$ after the point $R=1.0$. 

This means that at higher $\Pt^{\gamma}$
(or $\Pt^{Jet}$) the topology of \gpj events becomes more
pronounced and we get a clearer picture of an ``isolated'' jet. 
This feature clarifies the motivation of introducing following \cite{BKS_P1}--\cite{BKS_P5} the ``Selection 2'' criteria
in Section 3.2 (see point 9) for selection of events with ``isolated jets''.

\section{DEPENDENCE OF THE $\Pt$-DISBALANCE IN THE \gpj SYSTEM 
ON $\Pt^{clust}_{CUT}$ and $\Pt^{out}_{CUT}$ PARAMETERS.}

\it\small
\hspace*{9mm}
It is shown that with Selection 2 (that leads to about twice reduction of the number of events $Nevent$ for
$\Ptg\lt70~GeV/c$ and to about $30-40\%$ loss of them at $\Ptg\gt70~GeV/c$) one can select 
(at the particle level) the events with the value of a fractional $\Fptgj$ disbalance better than $1\%$.
Selection 3 leads to further about $25\%$ reduction of $Nevent$. The number of events 
(at $L_{int}=300~pb^{-1}$)
and other characteristics of \gpj events are presented in tables of Appendices 2--5 for interval
$40<\Pt^{\gamma}<140~ GeV/c$.
\rm\normalsize
\vskip4mm

In the previous sections we have introduced physical variables
for studying \gpj events (Section 3) and discussed what cuts
for them may lead to a decrease in the disbalance of $\Pt^{\gamma}$ and $\Pt^{Jet}$  
(Sections 6, 7). One can make these cuts to be tighter if more events would be collected
during data taking.

Here we shall study in detail the dependence of the $\Pt$ disbalance
 in the \gpj system on $\Pt^{clust}_{CUT}$ and $\Pt^{out}_{CUT}$  values. 
For this aim we shall use the same samples of events as in Section 5 that were generated
by using PYTHIA with 2 QCD subprocesses (1a) and (1b) and collected 
to cover four $\Ptg$ intervals: 40--50, 50--70, 70--90, 90--140 $GeV/c$.
These events were selected with\\[-6mm]
\begin{eqnarray}
\Ptg\geq 40~ GeV/c, ~~~
\Pt^{jet}\geq 30~ GeV/c ~~~
\nonumber
~\\[-9mm]
\end{eqnarray}

\noindent
and with the use of the set of cut parameters defined by (29). 

The dependence of the number of events (selected with above-mentioned set of cut parameters)
on the value of $\Pt^{clust}_{CUT}$ is shown for the case of $\dphi\leq17^\circ$ and for four
$\Ptg$ intervals in Fig.~13 for Selection 1, in Fig.~15 for Selection 2 and
in Fig.~17 for Selection 3. Each of these plots is accompanied at the same page by four
additional plots that show the dependence of the fractional disbalance 
$\Fptgj$ on $\Pt^{clust}_{CUT}$ in different $\Ptg$ intervals. The dependence of this 
ratio is presented for three different jetfinders LUCEL, UA1 and UA2 used to determine
a jet in the same event. It is worth mentioning that in contrast to UA1 and LUCELL
algorithms that use a fixed value of jet radius $R^{jet}(=0.7)$, the value of
$R^{jet}$  is not restricted directly for UA2
\footnote{The only radii defining in UA2 algorithm are cone radius for preclusters search
($=0.4$) and cone radius for subsequent precluster dressing ($=0.3$) (see \cite{CMJ}).}
and, thus, it may take different values (see \cite{BKS_P2} and $R$ values in Appendices 1).
More details about the differences in the results of these three jetfinders  application 
can be found in Section 6.2, Appendices 1--5 and in \cite{GPJ}.

\begin{figure}
\vskip-10mm
\hspace{-2mm} \includegraphics[width=16cm]{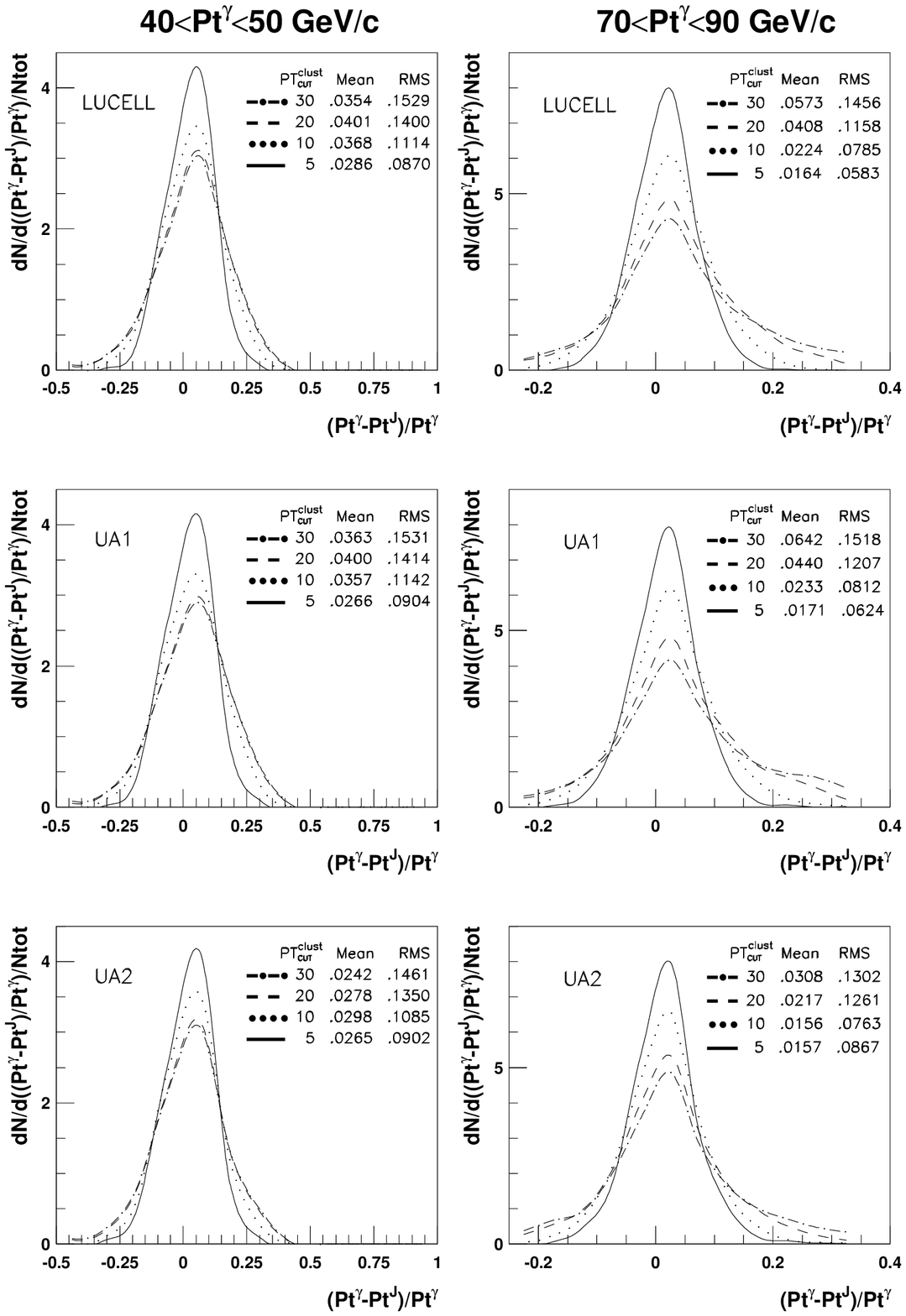} 
\caption{A dependence of $(\Pt^{\gamma}-\Pt^{J})/\Pt^{\gamma}$ on
$\Pt^{clust}_{CUT}$ for LUCELL, UA1 and UA2 jetfinding algorithms and two
intervals of \ptg. ~~The mean and RMS of the distributions are displayed on
the plots. $\dphi\lt17^\circ$. $\Pt^{out}$ is not limited. Selection 1.}
\label{fig:j-clu}
\vskip25mm 
\end{figure}

The normalized event distributions over $\Fptgj$ for two most illustrative $\Ptg$ intervals
$40\lt\Ptg\lt50$ and $70\lt\Ptg\lt90 ~GeV/c$ are shown for a case of $\dphi\leq17^\circ$ 
in Fig.~\ref{fig:j-clu} in different plots for three jetfinders. These plots demonstrate
the dependence of the mean square deviations on $\Pt^{clust}_{CUT}$ value, not shown in Fig.~14.
From the comparison of Figs.~14, 16 and 18 one can easily see that passing from Selection 1
to  Selection 2 and 3 allows to select events with a better balance of \Ptgj
(about $1\%$ and better) on the PYTHIA particle level. It is also seen that in events
with ``isolated jets'' there is no such a strong dependence on $\Pt^{clust}_{CUT}$ value
in the events with $\Ptg\gt50~GeV/c$.

More details on $\Pt^{clust}_{CUT}$ dependence of different important
features of \gpj events (as predicted by PYTHIA. i.e. without account of detector effects)
are presented in tables of Appendices 2 -- 5. They include the information about 
a topology of events and mean values of most important variables that characterize
$\Ptg-\Pt^{Jet}$ disbalance
\footnote{Please note that the information about averaged values of
jet radius as well as $\Pt^{miss}$ and non-detectable content of a jet is included in the tables of
Appendix 1 for the same $\Ptg$ intervals.}.
This information can be useful as a model guideline
while performing jet energy calibration procedure and also serve for fine tuning
of PYTHIA paramters while comparing its predictions with the collected data.

Appendix 2 contains the tables for events with $\Pt^{\gamma}$ varying from $40$ to
$50~GeV/c$. In these tables we present the values of
interest found with the UA1, UA2 and LUCELL jetfinders
\footnote{the first two are taken from CMSJET fast Monte Carlo program \cite{CMJ}}
for three different Selections
mentioned in Section 3.2. Each page corresponds to a definite value of $\Delta \phi$ 
(see (22)) as a measure of deviation from the absolute
back-to-back orientation of two $\vec{\Pt}^{\gamma}$ and $\vec{\Pt}^{Jet}$ vectors.

So, Tables 1 -- 3 on the first page of each of Appendices 2--5 correspond to 
$\dphi<180^{\circ}$, i.e. to the case when no restriction
on the back-to-back $\dphi$ angle is applied. Tables 4--6 on the second page
correspond to $\dphi<17^{\circ}$.
The third and fourth pages correspond to $\dphi<11^{\circ}$
and $\dphi<6^{\circ}$ respectively. 

The first four pages of each Appendix contain information
about variables that characterize the $\Pt^{\gamma}$ -- $\Pt^{Jet}$ balance
for Selection 1, i.e. when only cuts (17)--(24) of Section 3.2 are used.

On the fifth page of each of Appendices 2--5 we present Tables
13 -- 15 for the cut $\dphi<17^{\circ}$
that correspond to Selection 2 described in Section 3.2.
Selection 2 differs from Selection 1 presented in Tables 1 -- 12
by addition of cut (25). It allows one to select events with the ''isolated jet'', i.e. events
with the total $\Pt$ activity in the $\Delta R = 0.3$ ring around the jet not 
exceeding $3\%$ of jet $\Pt$
\footnote{In contrast to the case of LHC energies, where we required in Selection 2 
$\epsilon^{jet}\leq 6-8\%$ for $40\lt\Ptg\lt 50$ (see \cite{GPJ}),  
at FNAL energies, due to less $\Pt$ activity in the space beyond the jet,
one can impose the tighter cut $\epsilon^{jet}\leq 3\%$.}.
The results obtained with Selection 3
\footnote{Selection 3 (see Section 3.2, point 10) leaves
only those events in which jets are found simultaneously by UA1, UA2 and LUCELL jetfinders
i.e. events with jets having up to a good accuracy equal coordinates of
the center of gravity, $\Pt^{jet}$ and $\phigj$.}
are given on the sixth page of Appendices 2--5. 

The columns in all Tables 1 -- 18 correspond to five different
values of cut parameter $\Pt^{clust}_{CUT}=30,\ 20,\ 15,$ $10$ and $5 ~GeV/c$.
The upper lines of Tables 1 -- 15 in Appendices 2--5
contain the expected numbers $N_{event}$ of ``CC events''
(i.e. the number of signal \gpj events in which the jet is entirely fitted into the CC region of the calorimeter; 
see Section 5) for the integrated luminosity $L_{int}=300\;pb^{-1}$. 

In the next four lines of the tables we put the values of $\Pt56$,
$\Delta \phi$, $\Pt^{out}$ and $\Pt^{|\eta|>4.2}$
defined by formulae (3), (22), (24) and (5) respectively and
averaged over the events selected with a chosen $\Pt^{clust}_{CUT}$ value.

From the tables we see that the values of $\Pt56$, $\Delta \phi$, $\Pt^{out}$ decrease fast
with decreasing $\Pt^{clust}_{CUT}\,$, while the averaged values of
$\Pt^{|\eta|>4.2}$ show very weak dependence on it (practically constant)
\footnote{Compare also with Figs.~8 and 9.}.

The following three lines (from 6-th to 8-th) present the average values of the variables
$\gpart$,
$\Jpart$,
$\gJ$ (here J$\equiv$Jet)
that serve as the measures of the $\Pt$ disbalance in the \gpp and \gpj
systems as well as the measure of the parton-to-hadrons (Jet) fragmentation effect. 

The 9-th and 10-th lines include the averaged
values of $\Db/\Pt^{\gamma}$ and $\,(1-cos(\dphi))$ quantities that appear on 
the right-hand side of equation (28) that has the meaning of the scalar variant of vector
equation (16) for the total transverse momentum conservation in a physical event.

After application of the cut $\dphi\lt17^\circ$
the value of $\left<1-cos(\dphi)\right>$ becomes smaller than the value of
$\left<\Db/\Pt^{\gamma}\right>$ in the case of Selection 1 and tends to decrease faster with
growing energy$^{\small 22,29}$. So, we can conclude that the main contribution 
into the $\Pt$ disbalance in the \gpj system, as defined by equation (28), comes from
the term $\Db/\Pt^{\gamma}$, while in Selections 2 and 3 the contribution of
$\left<\Db/\Pt^{\gamma}\right>$ reduces with growing $\Pt^{clust}$
to the level of that of $\left<1-cos(\dphi)\right>$ and even to smaller values.

We have estimated separately the contributuions two terms  
$\vec{\Pt}^{O}\cdot \vec{n}^{Jet}$ and $\vec{\Pt}^{|\eta|>4.2}\cdot \vec{n}^{Jet}$ 
(with $\vec{n}^{Jet}=\vec{\Pt}^{Jet}/\Pt^{Jet}$, see (28)) that enter $\Db$. 
Firstly from tables it is easily seen that $\Pt^{|\eta|>4.2}$ has practically the same value
in all $\Ptg$ intervals and it does not depend neither on $\dphi$ nor on $\Pt^{clust}$ values
being equal to $2 ~GeV/c$ up to a good precision. Let us emphasize that it is a prediction 
of PYTHIA. $\vec{\Pt}^{|\eta|>4.2}\cdot \vec{n}^{Jet}$ contribution is also practically constant 
($\approx 0.6~ GeV/c$) and also does not depend on $\Ptg$ or $\Pt^{clust}$.
The value of the fraction $\vec{\Pt}^{|\eta|>4.2}\cdot \vec{n}^{Jet}/\Ptg$ 
is $0.015$ at $40\lt\Ptg\lt50~GeV/c$ and decreases to $0.008$ at $70\lt\Ptg\lt 90~GeV/c$
(and to $0.006$ at $90\lt\Ptg\lt140~GeV/c$). Among these two terms the first one,
$\vec{\Pt}^{O}\cdot \vec{n}^{Jet}$, is a measurable one (its value can be
found from the numbers in lines with $\Db$). Below in this section the cuts
on the value of $\Pt^{out}$ is applied to select events with better \Ptgj balance.

The following two lines contain the averaged values of the standard deviations
{\small $\sgmgj$} and {\small $\sgmgp$} of $\gJ (\equiv Db[\gamma,J])$ and
$\gpart(\equiv Db[\gamma,part])$ respectively.
These two variables drop approximately by about $50\%$ 
(and even more for $\Pt^{\gamma}>70 ~GeV/c$)
as one goes from $\Pt^{clust}_{CUT}=30 ~GeV/c$ to $5 ~GeV/c$
for all $\Pt^{\gamma}$ intervals and for all jetfinding algorithms.

The last lines of the tables present the number of generated events (i.e. entries) 
left after cuts.

Three features are clearly seen from these tables:\\
(1) after passing from tables with $\dphi\leq180^\circ$ to those with $\dphi\leq17^\circ$,
the $\dphi$ cut, supposed to 
\hspace*{5mm} be most effective in low $\Ptg$ intervals, does not 
affect the $\Fptgj$ disbalance strongly 
\hspace*{5mm} as compared with ``jet isolation'' criterion or cut on $\Pt^{clust}$;\\
(2) in events with $\dphi\lt17^\circ$ the fractional disbalance on the {\it parton-photon} level $\gpart$\\
\hspace*{5mm} reduces to about $1\%$ (or even less) after imposing $\Pt^{clust}\lt 10~GeV/c$. It means that 
$\Pt^{clust}_{CUT}=$ \\
\hspace*{5mm} $10~GeV/c$  is really effective for ISR suppression as it was supposed in Section 3.1.\\
(3) {\it parton-to-jet} hadronization/fragmentation effect, that includes also FSR, can be estimated by
\hspace*{5mm}  the value of the following ratio $\Jpart$. It always has a negative value. It means \\
\hspace*{5mm} that a jet loses some part of the parent parton transverse momentum $\Pt^{part}$. It is seen that in  \\ 
\hspace*{5mm} the case of Selection 1 this effect gives a big contribution into \Ptgj disbalance even  \\ 
\hspace*{5mm} after application of $\Pt^{clust}_{CUT}=10~GeV/c$. The value of the fractional $\Jpart$ disbal- \\
\hspace*{5mm} ance does not vary strongly with $\Pt^{clust}_{CUT}$ in the cases of Selections 2 and 3.

We also see from the tables  that more restrictive cuts on
the observable $\Pt^{clust}$ lead to a decrease in the values of $\Pt56$ variable
(non-observable one) that serves, according to (3), as
a measure of the initial state radiation transverse momentum $\Pt^{ISR}$,
i.e. of the main source of the $\Pt$ disbalance in
the fundamental $2\to 2$ subprocesses (1a) and (1b).
Thus, variation of $\Pt^{clust}_{CUT}$ from $30 ~GeV/c$ to
$5 ~GeV/c$ (for $\dphi<17^\circ$) leads to suppression of the $\Pt56$ value
(or $\Pt^{ISR}$) approximately by $40\%$ for $40<\Pt^{\gamma}<50 ~GeV/c$
and by $\approx 60\%$ for $\Pt^{\gamma} \geq 90 ~GeV/c$.

In the first three intervals with $\Ptg\lt90~GeV/c$~ the decrease in $\Pt^{clust}_{CUT}$ leads to
some decrease in the $(\Pt^{\gamma}-\Pt^J)/\Pt^{\gamma}$
ratio. In the case of $90<\Pt^{\gamma}<140 ~GeV/c$ (for $\dphi\lt17^\circ$) the mean value of
$(\Pt^{\gamma}\!-\!\Pt^J)/\Pt^{\gamma}$ drops from $3.9\!-\!4.2\%$ to
$1.1\!-\!1.3\%$ (see Tables 4, 6 of Appendix 5).
But the value of the fractional disbalance is higher than $1\%$.
After we pass to Selections 2 and 3 this disbalance becomes of the 
$1\%$ level and smaller but at the cost of statistics loss (by about $40-60\%$).
Tables 13--18 clearly show the prediction of PYTHIA about
the best level of jet calibration precision that can be achieved
after application of Selections 2 and 3 and with the use of the above-mentioned
jet finding algorithms. The difference in the fractional disbalance $\Fptgj$
caused by their applications defines one of parts of systematical error of the calibration
procedure.

{\normalsize \it Thus, to summarize the results presented in tables of  Appendices 2--5,
we want to underline that only
after imposing the jet isolation requirement (see Tables 13 -- 15 of Appendices 2--5)
the mean values of $\Pt^{\gamma}$ and $\Pt^{Jet}$ disbalance, i.e.
$(\Pt^{\gamma}\!-\!\Pt^J)/\Pt^{\gamma}$, for all $\Ptg$ intervals 
are contained inside the $1\%$ window for any $\Pt^{clust}\leq20~GeV/c$.
The reduction of $\Pt^{clust}$ leads to lower values of mean square deviations 
of the photon-parton $Db[\gamma,part]$ and of photon-jet $Db[\gamma,J]$ balances. 
}
The Selection 2  (with $\Pt^{clust}_{CUT}=10 ~GeV/c$, for instance) 
leaves after its application the following number of events 
with jets {\it entirely contained} (see Section 5) {\it in the CC region} 
(at $L_{int}=300 ~pb^{-1}$):\\
(1) about 4000 for  $40<\Pt^{\gamma}<50 ~GeV/c$,~~~~
(2) about 3000 for  $50<\Pt^{\gamma}<70 ~GeV/c$, \\
(3) about 850 for  $70<\Pt^{\gamma}<90 ~GeV/c$ and 
(4) about 500  for the $90<\Pt^{\gamma}<140 ~GeV/c$. 

The analogous results for Selection 3 are presented in Tables 16--18 of Appendices 2--5. 
This selection leads to approximately $25-30\%$ further reduction of the number of 
selected events as compared with Selection 2
and practically does not change values of the \ptgj balance and other variables, 
presented in Tables 13--15. The advantage of Selection 3 is that it includes only events
containing  jets simultaneously found by all three used jetfinders.

So, we can say
that Selections 2 and 3, besides improving the \ptgj balance value,
are also important for selecting events with a clean jet topology and for rising the confidence level 
of a jet determination.


Up to now we have been studying the influence of the $\Pt^{clust}_{CUT}$
parameter on the balance. Let us see, in analogy with Fig.~12,
what effect is produced by $\Pt^{out}_{CUT}$ variation
\footnote{This variable enters into the expression $\Db/\Pt^{\gamma}$,
which makes a dominant contribution to the right-hand side of $\Pt$ balance
equation (28), as we mentioned above.}.

If we constrain this variable to 5 $GeV/c$, keeping $\Pt^{clust}$
slightly restricted by $\Pt^{clust}_{CUT}=30~ GeV/c$ (practically
unbound), then, as can be seen from Fig.~19, the mean
and RMS values of the $(\Pt^{\gamma}\!-\!\Pt^J)/\Pt^{\gamma}$
in the case of the LUCELL algorithm in the case of $40<\Pt^{\gamma}<50~GeV/c$ 
decrease from $3.6\%$  to $1.3\%$ and from $14.5\%$
to $7.1\%$, respectively. For $70<\Pt^{\gamma}<90~GeV/c$ 
the mean and RMS values drop from $4.5\%$  to $0.7\%$ and from $11.5\%$ to $3.7\%$ respectively.
From these plots we also may conclude that variation of $\Pt^{out}_{CUT}$ improves the
disbalance, in fact, in the same way as the variation of
$\Pt^{clust}_{CUT}$. It is not surprising as the  cluster $\Pt$
activity is a part of the $\Pt^{out}$ activity. 

The influence of the $\Pt^{out}_{CUT}$ variation (with the fixed value
$\Pt^{clust}_{CUT}=10 ~GeV/c$) on the
distribution of $(\Pt^{\gamma}\!-\!\Pt^J)/\Pt^{\gamma}$ is shown in
Fig.~20 for Selection 1. In this case the mean value of
$(\Pt^{\gamma}\!-\!\Pt^J)/\Pt^{\gamma}$ drops
from $3.2\%$ to $1.3\%$ for LUCELL and from $2.7\%$ to $1.3\%$ for UA2
algorithms for the $40<\Pt^{\gamma}<50~ GeV/c$ interval. At the same time
the RMS value changes from $12\%$ to $7\%$ for all algorithms.
For interval $70<\Pt^{\gamma}<90~ GeV/c$ the mean value of fractional disbalance
$(\Pt^{\gamma}\!-\!\Pt^J)/\Pt^{\gamma}$ decrease to less then $1\%$ at
$\Pt^{out}_{CUT}=5 ~GeV/c$ (and to $1.1-1.4\%$ at $\Pt^{out}_{CUT}=10 ~GeV/c$).
Simultaneously, RMS decreases to about $3.7\%$ for all three jetfinders.
More detailed study of $\Pt^{out}_{CUT}$ influence on the $\Fptgj$ disbalance
will be continued in the following Section 8 (see also Appendix 6).

{\it 
So, we conclude basing on the analysis of PYTHIA simulation (as a model)
 that the new cuts $\Pt^{clust}_{CUT}$ and $\Pt^{out}_{CUT}$ introduced in Section 3
as well as introduction of a new object, the ``isolated jet'', are found as those that may be
very efficient tools to improve the jet calibration accuracy }
\footnote{We plan to continue this study on the level of the full event reconstruction
after D0GSTAR simulation.}. 
Their combined usage for this aim and for the background suppression will be a subject of a further
more detailed study in Section 8. 

The results of our preliminary estimation of the number of \gpj events
taken in D0  Run~II experiment during January 2002 and satisfying the discussed above cuts
on $\Pt^{clust}$ and $\Pt^{out}$ can be found in our talk at QCD group \cite{QCD_talk2}.

Some comments have to be added about the impact of $\dphi$ cut onto the $\Pt$ disbalance. 
Let us consider the case of Selection 1.
In this case, as one can easily see from Appendices 2--5, the restriction of $\dphi$ leads to improving of the
positive $\gpart$ disbalance for all jetfinders. For the low $\Ptg$ interval ($\Ptg\lt50~GeV/c$), as
one can see from Appendix 2, the application of $\dphi$ cut alone, i.e. for the fixed value of
$\Pt^{clust}_{CUT}=30~GeV/c$, allows to reduce the fractional disbalance $\Ptg-\Pt^{part}$ to the level less
than $1\%$ for LUCELL and UA1 jetfinders. In a case of UA2 this disbalance became less than $1\%$ already
after the first cut $\dphi\lt17^\circ$. As for higher $\Ptg$ intervals, the $\Pt^{clust}_{CUT}$ becomes
more effective than the $\dphi$ cut for the $\Ptg-\Pt^{part}$ disbalance improvement mainly
due to the presence of more energetic clusters in an event at higher $\Ptg$.

Now let us turn to a case of Selections 2 and 3. 
If we look at the corresponding Tables 13--18 of Appendices 2--5, we can find out 
that in events with isolated jet the sign of $\gpart$ disbalance has a negative value.
It was mentioned previously that for all $\Ptg$ intervals the hadronization effect $\Jpart$ has also negative value.
Thus, the ``negative sign'' parton-to-jet hadronization effect compensates partially the $\gpart$ disbalance
when passing to the photon--jet final state $\Pt$ disbalance $\Fptgj$
as $\Pt^{Jet}<\Pt^{part}$ (compare three corresponding lines with disbalance values in Tables 13--18).

So, diminishing $\gpart$ disbalance by restricting $\Pt^{clust}_{CUT}$ and keeping in mind
that the hadronization effect practically rather weakly depends on $\Pt^{clust}_{CUT}$ 
we can get in the  line $\Fptgj$ of Tables 13--18 the achieved values of final state photon--jet disbalance.

The extreamly strong cut $\Pt^{clust}_{CUT}=5~GeV/c$, as it is seen from Tables 13--18, leads to
a very small value of $\gpart$ disbalance. It means that we have chosen the events with a good
balance at the parton level, i.e. those really corresponding to LO diagrams, shown in Fig.~1 of Section 2.
In this case  the parton-to-jet hadronization effect, that practically is not affected by  $\Pt^{clust}_{CUT}$,
naturally leads to worsening $\Fptgj$ disbalance at $\Pt^{clust}_{CUT}\lt10~GeV/c$ as is seen from Figs.~16 and 18.

\begin{figure}[htbp]
 \vspace{-15mm}
 \hspace*{15mm} \includegraphics[height=11cm,width=15cm]{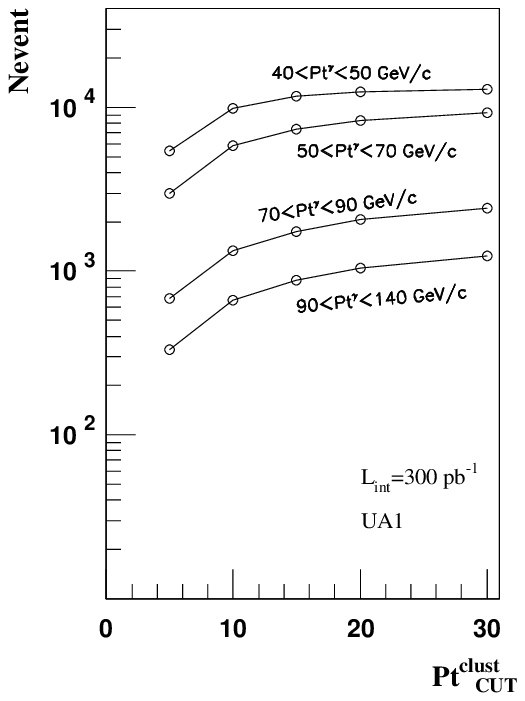}
    \label{fig01}
   \nonumber
  \end{figure}
\begin{figure}[htbp]
 \vskip-37mm
\hspace*{-3mm} \includegraphics[height=14cm,width=18cm]{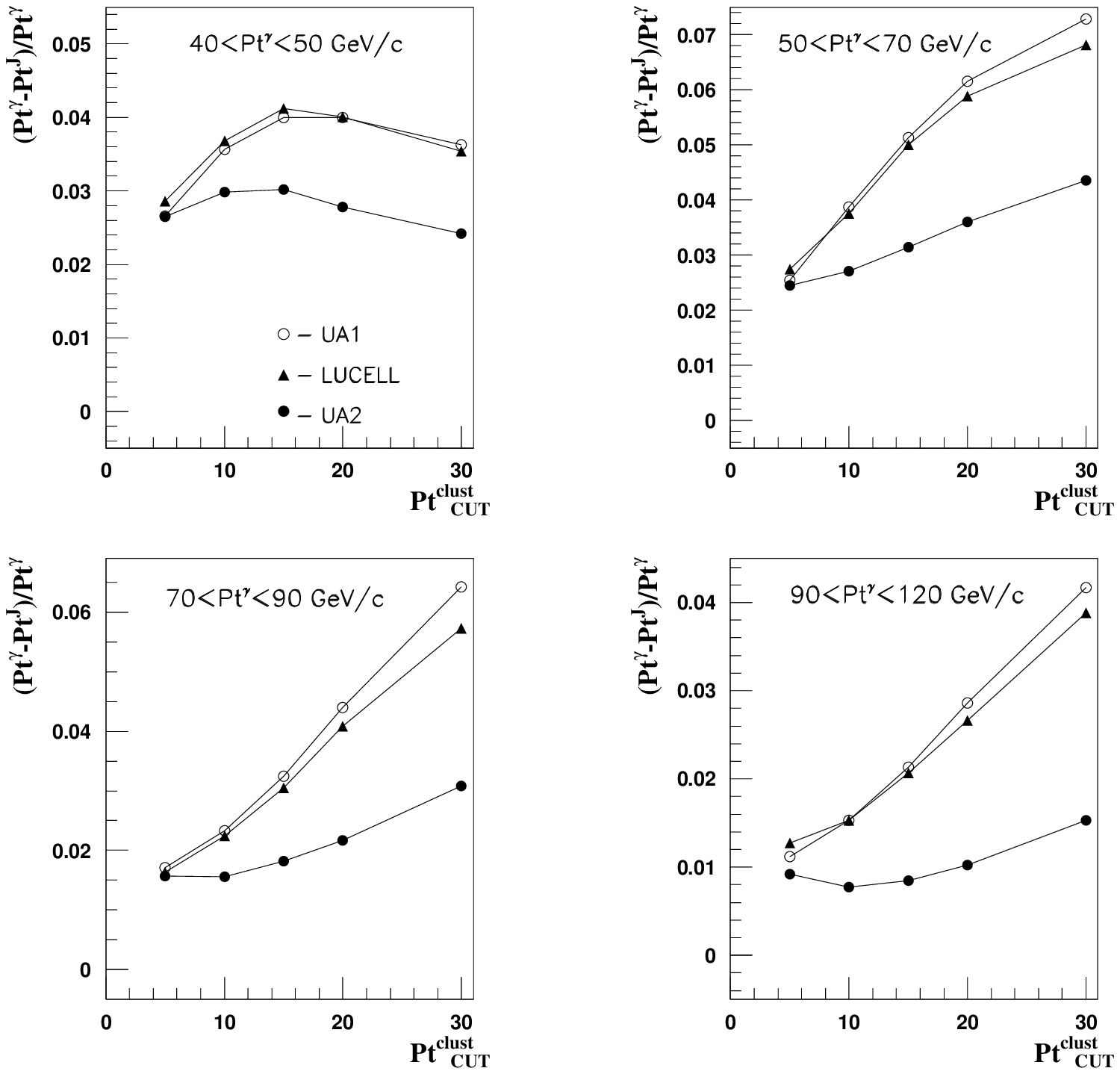}
    \label{fig02}
\nonumber
{\footnotesize Selection 1. Dependence of the
number of events for $L_{int}=300~ pb^{-1}$ (Fig.~13, top) and $\gJ$
(Fig.~14, four bottom plots) on $\Pt^{clust}_{CUT}$ in cases of LUCELL, UA1 and UA2 jetfinding algorithms. 
 $\dphi\leq17^{\circ}$. $\Pt^{out}$ is not limited.
}
\end{figure}

\begin{figure}[htbp]
\vspace{-15mm}
 \hspace{15mm}
\includegraphics[height=11cm,width=15cm]{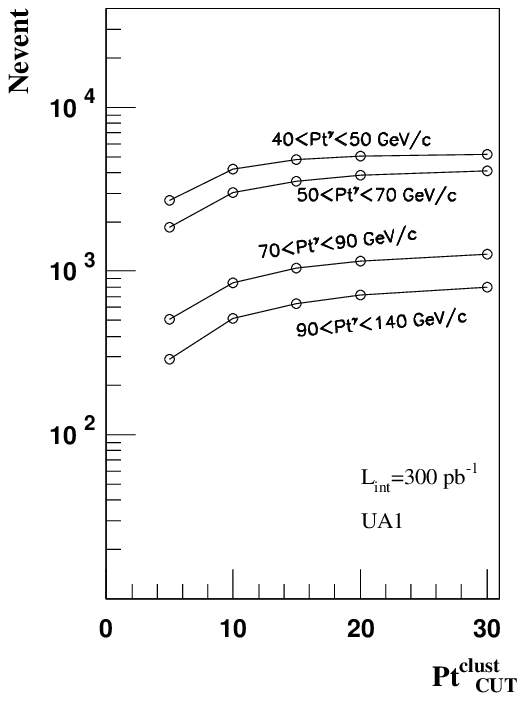}
 \label{fig03}
 \nonumber
 \end{figure}
\begin{figure}[htbp]
\vskip-25mm
\hspace*{-3mm} \includegraphics[height=14cm,width=18cm]{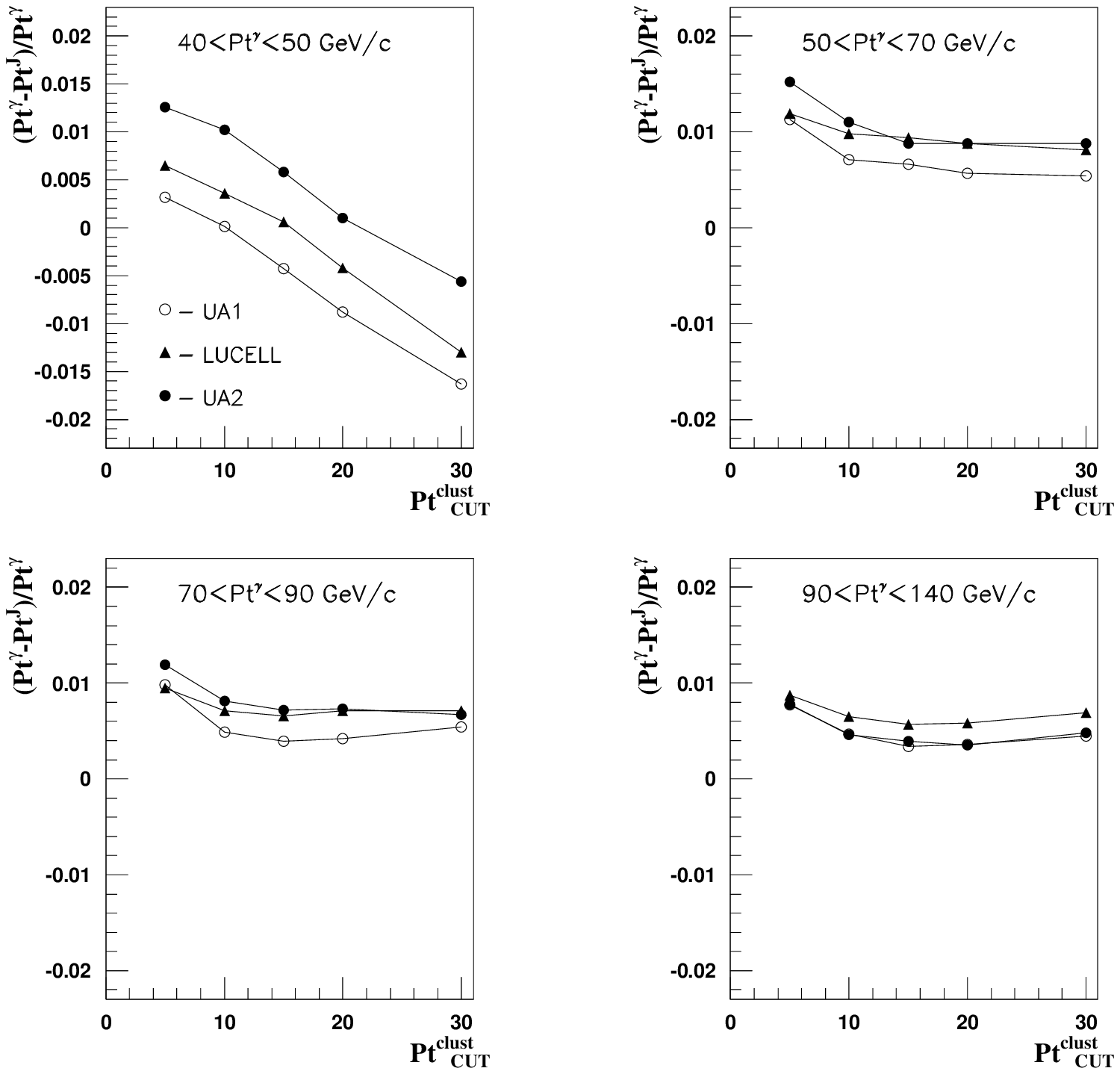}
\label{fig04}
\nonumber
\vskip-7mm
{\footnotesize Selection 2. Dependence of the
number of events for $L_{int}=300~ pb^{-1}$ (Fig.~15, top) and $\gJ$
(Fig.~16, four bottom plots) on $\Pt^{clust}_{CUT}$ in cases of LUCELL, UA1 and UA2 
jetfinding algorithms.  $\dphi\leq17^{\circ}$. $\Pt^{out}$ is not limited. $\epsilon^{jet}\lt3\%$.}
\end{figure}
\begin{figure}[htbp]
\vspace{-15mm}
 \hspace{15mm}
\includegraphics[height=11cm,width=15cm]{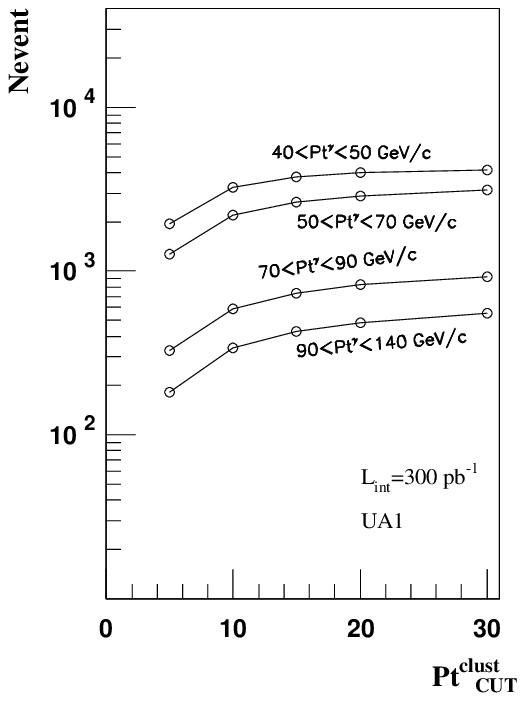}
 \label{fig03}
 \nonumber
 \end{figure}
\begin{figure}[htbp]
\vskip-25mm
\hspace*{-3mm} \includegraphics[height=14cm,width=18cm]{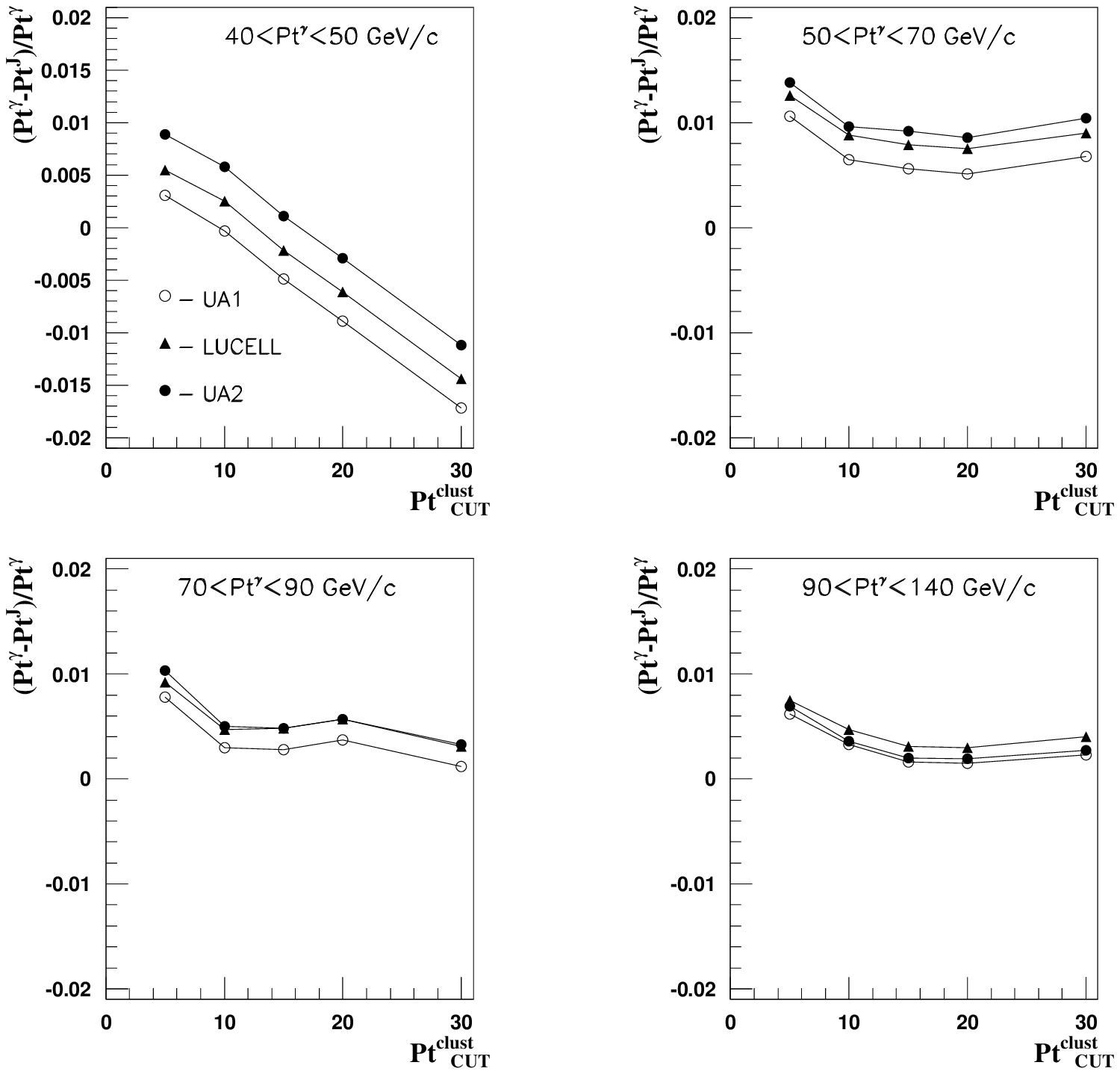}
\label{fig04}
\nonumber
\vskip-7mm
{\footnotesize Selection 3. Dependence of the
number of events for $L_{int}=300 pb^{-1}$ (Fig.~17, top) and $\gJ$
(Fig.~18, four bottom plots) on $\Pt^{clust}_{CUT}$ in cases of LUCELL, UA1 and UA2 
jetfinding algorithms.  $\dphi\leq17^{\circ}$. $\Pt^{out}$ is not limited. $\epsilon^{jet}\lt3\%$.}
\end{figure}

\setcounter{figure}{18}
\begin{figure}
\vskip-10mm
  \hspace*{-2mm} \includegraphics[width=16cm]{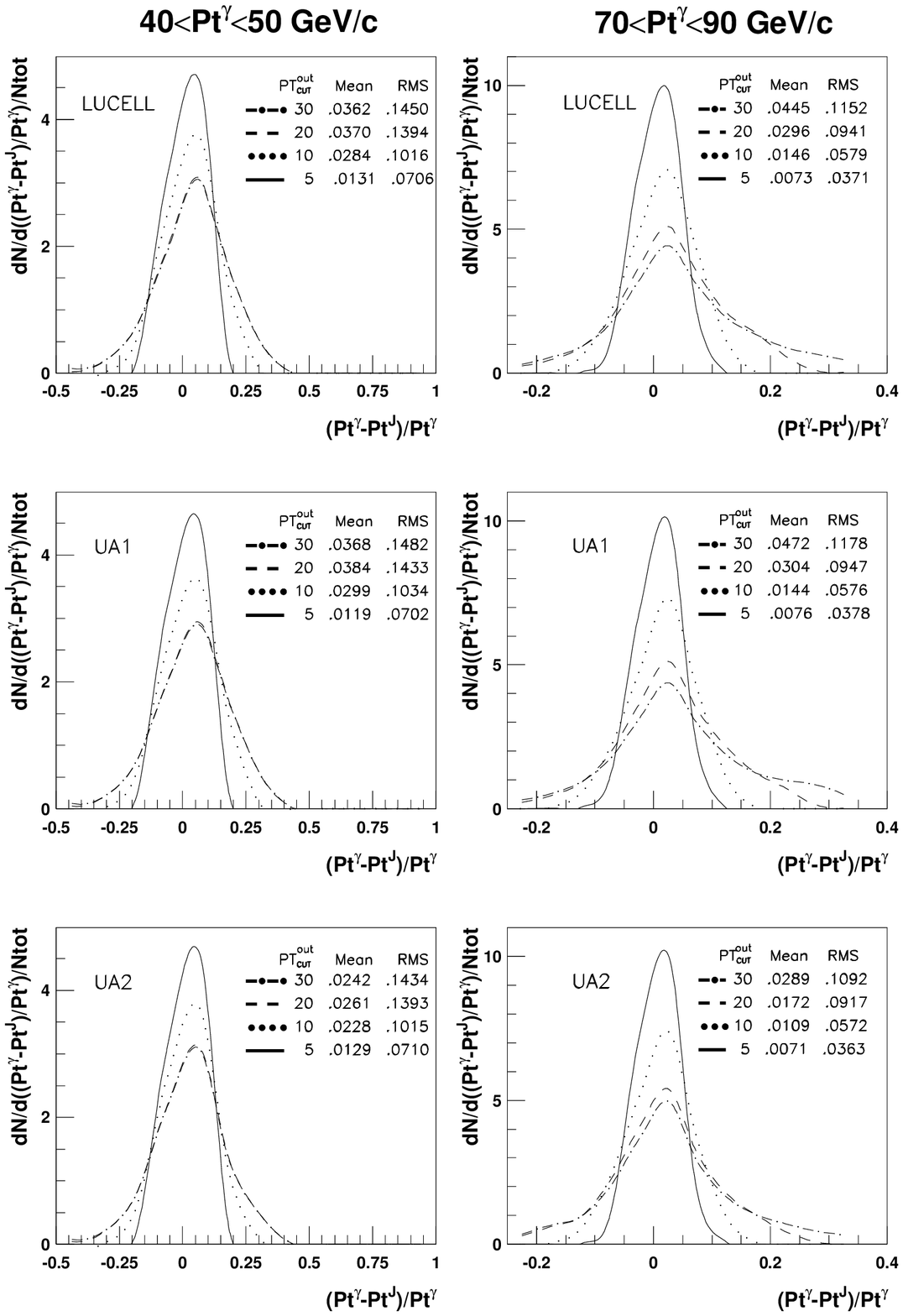} 
\caption{A dependence of $(\Pt^{\gamma}-\Pt^{J})/\Pt^{\gamma}$ on
$\Pt^{out}_{CUT}$ for LUCELL, UA1 and UA2 jetfinding algorithms and two
intervals of \ptg.  ~~The mean and RMS of the distributions are displayed on
the plots. $\dphi\leq17^\circ$, $\Pt^{clust}_{CUT}=30 ~GeV/c$. Selection 1.}
\label{fig:j-out}
\vskip25mm 
\end{figure}
\begin{figure}
\vskip-10mm
  \hspace*{-2mm} \includegraphics[width=16cm]{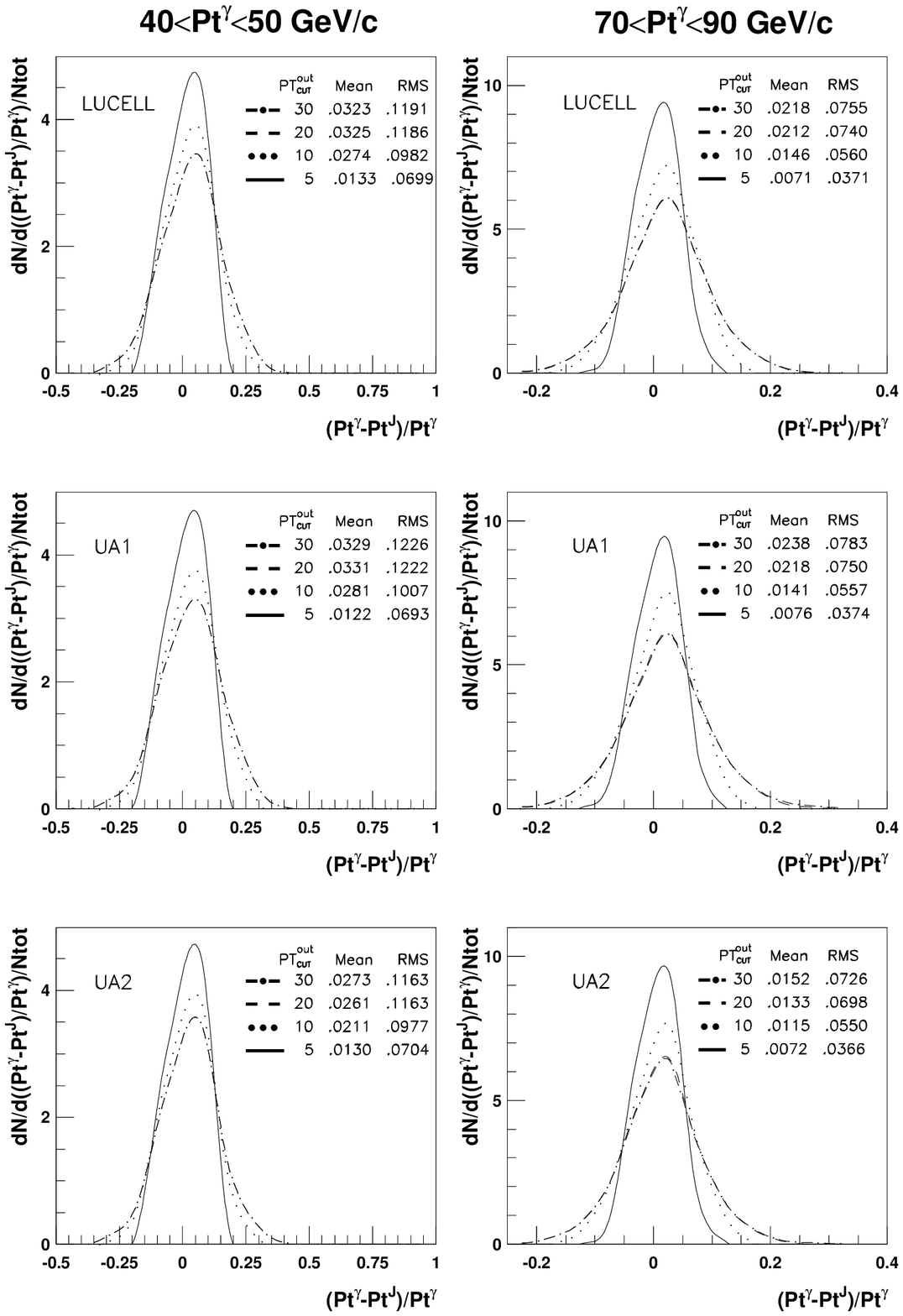} 
\caption{A dependence of $(\Pt^{\gamma}-\Pt^{J})/\Pt^{\gamma}$ on
$\Pt^{out}_{CUT}$ for LUCELL, UA1 and UA2  jetfinding algorithms and two
intervals of \ptg.  ~~The mean and RMS of the distributions are displayed on
the plots. $\dphi\leq17^\circ$, $\Pt^{clust}_{CUT}=10 ~GeV/c$. Selection 1.}
\label{fig:j-out-}
\vskip25mm 
\end{figure}

\newpage


\section{ESTIMATION OF BACKGROUND SUPPRESSION CUTS EFFICIENCY.}       

\it\small
\hspace*{9mm}
The relative efficiency of ``hadronic'' cuts that are added to ``photonic'' ones, used to suppress the background
in the case of inclusive photon measurement, is estimated at the particle level. It is shown that an imposing of new cuts
on $\Pt$ of ``clusters'' ($\Pt^{clust}_{CUT}$) and on $\Pt$ activity in the region out of \gpj system 
($\Pt^{out}_{CUT}$) and also an application of ``jet isolation'' criterion would allow  to achieve
further (after ``photonic'' cuts) fourteen-fold background suppression at the cost of four-fold loss of the signal 
``$\gamma^{\,dir}+jet$'' events.

It is also shown that the imposing of $\Pt^{out}_{CUT}$, $\Pt^{clust}_{CUT}$ together with a usage of jet 
isolation criterion would lead to a substantial improvement of \ptgj balance.

The potentially dangerous role of a new source of background 
to the signal ``$\gamma^{\,dir}+jet$'' events caused by hard
bremsstrahlung photons (``$\gamma-brem$'') is demonstrated. It is shown that at Tevatron energy this new
``$\gamma-brem$'' irreducible background may be compatible at low $\Ptg$ intervals with the $\pi^0$ contribution
and it may grow faster with $\Ptg$ increasing than the latter one.

\rm\normalsize
\vskip5mm

\noindent
\hspace*{8mm}
To$\!$ estimate$\!$ the$\!$ efficiency$\!$ of$\!$ the$\!$ cuts$\!$ proposed in Section 3.2 $\!$ we$\!$ 
carried$\!$ out$\!$ the$\!$ simulation
\footnote{ PYTHIA~5.7 version with default CTEQ2L parameterization
of structure functions is used here.}
with a mixture of all QCD and SM subprocesses with large cross sections existing in PYTHIA
(namely, in notations of PYTHIA, with
ISUB=1, 2, 11--20, 28--31, 53, 68). The events caused by this set of the subprocesses may give a large background 
to the ``$\gamma^{dir}+jet$'' signal events defined by  the subprocesses (1a) and (1b)
\footnote{A contribution of another possible NLO channel $gg\rrr g\gamma$
(ISUB=115 in PYTHIA) was found to be still negligible even at Tevatron energies.}
(ISUB=29 and 14) that were also included in this simulation.

Three generations  with the above-mentioned set of subprocesses
were performed. Each of them was done with a different value of
$CKIN(3)\equiv\pth$ PYTHIA parameter that defines the minimal value 
of $\Pt$ appearing in the final state of a hard $2\to 2$ parton level fundamental subprocess
in the case of ISR absence. These values were $\pth=40, 70$ and $100~GeV/c$. 
By 40 million events were generated for each of $\pth$ value. 
The cross sections of the above-mentioned subprocesses define the rates of corresponding physical
events and, thus, appear in simulation as weight factors.

We selected ``$\gamma^{dir}$-candidate +1 Jet'' events  containing 
one $\gamma^{dir}$-candidate (denoted in what follows as ${\tilde{\gamma}}$) and one jet (found by LUCELL)
with $\Pt^{Jet}> 30~ GeV/c$.
Here and below, as we work at the PYTHIA particle level of simulation, speaking about the $\gamma^{dir}$-candidate 
we actually mean, apart from  $\gamma^{dir}$,  a set of particles
like electrons, bremsstrahlung photons and also photons from neutral meson decays that may be registered in 
one D0 calorimeter tower of the $\Delta\eta\times\Delta\phi=0.1\times0.1$ size.

Here we consider a set of 17 cuts that are separated into 2 subsets: a set of ``photonic'' cuts and a set
of ``hadronic'' ones. The first set consists of 6 cuts used to select an isolated photon candidate
in some $\Pt^{\tilde{\gamma}}$ interval. The second one includes 11 cuts  applied after the first six cuts.
They are connected mostly with jets and clusters and are used to select events having
one ``isolated jet'' and limited $\Pt$ activity out of ``${\tilde{\gamma}}+jet$'' system.

The used cuts are listed in Table 13. To give an idea
about their physical meaning and importance we have done an estimation of their possible 
influence on the signal-to-background ratios $S/B$. The letter were calculated after application of each cut.
Their values are presented in Table 14 for a case of the most illustrative intermediate interval
of event generation with $\pth=70~GeV/c$. In Table 14 the number in each line corresponds
to the number of the cut in Table 13 (three important lines of Table \ref{tab:sb4} are darkened because they
will be often referenced to while discussing the following Tables \ref{tab:sb1}--\ref{tab:sb3}).
\\[-6mm]
\begin{table}[h]
\caption{List of the applied cuts (will be used also in Tables \ref{tab:sb4} -- \ref{tab:sb3}).}
\begin{tabular}{lc} \hline
\label{tab:sb0}
\hspace*{-2mm} {\bf 1}. $a)~ \Pt^{\tilde{\gamma}}\geq 40 ~GeV/c, ~~b)~|\eta^{\tilde{\gamma}}|\leq 2.5,
~~ c)~ \Pt^{jet}\geq 30 ~GeV/c,~~d) \Pt^{hadr}\!<7 ~GeV/c^{\;\ast}$;\\
{\bf 2}.  $\Pt^{isol}\!\leq 5~ GeV/c, ~\epsilon^{\tilde{\gamma}}<15\%$;
\hspace*{1.2cm} {\bf 10}. $\dphi<17^\circ$; \\
{\bf 3}. $\Pt^{\tilde{\gamma}}\geq\pth$;
\hspace*{4.08cm} {\bf 11}. $\Pt^{clust}<20 ~GeV/c$; \\
{\bf 4}. $\Pt^{isol}_{ring} \leq 1~ GeV/c^{\;\ast\ast}$;
\hspace*{2.58cm} {\bf 12}. $\Pt^{clust}<15 ~GeV/c$; \\
{\bf 5}.  $\Pt^{ch}_{max}(R=0.4)<2~ GeV/c$;
\hspace*{1.35cm} {\bf 13}. $\Pt^{clust}<10 ~GeV/c$; \\
{\bf 6}.  $\Pt^{isol}\!\leq 2~ GeV/c, ~\epsilon^{\tilde{\gamma}}<5\%$;
\hspace*{1.41cm}  {\bf 14}. $\Pt^{out}<20 ~GeV/c$; \\
{\bf 7}. $Njet\leq3$;
\hspace*{4.37cm}  {\bf 15}. $\Pt^{out}<15 ~GeV/c$;\\
{\bf 8}. $Njet\leq2$;
\hspace*{4.37cm}  {\bf 16}. $\Pt^{out}<10 ~GeV/c$;\\
{\bf 9}. $Njet=1$;
\hspace*{4.35cm} {\bf 17}. $\epsilon^{jet} \leq 3\%$.\\\hline
\footnotesize{${\;\ast}~\Pt$ of a hadron in the tower containing a $\gamma^{dir}$-candidate;}\\
\footnotesize{${\;\ast\ast}$ A scalar sum of $\Pt$ in the ring:
$\Pt^{sum}(R=0.4)-\Pt^{sum}(R=0.2)$.}\\[-3mm]
\end{tabular}
\end{table}

Line number 1 of  Table 13 makes primary preselection. It includes and specifies
our first general cut (17) of Section 3.2 as well as the cut connected with ECAL geometry and 
the cut (21) that excludes $\gamma^{dir}$-candidates accompanied by hadrons.

Line number 2 of  Table 13 fixes the values of $\Pt^{isol}_{CUT}$ and $\epsilon^{\gamma}_{CUT}$
that, according to (18) and (19), define the isolation parameters of ${\tilde{\gamma}}$.

The third cut selects the events with $\gamma^{dir}$-candidates having $\Pt$ higher than
$CKIN(3)\equiv\pth$ threshold. We impose the third cut to select the samples of events with
$\Pt^{\tilde{\gamma}}\geq40, 70$ and $100~GeV/c$ as ISR may smear the sharp kinematical cutoff defined by
$CKIN(3)$ \cite{PYT}. This cut reflects an experimental viewpoint when one is interested in
how many events with $\gamma^{dir}$-candidates are contained in some definite interval of $\Pt^{\tilde{\gamma}}$.

The forth cut restricts a value of $\Pt^{isol}_{ring}=\Pt^{isol}_{R=0.4}-\Pt^{isol}_{R=0.2}$,
where $\Pt^{isol}_{R}$ is a sum of $\Pt$ of all ECAL cells contained in the cone of radius 
$R$ around the tower fired by $\gamma^{dir}$-candidate \cite{D0_1}, \cite{D0_2}. Here it is taken
to be even stricter than that one used in \cite{D0_1}, \cite{D0_2}. As it is seen
from line 4 of Table 14 the fourth cut leads to about $25\%$ reduction of background contribution.
It makes not so big effect as compared with the transition from line 1 to line 2 because
in line 2 we have already imposed a strict enough isolation cut that covers a wide region of $R=0.7$ around
$\gamma^{dir}$-candidate. So, the restriction for $\Pt^{isol}_{ring}$, realized in the fourth line,
acts already on the events having a rather clean surrounding space near $\gamma^{dir}$-candidate.
The reduction  of the number of events happens mainly due to a passing from general isolation 
$\Pt^{isol}\!\leq 5~ GeV/c$ to a small value of $\Pt^{isol}_{ring}\leq1~GeV/c$.

The fifth cut corresponds to that one described in point 4 of Section 3.2 and it excludes 
events having the tracks of charged particles with $\Pt^{ch}\gt2~GeV/c$ that are contained in the cone of
$R=0.4$ around $\tilde{\gamma}$. It gives about $10\%$ reduction of background that is achieved
practically without any loss of signal events
\footnote{In our PYTHIA particle level simulation this cut stands for the effective account
of three criteria used in Section 6.1.1 of [1] for reduction of events with a track presence
(ECAL cluster--track matching, total charge in TRD and the ionization losses in the central tracking detectors).}.

Here we take the efficiency of each cut to be equal to $100\%$ as we study the results of simulation
at the particle level. The estimations of detector effects are given in \cite{D0_1}, \cite{D0_2}, where 
track finding efficiency was found to be $83\%$. 

Having in mind this value ($83\%$) we have not included the
$e^\pm$  events contribution (i.e. with $\tilde{\gamma}=e^\pm$) to the background values $B$ 
presented in Table \ref{tab:sb4}.
Nevertheless, let us mention that their number was estimated and presented in \cite{QCD_talk}
for a slightly different set of selection cuts. The expected contribution of events with $\tilde{\gamma}=e^\pm$
at the level of the last cut 17 of Table 14 to the value of total background was found to be about $10\%$. Now, accepting
the value of track finding efficiency for electrons as mentioned above, we conclude that
the contribution of $e^\pm$ events to the total background would not exceed $2\%$
\footnote{This number agrees with our estimation of the electron contribution to the total background
to $\gamma^{dir}+jet$ events done in \cite{BKS_P5}, \cite{GPJ} for LHC energies. It is also worth 
mentioning that a sizeable rejection  of $e^\pm$ events, that are characterized, as discussed in Section 4, 
by noticeable $\Pt^{miss}$ values,
can be achieved by applying $\Pt^{miss}_{CUT}\!=\!10~GeV/c$ (not included in present Table \ref{tab:sb0} 
but shown in analogous table in \cite{QCD_talk}). This cut reduces strongly the number of the events with 
$e^\pm$ produced in weak decays (see \cite{BKS_P5} and \cite{QCD_talk} for details).}.

The sixth cut makes tighter the isolation criterion of $\gamma^{dir}$-candidate 
(within $R=0.7$) than it was required by the second line of Table 13. It gives further
$30\%$ reduction of the background at the cost of $10\%$ loss of signal events.
It should be noted that this cut includes the restriction of ``infrared'' cut (20)
of Section 3.2 which was not included to this reason into Table \ref{tab:sb0}.

The data taken in D0  Run~II experiment during January 2002 were analyzed to 
select events with $\gamma^{dir}$-candidates satisfying the photonic cuts 1--5 of Table 13.
The very preliminary plot of the number of events dependence on $\Pt^{\tilde{\gamma}}$
can be found in the slides from our talk at QCD group \cite{QCD_talk2}.
\\[-18mm]
\begin{table}[htbp]
\begin{center}
\vskip0.0cm
\vspace*{11mm}
\caption{Values of significance and  efficiencies for $\pth=70~GeV/c$.}
\vskip0.5mm
\begin{tabular}{||c||c|c|c|c|c||}                  \hline \hline
\label{tab:sb4}
Cut& $S$ & $B$ & $Eff_S(\%)$ & $Eff_{B}(\%)$  & $S/B$ \\\hline \hline
\rowcolor[gray]{\coltab}%
\rowcolor[gray]{\coltab}%
 1 & 38890 &   1279436&  100.00$\pm$  0.00& 100.000$\pm$  0.000&  0.03  \\\hline 
2 & 35603&     46023&   94.56$\pm$  0.70&  54.733$\pm$  0.317&  0.77 \\\hline 
3 & 29235 &    16221&   77.64$\pm$  0.61&  19.291$\pm$  0.165&  1.80\\\hline 
4 & 27301 &    12228&   72.51$\pm$  0.58&  14.542$\pm$  0.141&  2.23\\\hline 
5 & 27293 &    10903&   72.49$\pm$  0.58&  12.966$\pm$  0.132&  2.50\\\hline 
6& 24657 &     7714&   65.48$\pm$  0.54&   9.174$\pm$  0.109&  3.20\\\hline 
7& 24610 &     7613&   65.36$\pm$  0.54&   9.054$\pm$  0.108&  3.23\\\hline 
8& 24129 &     6920&   64.08$\pm$  0.53&   8.230$\pm$  0.103&  3.49\\\hline 
9& 19905 &     4041&   52.86$\pm$  0.46&   4.806$\pm$  0.077&  4.93\\\hline 
10& 18781&      3401&   49.88$\pm$  0.45&   4.045$\pm$  0.071&  5.52 \\\hline 
11& 16225&      2474&   43.09$\pm$  0.40&   2.942$\pm$  0.060&  6.56\\\hline 
12& 14199&      1884&   37.71$\pm$  0.37&   2.241$\pm$  0.052&  7.54\\\hline 
13& 11023&      1232&   29.28$\pm$  0.32&   1.465$\pm$  0.042&  8.95\\\hline 
14& 10915&      1208&   28.99$\pm$  0.32&   1.437$\pm$  0.042&  9.04\\\hline 
15& 10481&      1128&   27.84$\pm$  0.31&   1.341$\pm$  0.040&  9.29\\\hline 
\rowcolor[gray]{\coltab}%
16& 8774&       896&   23.30$\pm$  0.28&   1.066$\pm$  0.036 & 9.79\\\hline 
\rowcolor[gray]{\coltab}%
17& 6264&       551&   16.64$\pm$  0.23&   0.655$\pm$  0.028& 11.37\\\hline 
\hline 
\end{tabular}
\end{center}
\vskip-4mm
\noindent
\end{table}
\normalsize

The cuts considered up to now, apart from general preselection cut $\Pt^{jet}\geq30~GeV/c$
used in the first line of Table \ref{tab:sb0},
were connected with photon selection (``photonic'' cuts). Before we go further, some words of caution 
must be said here.
Firstly, we want to emphasize that the starting numbers of the signal ($S$) and background ($B$)
events (first line of Table 14) may be specific only for PYTHIA generator and for the way of
preparing primary samples of the signal and background events described above. So, we want to underline here
that the starting values of $S$ and $B$ in the first columns of Table 14 are model dependent
\footnote{Let us notice that $S/B$ ratio, obtained after application of
photon isolation cuts $1-5$, is equal to $2.5$. This number is close 
to that one stemming from the value of photon purity $P\approx0.75-0.80$ found in inclusive photon measurement
\cite{D0_2} for interval $\Pt\geq70~GeV/c$ and for $CC$ region, but it is still lower, what is quite expectable
as we have not taken into account the detector effects.}.

But nevertheless, for our aim of investigation of new cuts 11--17 
(see \cite{9}--\cite{BKS_P5}) efficiency the important thing here is that we can use these starting model numbers of
$S$- and $B$-events  for studying the further relative  influence of these cuts on $S/B$ ratio
(also as for the conventional normalization to $100\%$ of the cut efficiencies
\footnote{In Table \ref{tab:sb4} the efficiencies $Eff_{S(B)}$ (with their errors) are defined as a ratio
of the number of signal (background) events that passed under a cut
(1--17) to the number of the preselected events (1st cut of this table).}
for $S$- and $B$-events in line 1).  

In spite of self-explaining nature of the cuts 7--10 let us mention, before passing to cuts 11--17, 
that the cuts 7--10 are connected with the selection of events having only one jet 
and the definition of jet-photon spatial orientation. Usage of these four cuts leads to the almost two-fold 
relative improvement of model $S/B$ ratio (compare lines 6 and 10 of Table 14).

Moving further we see from Table \ref{tab:sb4} that the
cuts 11--16 of Table \ref{tab:sb0} reduce the values of $\Pt^{clust}$ and $\Pt^{out}$ down to the values 
less than $10 ~GeV/c$. The 17-th cut of Table \ref{tab:sb0} imposes the jet isolation requirement. 
It leaves only the events
with jets having the sum of $\Pt$ in a ring surrounding a jet to be less than $3\%$ of $\Pt^{Jet}$.
From comparison of the numbers in 10-th and 17-th lines we make the important conclusion that all these
new cuts (11--17), despite of model dependent nature of starting $S/B$ value in line 10, may, in principle,
lead to the following about two-fold improvement of $S/B$ ratio. 
This improvement is reached by reducing the $\Pt$ activity out of ``$\tilde{\gamma}+1~jet$'' system. 

It is also rather interesting to mention that the total effect of ``hadronic cuts'' 7--17
consist of about fourteen-fold decrease of background contribution 
at the cost of four-fold loss of signal events. So, in this sense, we may conclude that 
from the viewpoint of $S/B$ ratio the study of \gpj events may be more preferable 
as compared with a case of inclusive photon production.

Below we shall demonstrate in some plots how new selection criteria 11--17 work to choose the events
with further almost two-fold improvement of $S/B$ ratio.
 For this reason we have built the distributions that correspond to the three above-mentioned 
values of $\pth$ and for the ``$\tilde{\gamma}+1~jet$''
events that have passed the set of cuts 1--9 defined in Table 13.
Thus, no special cuts were imposed on $\Delta\phi$, $\Pt^{out}$ and $\Pt^{clust}$
(the values of $\Pt^{clust}$ are automatically bounded from above since
we select ``$\tilde{\gamma}+1~jet$'' events with $\Pt^{jet}>30~ GeV/c$).

\begin{figure}[htbp]
 \vspace*{-0.0cm}
 \hspace*{-0.1cm}
  \includegraphics[width=17cm,height=20cm]{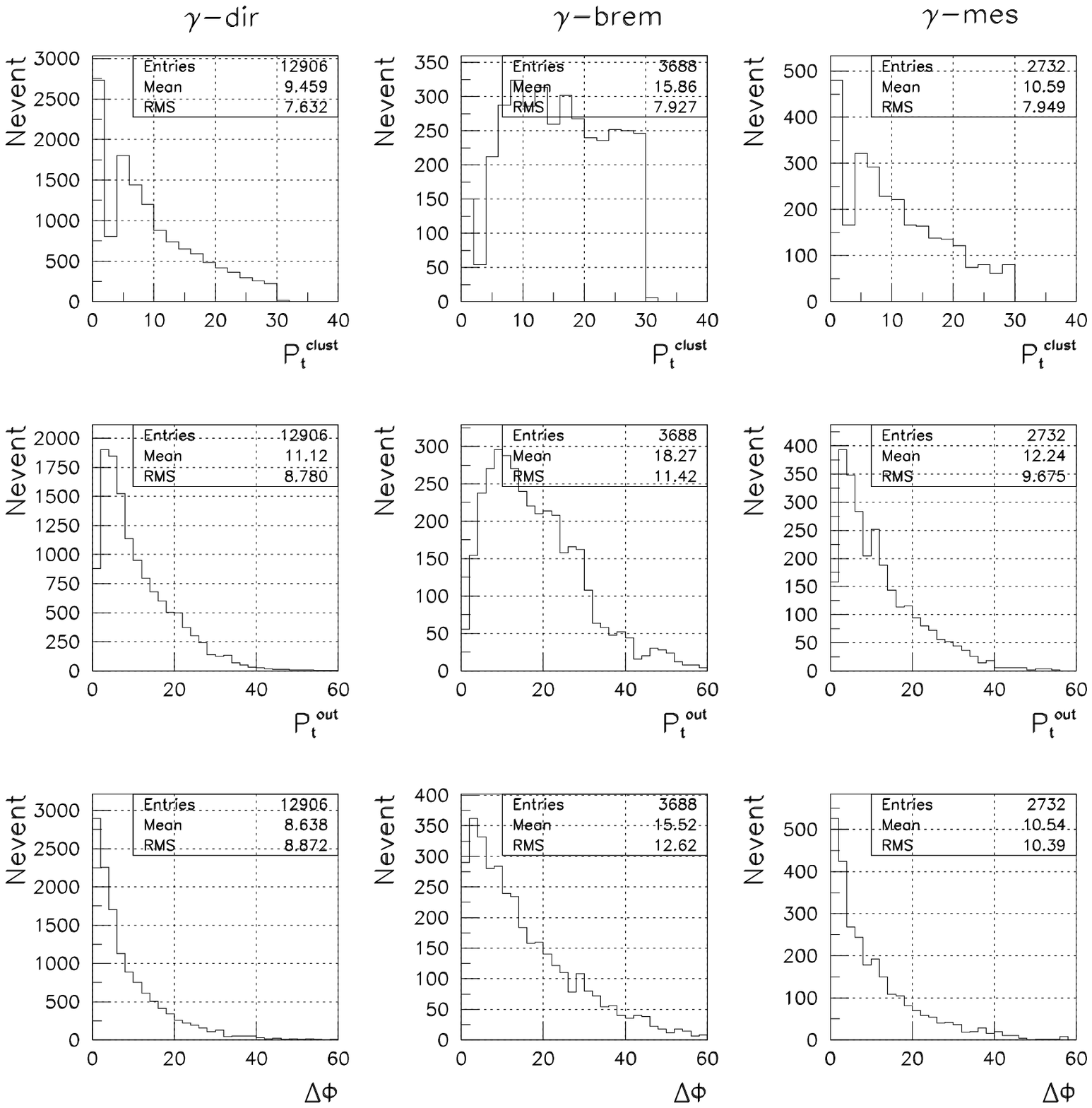}
  \vspace{-0.8cm}
    \caption{\hspace*{0.0cm}  Signal $\&$ Backgrounds: Number of events distribution
over $\Pt^{clust}$, $\Pt^{out}$, $\dphi$ ($\Pt^{\tilde{\gamma}}\geq40~GeV/c$).}
    \label{fig:1b40}
  \end{figure}

\begin{figure}[htbp]
 \vspace{-0.0cm}
 \hspace*{-0.1cm}
  \includegraphics[width=17cm,height=20cm]{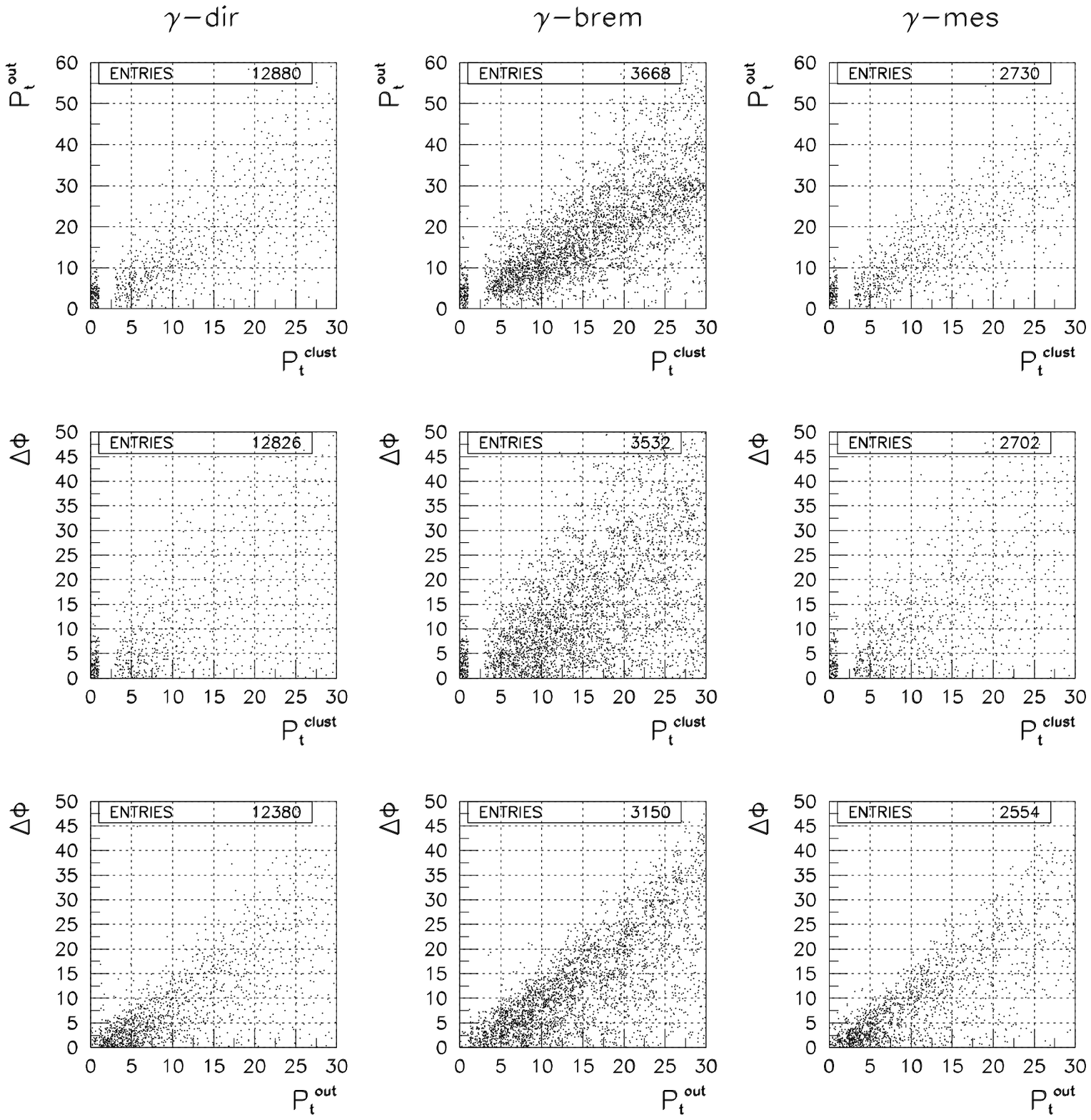}
  \vspace{-0.8cm}
    \caption{\hspace*{0.0cm} Signal $\&$ Backgrounds: $\Pt^{clust}$ vs. $\Pt^{out}$,
$\Pt^{clust}$ vs. $\dphi$, $\Pt^{out}$ vs. $\dphi$ ($\Pt^{\tilde{\gamma}}\geq 40~GeV/c$).}
    \label{fig:2b40}
  \end{figure}

\begin{figure}[htbp]
 \vspace{0.0cm}
 \hspace*{-0.1cm}
 \includegraphics[width=17cm,height=20cm]{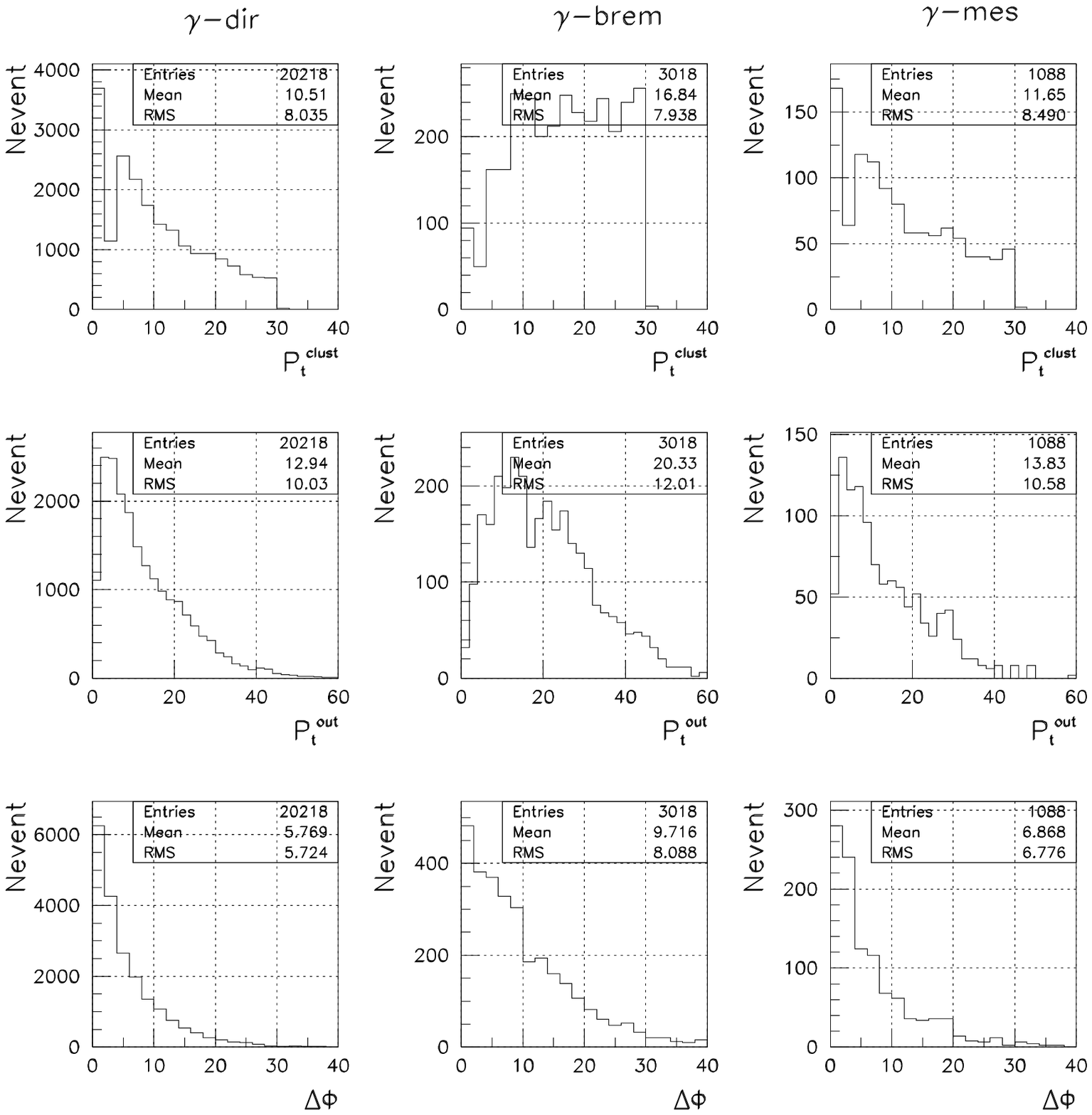}
  \vspace{-0.8cm}
    \caption{\hspace*{0.0cm}  Signal $\&$ Backgrounds: Number of events distribution
over $\Pt^{clust}$, $\Pt^{out}$, $\dphi$ ($\Pt^{\tilde{\gamma}}\geq 70~GeV/c$).}
    \label{fig:1b100}
  \end{figure}

\begin{figure}[htbp]
 \vspace{0.0cm}
 \hspace*{-0.1cm}
  \includegraphics[width=17cm,height=20cm]{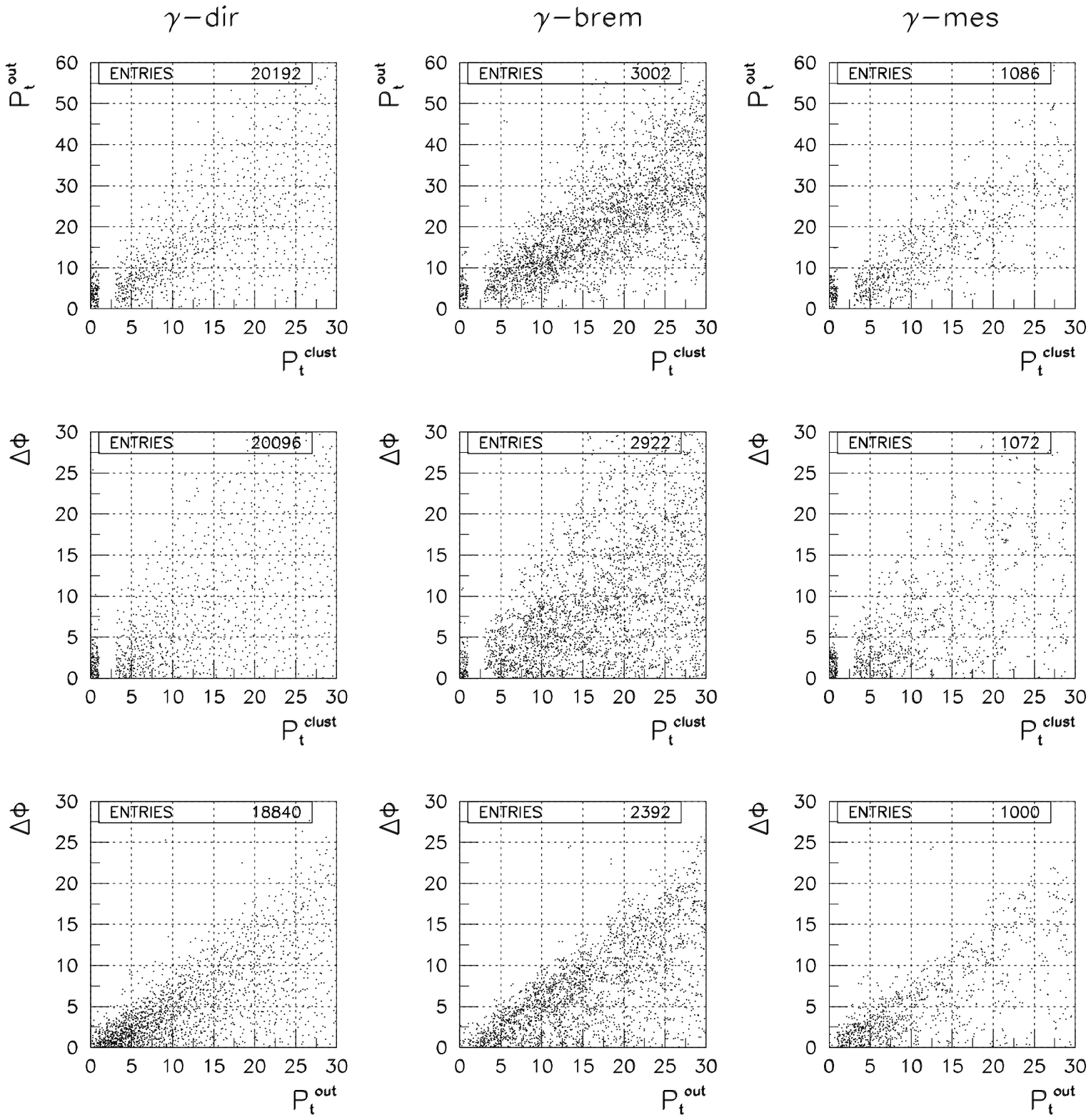}
  \vspace{-0.8cm}
    \caption{\hspace*{0.0cm}  Signal $\&$ Backgrounds: $\Pt^{clust}$ vs. $\Pt^{out}$,
$\Pt^{clust}$ vs. $\dphi$, $\Pt^{out}$ vs. $\dphi$ ($\Pt^{\tilde{\gamma}}\geq 70~GeV/c$).}
    \label{fig:2b100}
  \end{figure}

\begin{figure}[htbp]
 \vspace{0.0cm}
 \hspace*{-0.1cm}
 \includegraphics[width=17cm,height=20cm]{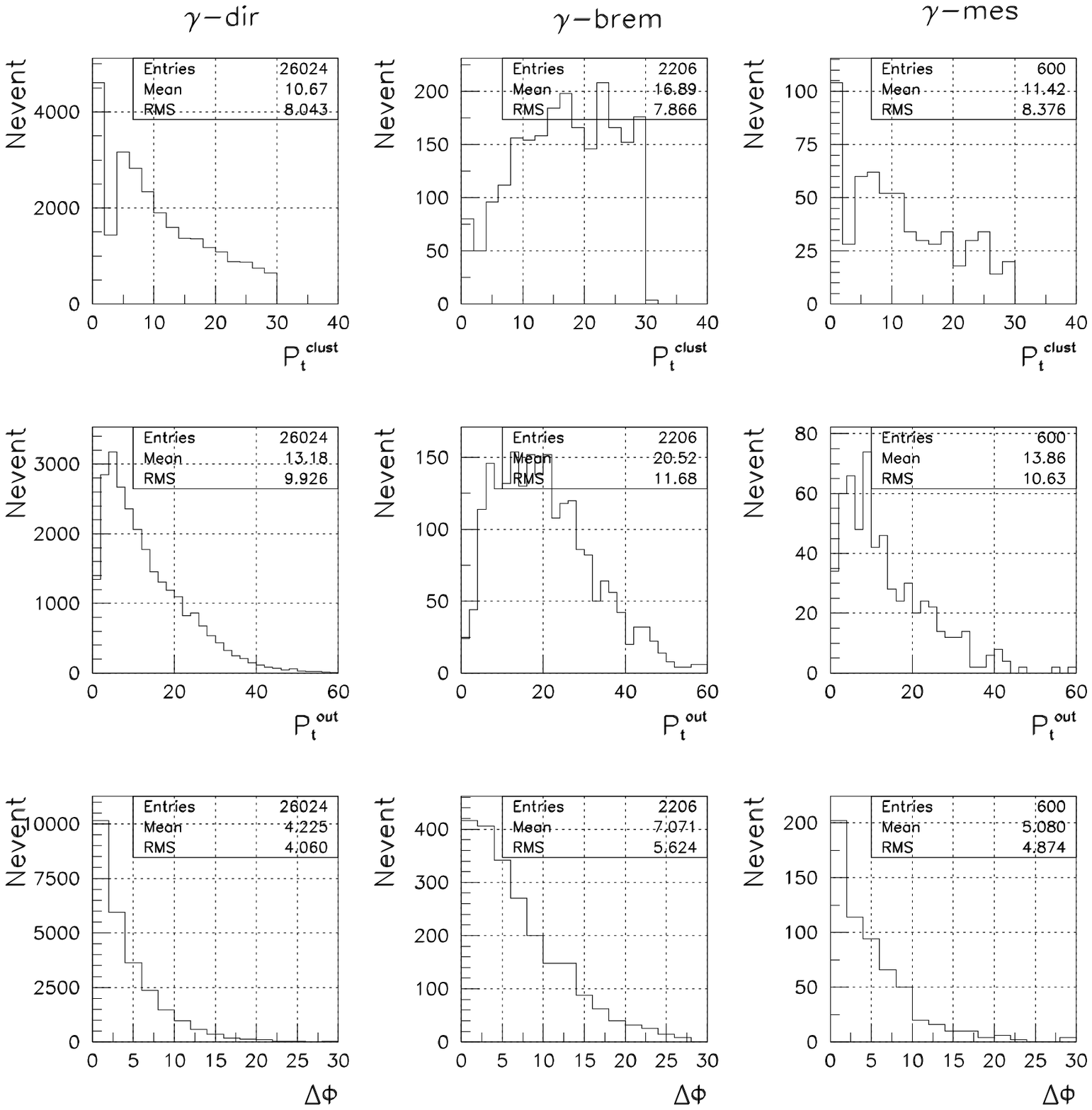}
  \vspace{-0.8cm}
    \caption{\hspace*{0.0cm} Signal $\&$ Backgrounds: Number of events distribution
over $\Pt^{clust}$, $\Pt^{out}$, $\dphi$ ($\Pt^{\tilde{\gamma}}\geq 100~GeV/c$).}
    \label{fig:1b200}
  \end{figure}

\begin{figure}[htbp]
 \vspace{0.0cm}
 \hspace*{-0.1cm}
  \includegraphics[width=17cm,height=20cm]{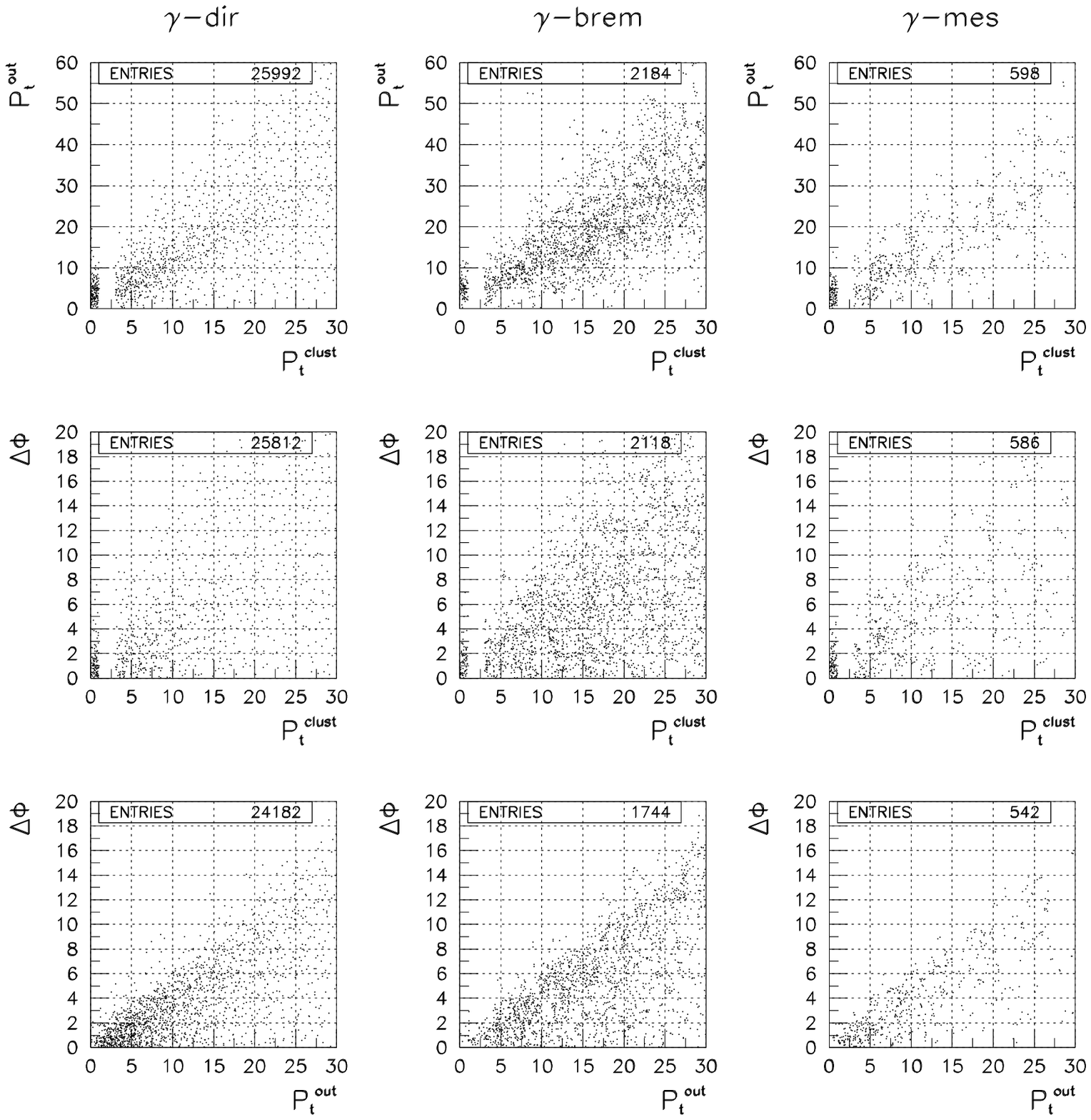}
  \vspace{-0.8cm}
    \caption{\hspace*{0.0cm} Signal $\&$ Backgrounds: $\Pt^{clust}$ vs. $\Pt^{out}$,
$\Pt^{clust}$ vs. $\dphi$, $\Pt^{out}$ vs. $\dphi$ ($\Pt^{\tilde{\gamma}}\geq 100~GeV/c$).}
    \label{fig:2b200}
  \end{figure}

These distributions are given here to show the dependence of the number of events on
the physical observables $\dphi, \Pt^{out}$ and $\Pt^{clust}$
introduced in Sections 3.1 and 3.2. We present them
separately for the signal ``$\gamma$-dir'' and background events contained in each of three generated
samples. The distributions are given for three different $\Pt^{\tilde{\gamma}}$ intervals
in Figs.~\ref{fig:1b40}, \ref{fig:1b100}, \ref{fig:1b200} and are accompanied by 
scatter plots \ref{fig:2b40}, \ref{fig:2b100}, \ref{fig:2b200}. So, each pair of a figure
and a scatter plot does correspond to one $\Pt^{\tilde{\gamma}}$ interval. 
Thus, Fig.~\ref{fig:1b40} and scatter plot \ref{fig:2b40} correspond to $\Pt^{\tilde{\gamma}}\geq40~GeV/c$ and so on.

The first columns in these figures, denoted by ``$\gamma$ - dir'', show the distributions 
in the signal events, i.e. in the events corresponding to processes (1a) and (1b).
The second columns, denoted as
``$\gamma$ - brem'', correspond to the events in which the photons
were emitted from quarks (i.e. bremsstrahlung photons).
 The distributions in the third columns were built on the basis of the events containing 
``$\gamma$-mes'' photons, i.e. those photons which originate from multiphoton decays of mesons
($\pi^0$, $\eta$, $\omega$ and $K^0_S$). 




First, we see that in the case of $\Pt^{\tilde{\gamma}} \geq 100 ~GeV/c$
(see Fig.~\ref{fig:1b200}) practically all ``signal events'' are within
$\Delta \phi<17^{\circ}$. In the case of $\Pt^{\tilde{\gamma}} \geq 70 ~GeV/c$
(see Fig.~\ref{fig:1b100}) most of them are also within $\Delta \phi<17^{\circ}$.
It is seen from Fig.~\ref{fig:1b40} that for $\Pt^{\tilde{\gamma}} \geq 40 ~GeV/c$ there is still
a large number  of signal events (about $70\%$) belonging to the
$\Delta \phi<17^{\circ}$ interval. From here and from the comparison of plots in the
``$\gamma$-dir'' and ``$\gamma$-brem'' columns  (showing the $\Delta\phi$ dependence) in the same figures
\ref{fig:1b40}--\ref{fig:2b200} we conclude that the upper cut $\Delta\phi<17^{\circ}$, used in previous sections,
 is reasonable, and moreover, it does discard a lot of ``$\gamma$-brem'' background events 
in the intervals with $\Pt^{\tilde{\gamma}}\lt100 ~GeV/c$.

From the second ``$\gamma$-brem'' columns of Figs.~\ref{fig:1b40},
~\ref{fig:1b100} and ~\ref{fig:1b200} one can also see that $\Pt^{clust}$ spectra 
of the events with bremsstrahlung photons look quite different from the analogous $\Pt^{clust}$ 
distributions of the signal ``$\gamma$-dir'' photons. The latter distributions have the most of the events
in the region of small $\Pt^{clust}$ values (there is a deep in spectra between $\Pt^{clust}\!\!=0$,
caused by events having no clusters, and the second peak appearing due to cluster definition as of the
object with $3\lt\Pt^{clust}\lt30~GeV/c$).

Since the bremsstrahlung (``$\gamma$-brem'') photons give the most sizeable background 
\footnote{The numbers in Table 15 below supports this remark. But it is also necessary to keep in mind the results 
obtained in \cite{Kirk} that the PYTHIA/JETSET fragmentation may underestimate the $\pi^0, \eta$ contribution to the isolated 
photon background.}, 
(compare the numbers of entries in the second ``$\gamma$-brem'' and the third ``$\gamma$-mes'' columns
of Figs.~21--26) the observed difference of the spectra prompts an idea of using
an upper cut for the value of $\Pt^{clust}$ to reduce
the ``$\gamma$-brem'' background which dominates at large $\Pt^{clust}$ values
(that was not a primary guideline for introduction of $\Pt^{clust}$ in Sections 2 and 3 
as a physical variable and a cut on it).

The analogous difference of $\Pt^{out}$ spectra of signal
``$\gamma$-dir'' events (which are concentrated at low  $\Pt^{out}$ values)
from those of the background ''$\gamma$-brem'' events 
having longer tails at high $\Pt^{out}$ enables us to impose an upper cut on the $\Pt^{out}$ value.

Now from the scatter plots in Figs.~\ref{fig:2b40},
\ref{fig:2b100} and \ref{fig:2b200} as well as from Figs.~\ref{fig:1b40},
\ref{fig:1b100} and \ref{fig:1b200} we can conclude that the use of cuts
\footnote{rather soft here, but the results of their further restriction were already shown
in tables of Appendices 2--5 and Figs.~12--20 and will be discussed below}:
\noindent
$\Delta\phi\lt17^{\circ}$, $\Pt^{clust}_{CUT}=10\;GeV/c$, $\Pt^{out}_{CUT}=10\;GeV/c$
\noindent
would allow to keep a big number of the signal ``$\gamma$-dir'' events and
to reduce noticeably the contribution from the background ``$\gamma$-brem'' and
``$\gamma$-mes'' events in all intervals of $\Pt^{\tilde{\gamma}}$. At the same time
the Figs.~21--26 give the information about what parts of different spectra are lost with the
imposed cuts.

So, Figs.~\ref{fig:1b40}--\ref{fig:2b200} illustrate well that the new physical variables $\Pt^{clust}$ 
and $\Pt^{out}$ \cite{9}--\cite{BKS_P5}, described in Sections 3.1 and 3.2
may be useful for separation of the ``$\gamma^{dir}+jet$'' events from the background ones 
(the latter, in principle, are not supposed to have the well-balanced $\Pt^{\tilde{\gamma}}$ and $\Pt^{Jet}$).


Table \ref{tab:sb1} includes the numbers of signal and background events left in three generated event samples
after application of cuts 1--16 and 1--17. They are given for all three intervals of $\Pt^{\tilde{\gamma}}$.
Tables \ref{tab:sb1} and \ref{tab:sb4} are complementary to each other.
The summary of Table \ref{tab:sb4} is presented in the middle section ($\pth=70 ~GeV/c$)
of Table \ref{tab:sb1} where the line ``Preselected'' corresponds to the cut 1 of Table \ref{tab:sb0} 
and, respectively, to the line number 1 of  Table  \ref{tab:sb4} presented above.
The line ``After cuts'' corresponds to the line 16 of  Table \ref{tab:sb4} and 
line ``+jet isolation'' corresponds to the line 17 of  Table \ref{tab:sb4}. 

Table \ref{tab:sb1} is done to show in more detail the origin of $\gamma^{dir}$-candidates.
The numbers in the  ``$\gamma-direct$'' column correspond to the respective  numbers  of 
signal events left in each of $\Pt^{\tilde{\gamma}}$ intervals after application of the cuts defined
in lines 1, 16 and 17 of Table \ref{tab:sb0} (and column ``$S$'' of Table \ref{tab:sb4}). Analogously
the numbers in the ``$\gamma-brem$'' column of Table  \ref{tab:sb1} correspond to the numbers
of events with the photons radiated from quarks
participating in the hard interactions. Their $\Pt^{clust}$ and $\Pt^{out}$ distributions
were presented in the central columns of Figs.~\ref{fig:1b40} -- \ref{fig:2b200}.
Columns 5 -- 8 of Table \ref{tab:sb1} illustrate the numbers of the ``$\gamma-mes$''  events with photons
originating from $\pi^0,~\eta,~\omega$ and $K^0_S$ meson decays.
Their distributions were shown in the right-hand columns of 
Figs.~\ref{fig:1b40} -- \ref{fig:2b200}. In a case of $\Pt^{\tilde{\gamma}}\gt70~GeV/c$ 
the total numbers of background events,
i.e. a sum over the numbers presented in columns 4 -- 8 of Table \ref{tab:sb1}, 
are shown in the lines 1, 16 and 17 of column ``$B$'' of Table \ref{tab:sb4}.
The other lines of Table \ref{tab:sb1} for $\pth\!=40$ and
$~100 ~GeV/c~$ have the meaning analogous to that described above for $\pth=70 ~GeV/c$.

~\\[-11mm]
\normalsize
\begin{table}[h]
\begin{center}
\caption{Number of signal and background events remained after cuts.}
\vskip.5mm
\begin{tabular}{||c|c||c|c|c|c|c|c|c||}                  \hline \hline
\label{tab:sb1}
\hmm$\pth$\hmm& &$\gamma$ & $\gamma$ &\multicolumn{4}{c|}{  photons from the mesons}  &
\\\cline{5-8}
\Gvc& Cuts&\hmm direct\hmm &\hmm brem\hmm & $\;\;$ $\pi^0$ $\;\;$ &$\quad$ $\eta$ $\quad$ &
$\omega$ &  $K_S^0$ &\hmm $e^{\pm}$\hmm \\\hline \hline
    &Preselected&\hmm18550&\hmm 14054& 151254& 55591& 18699& 15257&\hmm 2890\hmm  \\\cline{2-9}
 40 &After cuts &\hmm 6814&\hmm 711&     660&    369&   101& 111&\hmm   0\hmm\\\cline{2-9}
    &+ jet isol.  &\hmm 3529&\hmm 272&   283&    136&    45& 56&   0\\\hline  \hline
    &Preselected &\hmm38890&\hmm63709&773208&275524&93967&  73028 &\hmm 18510 \hmm\\\cline{2-9}
 70 &After cuts&\hmm  8774 &\hmm  445& 230 &144 &45  &32 &\hmm 0\hmm \\\cline{2-9}   
    &+ jet isol. &\hmm6264 &\hmm 289 & 132 &75 &33  & 22 &\hmm 0\hmm \\\hline \hline
    &Preselected&\hmm54007 &\hmm105715 &919932 &328259 &112553 & 86327 &\hmm38874\hmm\\\cline{2-9}
100 &After cuts&\hmm 11038 &\hmm  300& 116 &76 & 24  & 20 &\hmm 0\hmm\\\cline{2-9}
    &+ jet isol. &\hmm 9188 &\hmm 226& 84 & 52 & 22  & 18 &\hmm 0\hmm\\\hline \hline
\end{tabular}
\vskip0.2cm
\caption{Efficiency, $S/B$ ratio and significance values in the selected events without jet isolation cut.}
\vskip0.1cm
\begin{tabular}{||c||c|c|c|c|>{\columncolor[gray]{\coltab}}c|c||} \hline \hline
\label{tab:sb2}
$\pth$ \Gvc& $S$ & $B$ & $Eff_S(\%)$  & $Eff_B(\%)$  & $S/B$& $S/\sqrt{B}$
\\\hline \hline
40  & 6814& 1952 & 36.91 $\pm$ 0.52 & 4.767 $\pm$ 0.110&  3.5 & 154.2 \\\hline
70 & 8774&  896 & 23.30 $\pm$ 0.28 & 1.066 $\pm$ 0.036&  9.8 & 293.1 \\\hline 
100 & 11038& 536 & 20.58 $\pm$ 0.22 & 0.571 $\pm$ 0.025& 20.6 & 476.8 
\\\hline \hline
\end{tabular}
\vskip0.2cm
\caption{Efficiency, $S/B$ ratio and significance values in the selected events with jet isolation cut.}
\vskip0.1cm
\begin{tabular}{||c||c|c|c|c|>{\columncolor[gray]{\coltab}}c|c||}  \hline \hline
\label{tab:sb3}
$\pth$ \Gvc& ~~$S$~~ & ~~$B$~~ & $Eff_S(\%)$ & $Eff_B(\%)$  & $S/B$& $S/\sqrt{B}$
 \\\hline \hline
40  & 3529& 792 & 19.12 $\pm$ 0.35 & 1.934 $\pm$ 0.069&  4.5 & 125.4 \\\hline
70 & 6264& 551 & 16.64 $\pm$ 0.23 & 0.655 $\pm$ 0.028& 11.4 & 266.9 \\\hline
100 & 9188& 402 & 17.13 $\pm$ 0.19 & 0.428 $\pm$ 0.021& 22.9 & 459.3 
\\\hline \hline
\end{tabular}
\end{center}
\vskip-3mm
\end{table}

The last column of Table \ref{tab:sb1} shows the number of preselected events with
$e^\pm$ (see our notes above while discussing the fifth cut of Table \ref{tab:sb0}).

The numbers in Tables \ref{tab:sb2} (without jet isolation cut) and \ref{tab:sb3}
(with jet isolation cut) accumulate in a compact form the final information of 
Tables \ref{tab:sb0} -- \ref{tab:sb1}. 
Thus, for example, the columns $S$ and $B$ of the  line that corresponds to $\pth=70 ~GeV/c$ 
contain the total numbers of the selected signal and background events taken at the level of 16-th  (for Table
\ref{tab:sb2}) and 17-th (for Table \ref{tab:sb3}) cuts from Table \ref{tab:sb4}. 

It is seen from Table \ref{tab:sb2}  that in the case of Selection 1 the ratio $S/B$ grows   
from 3.5 to 20.6 while $\Pt^{\tilde{\gamma}}$ increases from
$\Pt^{\tilde{\gamma}}\geq 40 ~GeV/c$ to $\Pt^{\tilde{\gamma}}\geq 100 ~GeV/c$ interval.

The jet isolation requirement (cut 17 from Table \ref{tab:sb0})
noticeably improves the situation at low $\Pt^{\tilde{\gamma}}$ (see Table \ref{tab:sb3}).
After application of this criterion the value of $S/B$ increases from 3.5 
to 4.5 at $\Pt^{\tilde{\gamma}}\geq 40 ~GeV/c$ 
and from 20.6 to 22.9 at $\Pt^{\tilde{\gamma}}\geq 100 ~GeV/c$.
Remember on this occasion the conclusion  that the sample of events
selected with our criteria has a tendency to contain more events with an isolated jet
as $\Pt^{\tilde{\gamma}}$ increases (see Sections 5--7 and Appendices 2--5).
Thus, from Appendices 4 and 5 it can be seen that the main part of jets with $\Pt^{jet}\geq 70 ~GeV/c$
appears to be  isolated (compare also the last two lines in each $\pth$ section of Table \ref{tab:sb1})
\footnote{see also Fig.~11 for $\Pt^{\gamma}\geq 70 ~GeV/c$}.

Let us underline here that, in contrast to other types of background, ``$\gamma-brem$'' background
has an irreducible nature. So, the number of ``$\gamma-brem$'' events
should be carefully estimated for each $\Pt^{\tilde{\gamma}}$ interval using the particle level
of simulation in the framework of event generator like PYTHIA.
They are also have to be taken into account in experimental analysis
of the prompt photon production data at high energies.

Table \ref{tab:bg_or_gr} 
shows  the relative contributions of fundamental QCD subprocesses  (having the largest cross sections)
with ISUB=11, 12, 28, 53 and 68 (see \cite{PYT})
that define the main production of ``$\gamma\!-\!brem$'' background in event samples
selected with  criteria 1--13 of Table 13 in three $\Pt^{\tilde{\gamma}}$ intervals.

Accepting the results of simulation with PYTHIA, we found from the event listing analysis that
in the main part of selected ``$\gamma\!-\!brem$'' events
these photons are produced in the final state of the fundamental $2\to2$ subprocess
\footnote{i.e. from lines 7, 8 in Fig.~3}.
Namely, they are mostly radiated from the outgoing quarks 
in the case of the first three sets of subprocesses (ISUB=28, 11, 12 and 53).
They may also appear as a result of string breaking in a final state of 
$gg\to gg$ scattering (ISUB=68). But this subprocess,
naturally, gives a small contribution into ``$\tilde{\gamma}+jet$'' events production.

~\\[-12mm]
\begin{table}[h]
\begin{center}
\vskip-3mm
\caption{Relative contribution (in per cents) of different QCD subprocesses into
the ``$\gamma\!-\!brem$'' events production.}
\normalsize
\vskip.1cm
\begin{tabular}{|c||c|c|c|c|}                  \hline \hline
\label{tab:bg_or_gr}
$\Ptg$& \multicolumn{4}{c|}{fundamental QCD subprocess} \\\cline{2-5}
 \Gvc & { ISUB=28} & ISUB=11,12 & ISUB=53 & ISUB=68  
\\\hline \hline
 40--70   & 62.1$\pm$6.6 & 31.8$\pm$4.0 &  3.3$\pm$1.0 &  2.8$\pm$0.9  \\\hline 
 70--100  & 52.3$\pm$7.7 & 42.4$\pm$6.4 &  3.8$\pm$1.4 &  1.5$\pm$0.9 \\\hline 
 $>100$   & 41.8$\pm$6.0 & 56.9$\pm$7.2 &  1.3$\pm$0.7 &  ---  \\\hline\hline  
\end{tabular}
\end{center}
\vskip-7mm
\end{table}


It may be noted also from the first two columns of Table \ref{tab:bg_or_gr} 
that the most of ``$\gamma\!-\!brem$'' background events ($92\%$ at least) originate from  the
ISUB=28 ($fg\to fg$) and ISUB=11, 12 ($f_if_j\to f_if_j$, $f_i\bar{f_i}\to f_j\bar{f_j}$) subprocesses.
Table  \ref{tab:bg_or_gr} shows also a tendency of 
increasing the contribution from the sum of two subprocess ``11+12'' 
(given in the second column of Table \ref{tab:bg_or_gr}) with growing $\Pt^{\tilde{\gamma}}$.

From  Tables \ref{tab:sb1} -- \ref{tab:sb3} we have seen that the cuts
listed in Table \ref{tab:sb0} (having rather moderate values of
$\Pt^{clust}_{CUT}$ and  $\Pt^{out}_{CUT}$) allow to suppress
the major part of the background events.
The influence of these two cuts  on: \\[1pt]
\hspace*{10mm} (a) the number of selected events (for $L_{int}=300\,pb^{-1}$);\\
\hspace*{10mm} (b) the signal-to-background ratio $S/B$;\\
\hspace*{10mm} (c) the mean value of 
$(\Pt^{\tilde{\gamma}}\!-\!\Pt^{Jet})/\Pt^{\tilde{\gamma}}\equiv F$ and
its  standard deviation value $\sigma (F)$\\[1pt]
is presented in Tables 1 -- 12 of Appendix 6 for their variation in a wide range.

Let us emphasize that the tables of Appendix 6 include, in contrast to Appendices 2--5, the results
obtained after analyzing three generated samples (described in the beginning of this section)
of {\it signal and background} events. 
These events were selected with the cuts of Table \ref{tab:sb0}.

Namely, the cuts (1) -- (10) of Table \ref{tab:sb0} were applied for preselection of 
``$\tilde{\gamma}+1~jet$'' events. 
The jets in these events as well as clusters were found by use of only one jetfinder LUCELL
(for the whole $\eta$ region $|\eta^{jet}|<4.2$).

Tables 1 -- 4 of Appendix 6 correspond to the simulation with
$\pth=40 ~GeV/c$. Analogously, the values of $\pth=70 ~GeV/c$ and $\pth=100 ~GeV/c$ were used for
Tables 5 -- 8  and Tables  9 -- 12 respectively.
The  rows and  columns of Tables 1 -- 12 illustrate, respectively, the influence of
$\Pt^{clust}_{CUT}$ and $\Pt^{out}_{CUT}$ on the quantities
mentioned above (in the points (a), (b), (c)).

First of all, we see from Tables 2, 6 and 10 that
a noticeable reduction
of the background take place while moving along the table diagonal from the right-hand bottom corner to the
left-hand upper one, i.e. with reinforcing $\Pt^{clust}_{CUT}$ and $\Pt^{out}_{CUT}$. 
So, we see that for $\pth=40 ~GeV/c$  the value of 
$S/B$ ratio changes in the table cells along the diagonal
from $S/B=2.5$ (in the case of no limits on these two variables), to $S/B=3.5$ for the
cell with $\Pt^{clust}_{CUT}=10~ GeV/c$ and $\Pt^{out}_{CUT}=10 ~GeV/c$.
Analogously, for $\pth=100 ~GeV/c\,$ the value of $S/B$ changes in the same table cells
from 9.8 to 20.6 (compare with the numbers in Table 10 of Appendix 6).

The second observation from Appendix 6. The restriction of $\Pt^{clust}_{CUT}$ and
$\Pt^{out}_{CUT}$ improves the calibration accuracy. Table 3 shows that in the interval $\Pt^{\tilde{\gamma}}\gt40~GeV/c$
the mean value of the fraction $F(\equiv (\Pt^{\tilde{\gamma}}\!-\!\Pt^{Jet})/\Pt^{\tilde{\gamma}})$
decreases from 0.049 (the bottom right-hand corner) to 0.024
for the table cell with $\Pt^{clust}_{CUT}=10~ GeV/c$ and $\Pt^{out}_{CUT}=10 ~GeV/c$.
At the same time, the both cuts lead to a noticeable decrease of
the gaussian width $\sigma (F)$ (see Table 4 and also Tables 8 and 12).  
For instance, for $\pth=40 ~GeV/c$ ~$\sigma (F)$ drops by about a factor of two: from 0.159 to 0.080.
It should be also noted that
Tables 4, 8 and 12 demonstrate that for any fixed
value of $\Pt^{clust}_{CUT}$  
further improvement in $\sigma (F)$ can be achieved  by limiting $\Pt^{out}$ 
(e.g. in line with $\Pt^{clust}_{CUT}=10~GeV/c$
$\sigma (F)$ drops by a factor of 2 with variation of $\Pt^{out}$ from $1000$ to $5~GeV/c$).

The explanation is simple. The balance~ equation (28) contains 2 terms on the right-hand
side ($1-cos\dphi$) and $\Db/\Pt^{\tilde{\gamma}}$.
The first one is negligibly small in a case of Selection 1 and tends to decrease with growing 
$\Pt^{\tilde{\gamma}}$ (see tables in Appendices 2--5). So, we see that in this case
the main source of the disbalance in  equation (28) is the term $\Db/\Pt^{\tilde{\gamma}}$.
This term can be diminished by decreasing $\Pt$ activity beyond the jet,
i.e. by decreasing $\Pt^{out}$.

\begin{figure}[htbp]
\vspace{-2.6cm}
\hspace{-.7cm} \includegraphics[width=17cm,height=20cm]{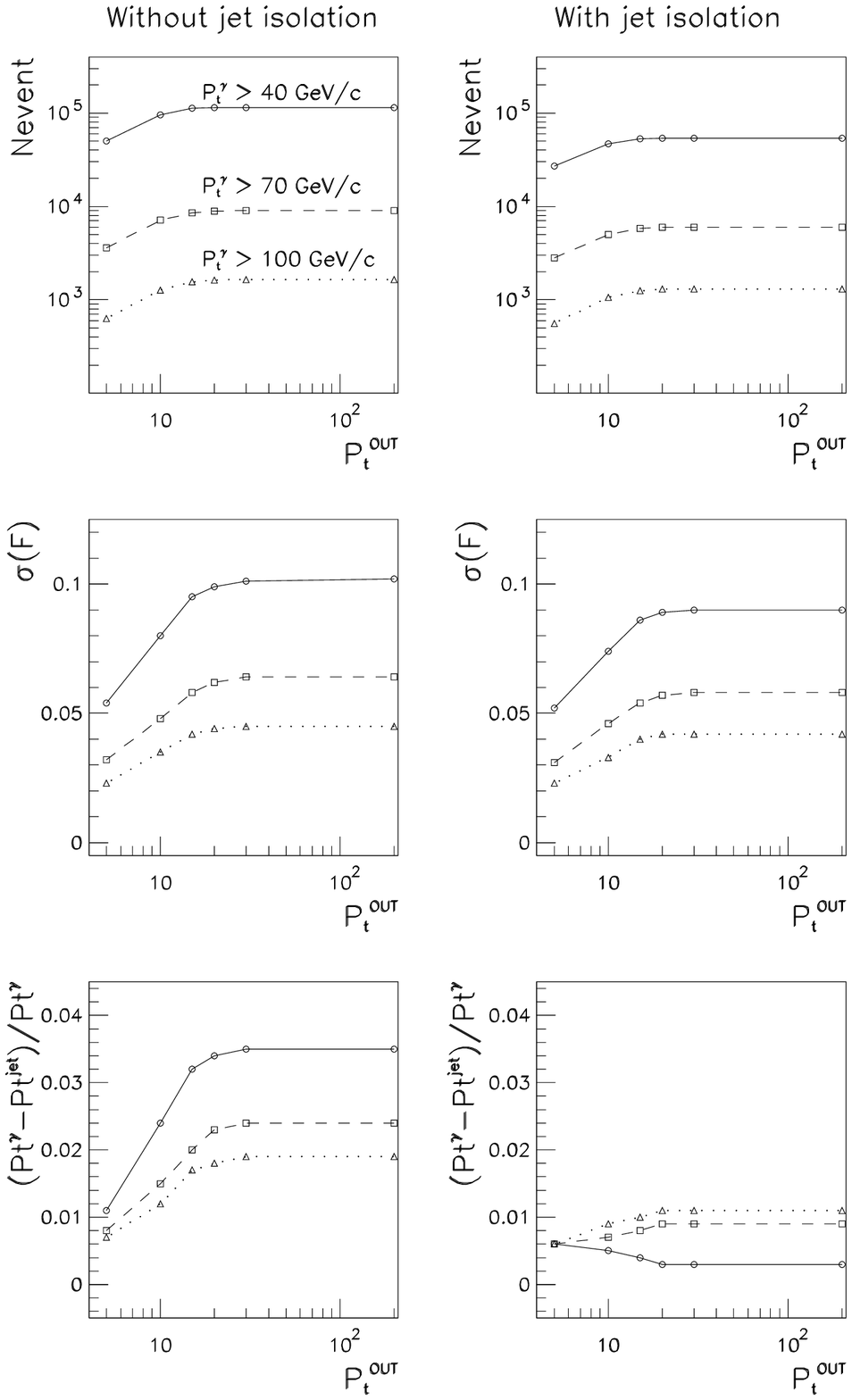}
\vspace{-0.7cm}
\caption{\hspace*{0.0cm} Number of events (for $L_{int}=300\,pb^{-1}$),
 mean value of $(\Pt^{\tilde{\gamma}}-\Pt^{Jet})/\Pt^{\tilde{\gamma}}$
($\equiv F$) and its standard deviation $\sigma(F)$ distributions over $\Pt^{out}$
for the cases of nonisolated (left-hand column) and isolated (right-hand column) jet and for
three  intervals: $\Pt^{\tilde{\gamma}} > 40, 70$ and $100 ~GeV/c$. $\Pt^{clust}_{CUT}=10 ~GeV/c.$
}
\label{fig:mu-sig}
\end{figure}

The behavior of the number of selected events (for $L_{int}=300\,pb^{-1}$),
 the mean values of $F=(\Pt^{\tilde{\gamma}}\!-\!\Pt^{Jet})/\Pt^{\tilde{\gamma}}$ and
its standard deviation $\sigma (F)$ as a function of $\Pt^{out}_{CUT}$ 
(with fixed $\Pt^{clust}_{CUT}=10~GeV/c$)
are also displayed in Fig.~\ref{fig:mu-sig} for events with non-isolated 
(left-hand column) and isolated jets (right-hand column, see also Tables 13--24 of Appendix 6).

{\it Thus, we can conclude that application of two criteria introduced
in Section 3.2, i.e. $\Pt^{clust}_{CUT}$ and $\Pt^{out}_{CUT}$,
results in two important consequences: significant background reduction
and essential improvement of the calibration accuracy.
}

The numbers of events (for $L_{int}= 300 ~pb^{-1}$)
for different $\Pt^{clust}_{CUT}$ and $\Pt^{out}_{CUT}$
are given in the cells of Tables 1, 5 and 9 of Appendix 6. One can see that even with such strict
$\Pt^{clust}_{CUT}$ and $\Pt^{out}_{CUT}$ values as, for example, $10 ~GeV/c$ for both
we would have a sufficient number of events
(about 100 000, 7 000 and 1 300 for $\Pt^{\tilde{\gamma}}\geq40 ~GeV/c$,
$\Pt^{\tilde{\gamma}}\geq70 ~GeV/c$ and  $\Pt^{\tilde{\gamma}}\geq100 ~GeV/c$, respectively)
with low background contamination ($S/B=3.5,~9.8$ and $20.6$)
and a good accuracy of the $\Pt^{\tilde{\gamma}}-\Pt^{Jet}$ balance:
$F=2.4\%, 1.5\%$ and $1.2\%$, respectively, for the case of Selection 1.

In addition, we also present Tables 13--24 of Appendix 6.
They contain the information analogous to that in Tables 1 -- 12
but for the case of isolated jets with $\epsilon^{jet}<3\%$.
From these tables we see that with the same cuts
 $\Pt^{clust}_{CUT}=\Pt^{out}_{CUT}=10 ~GeV/c$ one can expect about\\
47 000, 5 000 and 1000 events for $\Pt^{\tilde{\gamma}}\geq40 ~GeV/c$,
$\Pt^{\tilde{\gamma}}\geq70 ~GeV/c$ and  $\Pt^{\tilde{\gamma}}\geq100 ~GeV/c$, 
respectively, with a much more better fractional $\Pt^{\tilde{\gamma}}-\Pt^{Jet}$ balance:
$F=0.5\%, 0.7\%$ and $0.1\%$.

Let us mention that all these PYTHIA results give us an indication of a tendency
and may serve as a guideline for further full GEANT simulation that would allow to come to 
a final conclusion. 

To conclude this section we would like to stress, firstly, that, as is seen from
Tables~\ref{tab:sb1}, the  ``$\gamma-brem$'' background
defines a dominant part of the total background. 
One can see from Table ~\ref{tab:sb1}
that $\pi^0$ contribution being about the same as
``$\gamma-brem$'' at $\pth>40~GeV/c$ becomes three times less than
``$\gamma-brem$'' contribution at $\pth>100~GeV/c$. We would like to emphasize here
that this is a strong prediction of the PYTHIA generator that has to be compared with
predictions of another generator like HERWIG, for example.
%
%

Secondly, we would like to mention also that, as it is seen from Tables \ref{tab:sb4} and
\ref{tab:sb1},  the photon isolation and selection cuts 1--6, usually used in 
the study of inclusive  photon production (see, for instance, \cite{CDF1}, \cite{D0_1}, 
\cite{D0_2}), increase the $S/B$ ratio up to 3.20 only (for $\Pt^{\tilde{\gamma}}\geq70 ~GeV/c$).
The other ``hadronic'' cuts 7--17, that select events with a clear \gpj topology and
limited $\Pt$ activity beyond \gpj system, lead to quite a significant improvement of
$S/B$ ratio by a factor of four (to $S/B=11.37$) as they suppress the background events contamination 
by a factor of about fourteen at the cost of four-fold loss of signal events.

The numbers in the tables of Appendix 6 were obtained with inclusion of the contribution
from the background events. The tables show that their account does not spoil the \ptgj
balance in the event samples preselected with the cuts 1--10 of Table \ref{tab:sb0}. 
The estimation of the number of these background events would be important
for the gluon distribution determination (see Section 10).

\normalsize
\def\baselinestretch{1.0}

\section{STUDY OF DEPENDENCE OF THE \Ptgj BALANCE ON PARTON $k_t$.}

\it\small
\hspace*{9mm}
It is shown that in the case of ISR presence the value of fractional disbalance $\Fptgj$
depends weakly on the variation of the average value of intrinsic parton transverse momentum 
$\la k_{~t}\ra$.
\rm\normalsize
\vskip3mm

This section is dedicated to the study (within PYTHIA simulation) of a possible influence of the intrinsic parton
 transverse momentum $k_{~t}$ on the $\Pt$ balance of the \gpj system. For this aim we consider
two samples of signal events gained by simulation  with subprocesses
(1a) and (1b) in two different ranges of $\pth$: $\pth \geq 40~ GeV/c$ and  $\pth \geq 100~ GeV/c$.
For these two $\pth$  intervals Tables \ref{ap1:tab1} and \ref{ap1:tab2}
demonstrate the average values of $\Pt56$ (defined by (3)) for two different cases of generation: 
without initial state radiation 
(``ISR is OFF'') and with it (``ISR is ON''). Four different generations were done for each
$\pth$ interval. They correspond to four 
~\\[-8mm]
\begin{table}[htbp]
\small
\begin{center}
\caption{Effect of $k_t$ on the $\Pt^{\gamma}$ - $\Pt^{Jet}$ balance with
$\pth\! \!=\!\! 40 ~GeV/c ~~ (F=\Fptgj)$.}
\vskip0.1cm
\begin{tabular}{||c||c|c|c||c|c|c||}   \hline  \hline
\label{ap1:tab1}
$\la k_t\ra$& \multicolumn{3}{|c||}{ ISR is OFF}&\multicolumn{3}{c||}{ ISR is ON} \\\cline{2-7}
$(GeV/c)$ &$\la \Pt56\ra$&$\la F\ra$&$\sigma(F)$
&$\la \Pt56\ra$&$\la F\ra$&$\sigma(F)$   \\\hline  \hline
 0.0 &0.0 &0.021 &0.050&6.4 &0.022 &0.080\\\hline
 1.0 &1.8&0.023 &0.053 &6.7 &0.024 &0.082  \\\hline
 2.0 &3.5&0.024 &0.062 &7.2 &0.024 &0.084  \\\hline
 5.0 &8.4&0.027 &0.096 &9.0 &0.026 &0.100  \\\hline
\end{tabular}
\end{center}
\vskip-1mm
\end{table}
\begin{table}[htbp]
\small
\begin{center}
\vskip-9mm
\caption{Effect of $k_t$ on $\Pt^{\gamma}$ -$\Pt^{Jet}$ balance with
$\pth \!\!=\!\! 100~ GeV/c ~~ (F=\Fptgj)$.}
\vskip0.1cm
\begin{tabular}[h]{||c||c|c|c||c|c|c||}                  \hline  \hline
\label{ap1:tab2}
$\la k_t\ra$& \multicolumn{3}{|c||}{ ISR is OFF}&\multicolumn{3}{c||}{ ISR is ON}
\\\cline{2-7}
$(GeV/c)$ &$\la \Pt56\ra$&$\la F\ra$&$\sigma(F)$
&$\la \Pt56\ra$&$\la F\ra$&$\sigma(F)$   \\\hline  \hline
 0.0 &0.0 &0.011 &0.033 &7.2 &0.012 &0.045\\\hline
 1.0 &1.8 &0.012 &0.035 &7.5 &0.013  &0.045 \\\hline
 2.0 &3.6 &0.014 &0.039 &8.1 &0.013  &0.046 \\\hline
 5.0 &8.5 &0.014 &0.050 &10.3&0.014  &0.054 \\\hline
\end{tabular}
\end{center}
~\\[-12pt]
\hspace*{25mm}$\ast$ \footnotesize{All numbers in the tables above are given in
 $GeV/c$}.
\vskip-2mm
\end{table}
\normalsize

\noindent
values of parton $\la k_t\ra$:
\footnote{$\equiv$ PARP(91) parameter in PYTHIA}
$\la k_t\ra\!=\!0.0, 1.0, 2.0$ and $5.0~GeV/c$
(the values $\la k_t\ra > 1~ GeV/c$ are given
  here only for illustration of a tendency). 

Let us consider firstly the case with ISR switched off during the simulation.
The numbers in Tables \ref{ap1:tab1} and \ref{ap1:tab2} (obtained from the set of events selected by 
the cuts $\dphi<17^\circ$, $\Pt^{out}_{CUT}=10~GeV/c$ and $\Pt^{clust}_{CUT}=10~ GeV/c$) show 
that in the case when ``ISR is OFF'' the value of $\la\Pt56\ra$ 
grows rapidly with increasing $\la k_t\ra$ and does not depend on \ptg (or $\pth$).
In fact, the values  of $\la\Pt{56}\ra$  are proportional to the values of $\la k_t\ra$ in this case. 
\begin{figure}[t]
\begin{center}
\vspace{-3mm}
  \includegraphics[width=16.7cm,height=11.5cm]{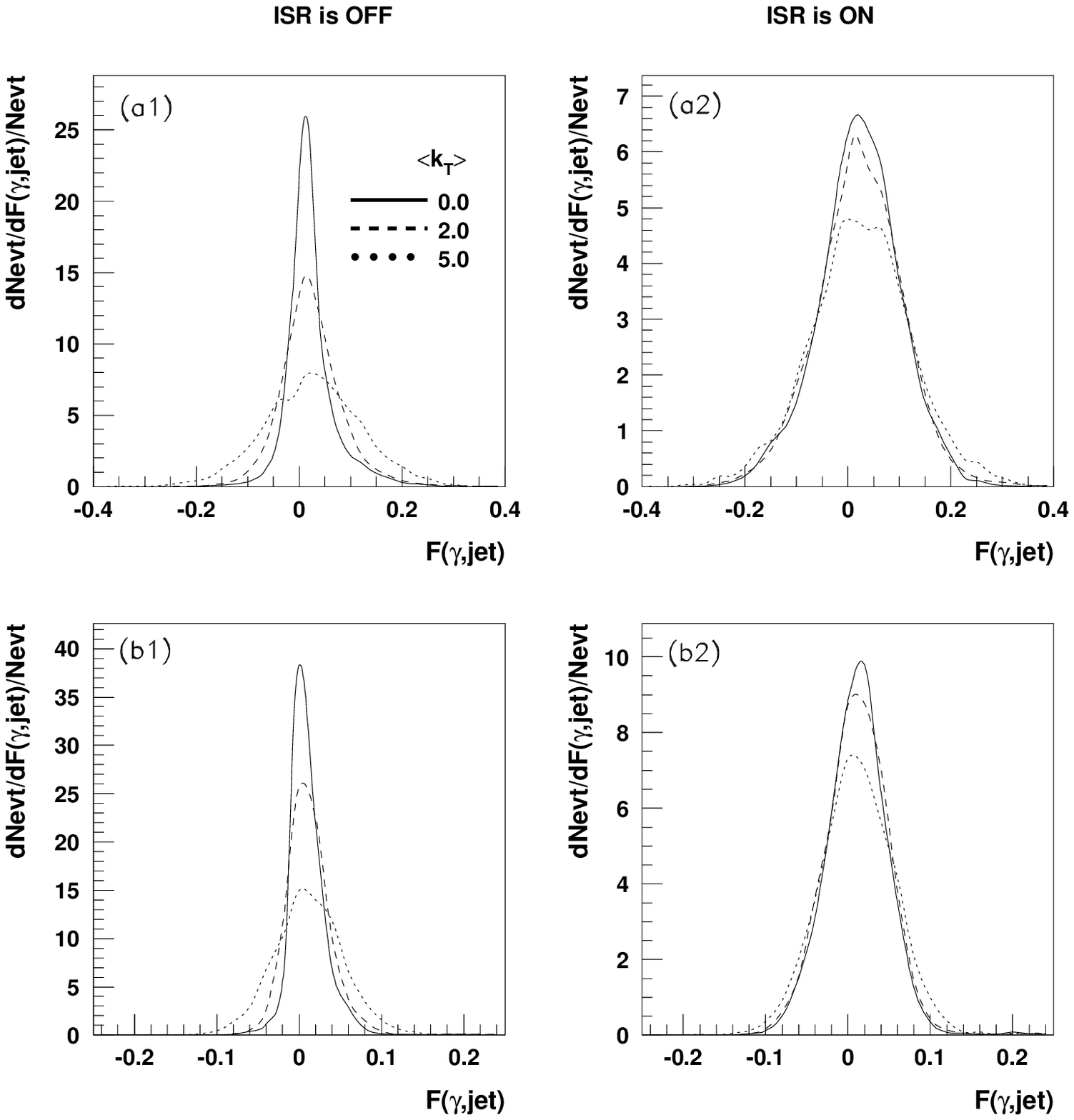}
  \vspace{-12mm}
    \caption{\hspace*{0.0cm} $\Fptgj\equiv F(\gamma,jet)$ as a function of primordial $k_t$ value
for the cases of switched on (a1, b1) and switched off (a2, b2) initial radiation and for
$\pth = 40$ (a1, a2) and $\pth = 100~~GeV/c$ (b1, b2).}
\label{fig:kt}
\end{center}
\vskip-8mm
\end{figure}

The picture changes when ISR is taken into accound. In this case the variable $\la\Pt56\ra$ initially gets large 
value at $\la k_t\ra=0$: $\la\Pt{56}\ra=6.4~ GeV/c$  and $\la\Pt{56}\ra=7.2 ~GeV/c$ 
for $\pth=40 ~GeV/c$ and $\pth=100 ~GeV/c$, respectively.
But at the same time, in contrast to the case ``ISR is OFF'',
the values of $\la\Pt56\ra$ grow more slowly with $\la k_t\ra$ when  ``ISR is ON''.
Indeed, they grow up 
from $6.4~(7.2)$ at $\la k_t\ra=0.0$ to $9.0~(10.3)$ at $\la k_t\ra=5~GeV/c$ for 
$\pth\!=\!40 ~GeV/c$~$(100 ~GeV/c)$.

The most remarkable thing, as it follows from Tables \ref{ap1:tab1} and \ref{ap1:tab2},
that $\la\Pt56\ra$ depends weakly on $\la k_t\ra$ in the range of its reasonable values 
$\la k_t\ra\leq 1~GeV/c$.

The variations of the fractional disbalance $F\!\equiv\!\Fptgj$  and its standard deviation $\sigma(F)$
with $\la k_t\ra$ are also shown in Tables \ref{ap1:tab1} and \ref{ap1:tab2} and in plots of
Fig.~\ref{fig:kt}. One can see that for reasonable values $\la k_t\ra\leq 1~GeV/c$ and for the case 
``ISR is ON'' the changes in the fractional disbalance $F$ with $k_t$ variation are very small. 
They are of order of $0.2\%$ for $\pth=40 ~GeV/c$ and of order of $0.1\%$ for $\pth=100 ~GeV/c$
\footnote{Recall that the numbers in Tables \ref{ap1:tab1} and \ref{ap1:tab2} may be 
compared with those in the tables of Appendix 6, where the same $\pth$ cuts are used,
rather than with the results of the tables of Appendices 2 -- 5, where  $\pth$ cuts 
were taken to be two times smaller (see for explanation the beginning of Section 7).}.
%

%
\section{\gpj EVENT RATE ESTIMATION FOR GLUON DISTRIBUTION DETERMINATION AT THE TEVATRON RUN~II.}

\it\small
\hspace*{9mm}
The number of \gpj events suitable for measurement of gluon distribution in different
$x$ and $Q^{\;2}$ intervals at Run~II is estimated. It is shown that with $L_{int}=3~fb^{-1}$
it would be possible to collect about one million of these events. This number
would allow to cover a new kinematical area not studied in any previous experiment
($10^{-3}\lt x\lt 1.0$ with $1.6\cdot 10^{3}\leq Q^2\leq2\cdot10^{4} ~(GeV/c)^2$).
This area in the region of small $x\geq10^{-3}$ has $Q^2$ by about one order of magnitude higher than 
reached at HERA now.
\rm\normalsize
\vskip4mm

As many of theoretical predictions for production of new particles
(Higgs, SUSY) at the Tevatron are based on model estimations of the gluon density behavior at
low $x$ and high $Q^2$, the measurement of the proton gluon density
for this kinematic region directly in Tevatron experiments
would be obviously useful. One of the promising channels for this measurement, as was shown in ~\cite{Au1},
is a high $\Pt$ direct photon production $p\bar{p}(p)\rightarrow \gamma^{dir} + X$.
The region of high $\Pt$, reached by UA1 \cite{UA1}, UA2 \cite{UA2}, CDF \cite{CDF1} and
D0 \cite{D0_1} extends up to $\Pt \approx 80~ GeV/c$ and recently up to $\Pt=105~GeV/c$ \cite{D0_2}. 
These data together with the later ones (see references in \cite{Fer}--\cite{Fr1} and recent
E706 \cite{E706} and UA6 \cite{UA6} results) give an opportunity for tuning the form of gluon 
distribution (see \cite{Au2}, \cite{Vo1}, \cite{Mar}). The rates and estimated cross sections 
of inclusive direct photon production at the LHC were given in \cite{Au1} (see also \cite{AFF}).

Here for the same aim we shall consider 
the process $p\bar{p}\rightarrow \gamma^{dir}\, +\, 1\,Jet\, + \,X$
defined in the leading order by two QCD subprocesses (1a) and (1b)
(for experimental results see \cite{ISR}, \cite{CDF2}).

Apart from the advantages, discussed in Section 8  in connection with the background suppression 
(see also \cite{Ber}--\cite{Hu2}),
the ``$\gamma^{dir}+1~Jet$'' final state may be easier for physical analysis
than inclusive photon production process ``$\gamma^{dir}+X$''  if we shall look at this problem from
the viewpoint of extraction of information on the gluon distribution in a proton.
Indeed, in the case of inclusive direct photon production the
cross section is given as an integral over the products of a fundamental $2\to2$ parton subprocess
cross sections and the corresponding parton distribution functions $f_a(x_a,Q^2)$ (a = quark or gluon), 
while in the case of $p\bar{p}\rightarrow \gamma^{dir}+1~Jet+X$ for $\Pt^{Jet}\,
\geq \,30\, GeV/c$ (i.e. in the region where ``$k_t$ smearing effects''
\footnote{This terminology is different from ours, used in Sections 2 and 9, as we denote by ``$k_t$''
only the value of parton intrinsic transverse momentum.}
are not important, see \cite{Hu3}) the cross section is
expressed directly in terms of these distributions (see, for example,
\cite{Owe}): \\[-16pt]
\begin{eqnarray}
\frac{d\sigma}{d\eta_1d\eta_2d\Pt^2} = \sum\limits_{a,b}\,x_a\,f_a(x_a,Q^2)\,
x_b\,f_b(x_b,Q^2)\frac{d\sigma}{d\hat{t}}(a\,b\rightarrow c\,d),
\label{gl:4}
\end{eqnarray}
\vskip-3mm
\noindent
where \\[-9mm]
\begin{eqnarray}
x_{a,b} \,=\,\Pt/\sqrt{s}\cdot \,(exp(\pm \eta_{1})\,+\,exp(\pm \eta_{2})).
\label{eq:x_def}
\end{eqnarray}
\vskip-1mm

The designation used above are as the following:
$\eta_1=\eta^\gamma$, $\eta_2=\eta^{Jet}$; ~$\Pt=\Pt^\gamma$;~ $a,b=q,\bar{q},g$; 
$c,d=q,\bar{q},g,\gamma$. Formula (\ref{gl:4}) and the knowledge of 
$q, \,\bar{q}$ distributions 
allow the gluon  distribution $f_g(x,Q^2)$
to be determined after account of selection efficiencies for jets and  $\gamma^{dir}-$candidates 
as well as after subtraction of the background contribution, 
left after the used selection cuts 1--13 of Table \ref{tab:sb0}
(as it was discussed in Section 8 keeping in hand this physical application).

In the previous sections a lot of details connected with the
structure and topology of these events and the features of objects appearing
in them were discussed. Now with this information in mind we are
in position to discuss an application of the \gpj
event samples, selected with the previously proposed cuts, for estimating
the rates of gluon-based  subprocess (1a) in different $x$ and $Q^2$ intervals.

Table~\ref{tab:q/g-1} shows percentage of ``Compton-like'' subprocess
(1a) (amounting to $100\%$ together with (1b)) in the samples of events selected with cuts
(17) -- (23) of Section 3.2 for $\Pt^{clust}_{CUT}=10~GeV/c$ for different $\Ptg$
and $\eta^{Jet}$ intervals: Central (CC) ($|\eta^{Jet}|\lt0.7$)
\footnote{see also tables of Appendix 1},
Intercryostat (IC) $0.7\lt|\eta^{Jet}|\lt1.8$ and End (EC) $1.8\lt|\eta^{Jet}|\lt2.5$ parts of calorimeter.
We see that the contribution of Compton-like subprocess grows by about $5-6\%$ with $|\eta^{Jet}|$
enlarging and drops with growing $\Pt^{Jet}(\approx\Ptg$ in the sample of the events collected with the cuts 
$1-13$  of Table 13).
\\[-10mm]
\begin{table}[h]
\begin{center}
\caption{The percentage of Compton-like process  $q~ g\rrr \gamma +q$.}
\normalsize
\vskip.1cm
\begin{tabular}{||c||c|c|c|c|}                  \hline \hline
\label{tab:q/g-1}
Calorimeter& \multicolumn{4}{c|}{$\Pt^{Jet}$ interval ($GeV/c$)} \\\cline{2-5}
    part   & 40--50 & 50--70 & 70--90 & 90--140   \\\hline \hline
CC         & 84     &  80   &  74&  68  \\\hline
IC         & 85     &  82   &  76&  70  \\\hline
EC         & 89     &  85   &  82&  73  \\\hline
\end{tabular}
\end{center}
\end{table}
\normalsize
~\\[-10mm]


In Table~\ref{tab:q/g-2} we present the $Q^2 (\equiv(\Ptg)^2)$
\footnote{see \cite{PYT}}
and $x$ (defined according to (\ref{eq:x_def}))
distribution of the number of events that are caused by the $q~ g\rrr \gamma +q$ subprocess, 
and passed the following cuts ($\Pt^{out}$ was not limited):\\[-8mm]
\begin{eqnarray}
\Ptg>40~ GeV/c,\quad |\eta^{\gamma}|<2.5,\quad \Pt^{Jet}>30~ GeV/c,\quad |\eta^{Jet}|<4.2,\quad \Pt^{hadr}>7~ GeV/c,
\label{l1}
\nonumber
\end{eqnarray}
~\\[-15mm]
\begin{eqnarray}
\Pt^{isol}_{CUT}=4\;GeV/c, \;
{\epsilon}^{\gamma}_{CUT}=7\%, \;
\dphi<17^{\circ}, \; 
\Pt^{clust}_{CUT}=10\;GeV/c. \;
\label{l2}
\end{eqnarray}

\begin{table}[h]
\begin{center}
\vskip-0.2cm
\caption{Number of~ $g\,q\to \gamma^{dir} \,+\,q$~
events at different $Q^2$ and $x$ intervals for $L_{int}=3~fb^{-1}$.}
\label{tab:q/g-2}
\vskip0.2cm
\begin{tabular}{|lc|r|r|r|r|r|r|r|}                  \hline
 & $Q^2$ &\multicolumn{6}{c|}{ \hspace{-0.9cm} $x$ values of a parton} &All $x$  \\\cline{3-9}
& $(GeV/c)^2$  &$.001-.005$ & $.005-.01$ & $.01-.05$ &$.05-.1$ & $.1-.5$ &$.5-1.$
&$.001-1.$     \\\hline
&\hmm\hmm 1600-2500\hmm  & 8582  & 56288  &245157  &115870  &203018  &  3647  &632563  \\\hline
&\hmm\hmm 2500-4900\hmm  &  371  & 13514  &119305  & 64412  &119889  &  3196  &320688 \\\hline
&\hmm\hmm 4900-8100\hmm  &    0  &   204  & 17865  & 13514  & 26364  &  1059  & 59007\\\hline
&\hmm\hmm 8100-19600\hmm &    0  &     0  &  3838  &  5623  & 11539  &   548  & 21549\\\hline
\multicolumn{8}{c|}{}&{\bf 1 033 807}\\\cline{9-9}
\end{tabular}
\end{center}
\end{table}

\begin{table}[h]
\begin{center}
\vskip-.2cm
\caption{Number of~ $g\,c\to \gamma^{dir} \,+\,c$~
events at different $Q^2$ and $x$ intervals for $L_{int}=3~fb^{-1}$.}
\label{tab:q/g-3}
\vskip0.2cm
\begin{tabular}{|lc|r|r|r|r|r|r|r|}                  \hline
 & $Q^2$ &\multicolumn{6}{c|}{ \hspace{-0.9cm} $x$ values of a parton} &All $x$\\\cline{3-9}
& $(GeV/c)^2$  &$.001-.005$ & $.005-.01$ & $.01-.05$ &$.05-.1$ & $.1-.5$ &$.5-1.$
&$.001-1.$     \\\hline
&\hmm\hmm 1600-2500\hmm  &  264  &  2318  & 21236  & 11758  & 14172  &    58  & 49805 \\\hline
&\hmm\hmm 2500-4900\hmm  &   13  &   332  &  9522  &  6193  &  7785  &    40  & 23885  \\\hline
&\hmm\hmm 4900-8100\hmm  &    0  &     4  &   914  &  1055  &  1648  &    16  &  3637\\\hline
&\hmm\hmm 8100-19600\hmm &    0  &     0  &   142  &   329  &   612  &     8  &  1092 \\\hline
\multicolumn{8}{c|}{}&{\bf 78 419}\\\cline{9-9}
\end{tabular}
\end{center}
\end{table}

The analogous information for events with the charmed quarks in the initial state
$g\,c\to \gamma^{dir} \,+\,c$ is presented in Table~\ref{tab:q/g-3} 
(see also tables of Appendix 1). 
The simulation of the process $g\,b\to \gamma^{dir}\,+\,b$ $\;$ has shown that the rates
for the $b$-quark are 8 -- 10 times smaller than for
the $c$-quark (these event rates are also given in tables of Appendix 1 for different $\Ptg$ intervals)
\footnote{Analogous estimation for LHC energy was done in \cite{BKS_GLU} and \cite{MD1}.}. 

\normalsize
~\\[10mm]
\begin{flushleft}
\begin{figure}[h]
   \vskip-33mm
   \hspace{-1mm} \includegraphics[width=.55\linewidth,height=9.8cm,angle=0]{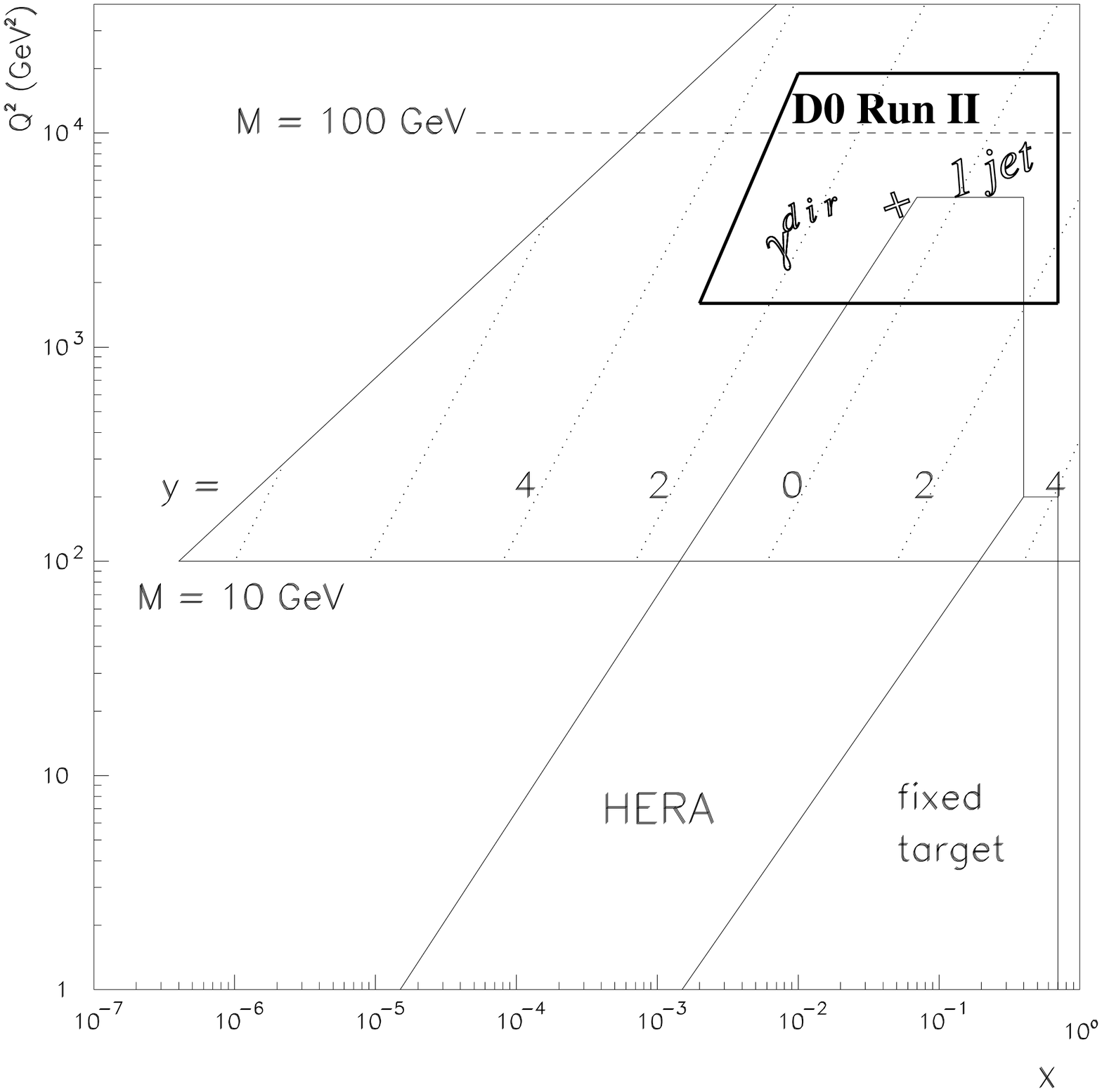}
\label{fig:q/g}
\end{figure}
\end{flushleft}
\begin{flushright}
\vskip-108mm
\parbox[r]{.42\linewidth}
{ \vspace*{0.1cm}
 \hspace{8mm} Fig.~29 shows in the widely used $(x,Q^2)$
kinematic plot (see \cite{Sti} and also in  \cite{Hu3})
what area can be covered by studying the process $q~ g\rrr \gamma +q$ at Tevatron.
The distribution of number of events in this area is given by Table~\ref{tab:q/g-2}.
From this figure and Table~\ref{tab:q/g-2} it becomes clear that with integrated luminosity
$L_{int}=3~fb^{-1}$ it would be possible to study the gluon distribution with a good statistics 
of \gpj events in the region of  $10^{-3}\lt x\lt 1.0$ with $Q^2$ by about 
one order of magnitude higher than reached at HERA now.
It is worth emphasizing that extension of the experimentally reachable
region at the Tevatron to the region of lower $Q^2$ overlapping with the area
covered by HERA would also be of great interest.}
\end{flushright}

{\vskip0.1cm
\hspace*{-.5cm} {Figure~29: \footnotesize {The  $(x,Q^2)$ kinematic region for
studying~ $p\bar{p}\to \gamma+Jet$ process at Tevatron Run~II.}}
}

~\\[2mm]

%
\section{SUMMARY.}

We have done an attempt here to consider, following 
\cite{9}--\cite{BKS_GLU}, the physics of high $\Pt$ direct photon and jet 
associative production in proton-antiproton collisions
basing on the predictions of PYTHIA generator and the models implemented there.
This work may be useful for two practical goals: for absolute jet energy scale determination
and for gluon distribution measurement at Tevatron energy.

The detailed information provided in the PYTHIA event listings allows to track the origin of
different particles (like photons) and of objects (like clusters and jets) 
that appear in the final state. 
So, the aims of this work was to explore at the particle level as much as
possible this information for finding out what effect may be produced
by new variables, proposed in \cite{9}--\cite{BKS_P5} for describing \gpj events,  
and the cuts on them for solution of the mentioned above practical tasks.

For the first problem of the jet energy determination an important task is to select the events
that may be caused (with a high probability) by the $q\bar{q}\to g+\gamma$ and 
$qg\to q+\gamma$~ fundamental parton subprocesses of direct photon production. 
To take into account a possible effect
of initial state radiation (its spectra are presented in different \ptg ~intervals 
in Tables 2--7 of Section 5) we used here the $\Pt$-balance equation (see (16)) written for
an event as a whole. It allows to express \ptgj fractional disbalance (see (28)) through new variables
\cite{9}--\cite{BKS_P5} that describe the $\Pt$ activity {\it out} of \gpj system. They are $\Pt^{out}$ and
$\Pt^{clust}$, i.e. $\Pt$  of mini-jets or clusters that are additional to the main jet in event.
The latter is the most ``visible'' part of $\Pt^{out}$.

These two sources of the \ptgj disbalance are investigated. It is shown that the limitation of $\Pt$
of clusters, i.e. $\Pt^{clust}$,  can help to decrease this disbalance
(see Figs.~12--18 and Tables 4--12 of Appendices 2--5). 

Analogously, the limitation of $\Pt$ activity of all detectable particles ($|\eta_i|\lt4.2$) beyond the \gpj system, 
i.e. $\Pt^{out}$, also leads to a noticeable \ptgj disbalance reduction (see Figs.~19 and 20).

It is demonstrated that in the events selected by means of simultaneous restriction  from above
of the $\Pt^{clust}$ and $\Pt^{out}$ activity the values of \Ptgj are well baalnced with each other
while considering the PYTHIA particle level of simulation. The samples of these \gpj events gained
in this way are of a large enough volume for jet energy scale determination in the interval
$40\lt\Ptg\lt140~GeV/c$ (see Tables 1--12 of Appendix 6). 

It is worth mentioning that the most effect for improvement of \Ptgj balance can be reached by applying 
additionally the jet isolation criterion defined in \cite{9}--\cite{BKS_P5}. As it can be seen from Figs.~15, 16 
(Selection 2) as well as from Figs.~17, 18 (Selection 3) and also from
Tables 13--18 of Appendices 2--5 and Tables 13--24 of Appendix 6, the application of this criterion
allows to select the events having the \ptgj disbalance at the particle level less than $1\%$
\footnote{The achieved disbalance value at the particle level of simulation shows the most
optimistic value of $\Fptgj$.}.
Definitely, the detector effects may worsen the balance determination
due to the limited accuracy of the experimental measurement.
We are planning to present the results of full GEANT simulation with 
the following digitization and reconstruction of signals by using the corresponding
D0 packages (like D0GSTAR) in the forthcoming papers.

We present also PYTHIA predictions for the dependence of the distributions of the number of 
selected \gpj events on \ptg~ and $\eta^{Jet}$
(see Tables 8--12 of Section 5 and also tables of Appendices 2--5 with account of $\Pt^{clust}$ 
variation). The features of \gpj events in the central region of the
D0 detector ($|\eta^{Jet}|<0.7$) are exposed (see Figs.~8, 9). The $\Pt$ structure of the region
in the $\eta-\phi$ space inside and beyond a jet is established (see Figs.~10, 11).

The corrections to the measurable values of $\Pt^{jet}$ that have take into account the
contribution from neutrinos belonging to a jet are presented for different $\Pt^{Jet}(\approx\Ptg$ for 
the selected events$)$ intervals
in the tables of Appendix 1. It is shown in Section 4 that a cut on $\Pt^{miss}\lt10~GeV/c$ allows 
to reduce this contribution down to the value of
$\Delta_\nu=\la\Pt^{Jet}_{(\nu)}\ra_{all\; events}=0.1~GeV/c$. At the same time, as it is shown in  
\cite{QCD_talk}, and discussed in Sections 4, 8 (see also \cite{BKS_P5}), 
this cut noticeably decreases the number of the background $e^\pm$-events 
in which $e^\pm$ (produced in the $W^\pm\to e^\pm\nu$ weak decay) may be registered as direct photon.

The study of the fractional disbalance $\Fptgj$ dependence on an intrinsic parton transverse momentum 
$\la k_{\,t}\ra$, performed in Section 9, has shown its weak impact on the disbalance
in the case of initial state radiation account.

The possibility of the background events (caused by QCD subprocesses of $qg, gg, qq$ scattering) 
suppression  was studied in Section 8. Basing on the introduced selection
criteria that include 17 cuts (see Table 13 of Section 8),
the background suppression relative factors and the values of signal event selection
efficiencies  are estimated (see Table 14). 

It is shown that after applying the first 6 ``photonic'' cuts 
(that may be used, for example, for selecting events with inclusive photon production and lead to
signal-to-background ratio $S/B=3.2$ in the interval $\Ptg>70 ~GeV/c$, see Table 14) 
the use of the next 11 ``hadronic'' cuts of Table 13 may lead to further essential improvement 
of $S/B$ ratio (by factor of 3.5 for the same $\Ptg>70 ~GeV/c$ where $S/B$ becomes $11.4$, see Table 14).

It is important to underline that this improvement is achieved by applying ``hadronic'' cuts
that select the events having clear \gpj topology at the particle level and also having rather ``clean'' area
(in a sense of limited $\Pt$ activity) beyond a \gpj system.
The  consideration of the cuts, connected with detector effects 
(e.g., based on an electromagnetic shower profile \cite{D0_1}, \cite{D0_2}),
may lead to  further improvement of $S/B$ ratio.
In this sense and taking into account the fact that these ``hadronic'' cuts lead 
to an essential improvement of \ptgj balance,
one may say that the cuts on $\Pt^{clust}$ and $\Pt^{out}$, considered here,
do act quite effectively to select the events caused by leading order diagrams (see Fig.~1) and do suppress the
contribution of NLO diagrams, presented in Figs.~2, 4.

Another interesting predictions of PYTHIA is about the dominant contribution of
``$\gamma$-brem'' events into the total background at Tevatron energy, as in was already mentioned in Section 8
(see also \cite{BKS_P5} and \cite{QCD_talk}).  As the ``$\gamma$-brem'' background has irreducible nature
its careful estimation is an important task and we plan to make the analogous estimation with HERWIG generator. 

To finish the discussion of the jet calibration study let us mention that the main results on this subject 
are summed up in Tables 1--12 (Selection 1) and 13--24 (Selection 2 with jet isolation criterion)
of Appendix 6 and Fig.~\ref{fig:mu-sig}.

It should be emphasized that numbers presented in all mentioned tables and figures were found
within the PYTHIA particle level of simulation. They may depend on the used generator and on the particular
choice of a long set of its parameters
\footnote{We have already mentioned that we are planning
to perform analogous analysis by help of another generator like HERWIG, for example.
The comparison of predictions of different generators (PYTHIA, HERWIG, etc.) with the experimental
results is a part of a work in any experiment.}
as well as they may change after account of the results of the full GEANT-based simulation.

The preliminary D0 Run~II data taken during January 2002  were used to demonstrate how do
the cuts (1--16) of Table 13 work to select $\gamma$-candidate and 
$\gamma$-candidate + 1 jet events (see the slides of already mentioned our  talk at QCD group 
\cite{QCD_talk2}). It is interesting to note that the final number of selected events with
$\Ptg\gt 40 ~GeV/c$ approximately agrees (after correction to the values of
photon purity and selection efficiency taken from \cite{D0_2}) with the results of our estimation
presented in Table 8 and Table 4 of Appendix 2.

It is shown that the samples of the \gpj events, gained with the cuts used for the jet energy calibration,
can provide an information suitable also for determining the
gluon distribution inside a proton in the kinematic region (see Fig.~29) that includes
$x$ values as small as accessible at HERA  \cite{H1}, \cite{ZEUS}, but
at much higher $Q^2$ values (by about one order of magnitude):
$10^{-3}\leq x \leq 1.0$ with $1.6\cdot10^3\leq Q^2\leq2\cdot10^4 ~(GeV/c)^2$.
The number of  events, based on the gluonic process (1a), that may be collected 
with $L_{int}=3~fb^{-1}$ in different $x$- and $Q^2$- intervals of this new kinematic region
for this goal are presented in Table 22
(all quarks included) and in Table 23 (only for charm quarks) 
\footnote{see also tables of Appendix 1}.

~\\[1mm]
\section{ACKNOWLEDGMENTS.}                                         
\normalsize
\rm
We are greatly thankful to D.~Denegri who initiated our interest to study the physics of
\gpj processes, for his permanent support and fruitful suggestions.
It is a pleasure for us to express our deep recognition for helpful discussions to P.~Aurenche,
M.~Dittmar, M.~Fontannaz, J.Ph.~Guillet, M.L.~Mangano, E.~Pilon,
H.~Rohringer, S.~Tapprogge, H.~Weerts and J.~Womersley. Our special gratitude is
to J.~Womersley also for supplying us with the preliminary version of paper [1],
interest in the work and encouragements.

~\\[5mm]


\end{document}